\documentclass[10pt,journal,compsoc]{IEEEtran}

\usepackage{lipsum}

\usepackage{tikz}
\usepackage{pgfplots}
\usepackage{filecontents}
\usepackage{amsmath}
\usepackage{graphicx}
\usepackage{adjustbox}
\usepackage{url}
\usepackage{hyperref}
\usepackage{amsfonts}
\usepackage{color, soul}
\usepackage{mathrsfs}
\usepackage{cleveref}
\usepackage{booktabs}
\usepackage{multirow}
\usepackage{float}
\usepackage{wrapfig}
\usepackage{amsthm}
\usepackage{bm}
\usepackage{bbm}
\usepackage{algorithm}
\usepackage{algorithmic}
\usepackage{colortbl}
\usepackage{enumitem}
\usepackage{color}
\usepackage{balance}
\usepackage{stfloats}
\usepackage{makecell}
\usepackage{multirow}
\usepackage{amsmath}
\usepackage{amsfonts,amssymb}
\usepackage{bbm}

\usepackage{array}
\usepackage[caption=false,font=normalsize,labelfont=sf,textfont=sf]{subfig}
\usepackage{textcomp}
\usepackage{verbatim}
\usepackage{graphicx}
\usepackage{cite}
\usepackage{bibentry}
\usepackage{ulem}

\usepackage{caption2}

\usepackage{tikz}
\usepackage{forest}

\usepackage{xcolor}
\usepackage{hyperref}

\usepackage{ulem}

\usepackage{colortbl}
\usepackage{tabularx}

\usepackage{longtable}
\usepackage{multirow}
\usepackage{makecell}

\definecolor{tablebg_gray}{HTML}{F2F2F2}  
\definecolor{tablealt_gray}{HTML}{E0E0E0}
\definecolor{tablehead_gray}{HTML}{CCCCCC}

\definecolor{mid-green}{RGB}{64,224,208} 
\definecolor{light-green}{RGB}{240,255,240} 
\definecolor{light-orange}{RGB}{255,245,204} 
\definecolor{dark-orange}{RGB}{255,165,0} 
\definecolor{light-blue}{RGB}{224, 255, 255} 
\definecolor{mid-blue}{RGB}{84, 140, 193} 
\definecolor{light-pink}{RGB}{255, 220, 220} 
\definecolor{mid-pink}{RGB}{253, 154, 155} 
\definecolor{light-purple}{RGB}{222, 222, 255} 
\definecolor{mid-purple}{RGB}{204, 153, 204} 

\tikzset{
  forked edges/.style={},
  grey/.style={fill=gray!30},
  extraction/.style={draw=blue, thick},
  inversion/.style={draw=red, thick},
  others/.style={draw=green, thick},
  reliability/.style={draw=purple, thick},
  generalizability/.style={draw=orange, thick},
  extraction-middle/.style={draw=extraction, fill=middle-color!40, text opacity=1, align=center, fill opacity=.5, text=black, font=\scriptsize, inner sep=3pt},
  extraction-leaf/.style={draw=extraction, fill=leaf-color!40, text opacity=1, align=left, fill opacity=.5, text=black, font=\scriptsize, inner sep=3pt},
  inversion-middle/.style={draw=inversion, fill=middle-color!40, text opacity=1, align=center, fill opacity=.5, text=black, font=\scriptsize, inner sep=3pt},
  inversion-leaf/.style={draw=inversion, fill=leaf-color!40, text opacity=1, align=left, fill opacity=.5, text=black, font=\scriptsize, inner sep=3pt},
  others-middle/.style={draw=others, fill=middle-color!40, text opacity=1, align=center, fill opacity=.5, text=black, font=\scriptsize, inner sep=3pt},
  others-leaf/.style={draw=others, fill=leaf-color!40, text opacity=1, align=left, fill opacity=.5, text=black, font=\scriptsize, inner sep=3pt},
  reliability-middle/.style={draw=reliability, fill=middle-color!40, text opacity=1, align=center, fill opacity=.5, text=black, font=\scriptsize, inner sep=3pt},
  reliability-leaf/.style={draw=reliability, fill=leaf-color!40, text opacity=1, align=left, fill opacity=.5, text=black, font=\scriptsize, inner sep=3pt},
  generalizability-middle/.style={draw=generalizability, fill=middle-color!40, text opacity=1, align=center, fill opacity=.5, text=black, font=\scriptsize, inner sep=3pt},
  generalizability-leaf/.style={draw=generalizability, fill=leaf-color!40, text opacity=1, align=left, fill opacity=.5, text=black, font=\scriptsize, inner sep=3pt}
}

\forestset{
  default preamble={
    for tree={
      align=center,
      parent anchor=south,
      child anchor=north,
    }
  },
}

\begin{document}

\title{Intellectual Property in Graph-Based Machine Learning as a Service: Attacks and Defenses}

\author{Lincan Li, Bolin Shen, Chenxi Zhao, Yuxiang Sun, Kaixiang Zhao,\\ Shirui Pan,~\IEEEmembership{Senior Member,~IEEE}, Yushun Dong$^{\dagger}$
\thanks{Lincan Li, Bolin Shen, and Yushun Dong are with Department of Computer Science, Florida State University, Tallahassee, Florida, US. (Emails: ll24bb@fsu.edu, blshen@fsu.edu, yd24f@fsu.edu) Chenxi Zhao is with Northeastern University, Boston, Massachusetts, US. (Email: zhao.chenxi1@northeastern.edu) Yuxiang Sun is with University of Wisconsin-Madison, Madison, Wisconsin, US. (Email: sun469@wisc.edu) Kaixiang Zhao is with University of Notre Dame, South Bend, Indiana, US. (E-mail: kzhao5@nd.edu) Shirui Pan is with the School of Information and Communication Technology, Griffith University, Nathan, QLD 4215, Australia (E-mail: s.pan@griffith.edu.au).}
\thanks{$^{\dagger}$Corresponding author: Yushun Dong (Email: yushun.dong@fsu.edu).}



}

\markboth{IEEE TRANSACTIONS ON KNOWLEDGE AND DATA ENGINEERING}%
{Shell \MakeLowercase{\textit{et al.}}: Bare Advanced Demo of IEEEtran.cls for IEEE Computer Society Journals}

\IEEEtitleabstractindextext{
\begin{abstract}

Graph-structured data, which captures non-Euclidean relationships and interactions between entities, is growing in scale and complexity. As a result, training state-of-the-art graph machine learning (GML) models have become increasingly resource-intensive, turning these proprietary models and data into invaluable Intellectual Property (IP) of their owners. To address the resource-intensive nature of training advanced GML models, graph-based Machine-Learning-as-a-Service (GMLaaS) has emerged as an efficient solution by leveraging third-party cloud services for model development and management. However, the deployment of such models in GMLaaS paradigm also exposes them to potential threats from attackers. Specifically, the APIs within a GMLaaS system provide interfaces for users to interact with the model by sending queries and receiving inference results. Nevertheless, attackers can exploit these exposed APIs to steal functionalities of GML models or sensitive information from the training graph data, posing severe threats to the safety of these proprietary models and data. In response to these challenges, this survey for the first time systematically presents a comprehensive taxonomy of threats and protective strategies at the level of both GML model and graph-structured data. Such a tailored taxonomy facilitates an in-depth understanding of IP protection for graph machine learning. Furthermore, we present a systematic evaluation framework to assess the effectiveness of IP protection methods, introduce a curated set of benchmark datasets across various domains, and discuss their application scopes and future challenges. Finally, we establish an open-sourced versatile library named PyGIP for researchers and practitioners, which evaluates various attack and defense techniques in GMLaaS scenarios and facilitates the implementation of a comprehensive set of benchmark methods. The library resource can be accessed at: \url{https://labrai.github.io/PyGIP}. We believe this survey will play a fundamental role in intellectual property protection for GML and provide practical recipes for the GML community.


\end{abstract}

\begin{IEEEkeywords}
Intellectual Property Protection, Attack and Defense, Privacy Preservation, Graph Machine Learning.
\end{IEEEkeywords}
}

\maketitle

\IEEEdisplaynontitleabstractindextext

%

\section{Introduction} \label{sec1}


In recent years, graph machine learning (GML) algorithms have achieved state-of-the-art performance in a wide range of graph-based predictive tasks, such as link prediction~\cite{luo2024cross,wang2023topological,zhang2023page}, anomaly detection~\cite{xu2022contrastive,han2022adbench,ding2021few}, and spatial-temporal forecasting~\cite{wu2019graph,shao2022pre,bai2020adaptive}. Thanks to these advancements, deep and large-scale GML models are now widely deployed across various industrial applications, including recommendation platforms~\cite{yang2023dgrec,wang2024distribution}, smart healthcare~\cite{sun2020disease,choi2020learning}, and autonomous systems~\cite{mo2023map,liu2023graph}. Nevertheless, despite their performance and broad application, training GML models usually requires significant computing resource investments. More recently, the continuously increasing complexity of GML model architectures and the growing scale of graph data have made it prohibitively challenging for practitioners to train advanced models locally from scratch~\cite{zeng2022gnn,philipp2020machine}. As a consequence, GML models and collected graph datasets have become highly valuable \textit{intellectual property} (IP) of their owners~\cite{peng2023intellectual,he2022protecting,chakraborty2020hardware}. The difficulty in obtaining well-optimized GML models has contributed to the growing popularity of Graph-based Machine-Learning-as-a-Service (GMLaaS)~\cite{10035510}, where a range of graph-based analytical functionalities and services are provided with cloud platforms as the backend. 
Specifically, GMLaaS providers offer clients convenient and on-demand access to pre-built, state-of-the-art GML models, enabling efficient inference without requiring users to build or train models from scratch. Such scalable and cost-effective solutions offered by GMLaaS democratize the access of well-trained GML models for industrial~\cite{hunt2018chiron}, governmental~\cite{zhao2021veriml}, and academic organizations~\cite{tanuwidjaja2020privacy} managing large-scale graph data.

Despite the fundamental advantage of GMLaaS, both the models and graph data in GMLaaS systems have been found to bear critical vulnerabilities~\cite{liang2024model,kesarwani2018model,hu2024learn}. These vulnerabilities make them susceptible to \textit{model-level attacks}, where the model parameters or architectures are targeted; and \textit{data-level attacks}, which involve inferring sensitive attributes of entities or reconstructing private information within the graph data. Consider a model-level attack scenario, where a FinTech company collaborates with a cloud service provider (e.g., AWS~\cite{ravindranathan2024cloud} and Google Cloud~\cite{bhol2024machine}), to deploy a proprietary GML model for financial fraud detection. The graph structure here denotes the relationships between financial accounts, transactions, and users~\cite{duan2024cat}. The GML model analyzes graph-based relationships and detects fraudulent patterns~\cite{li2022internet}. Client companies access the GML model hosted on cloud infrastructures through API calls. However, malicious attackers may also subscribe to the service and submit crafted transaction queries to the API~\cite{an2024finsformer,zhao2025surveya}. By analyzing and exploiting the model predictions, the attacker can reconstruct a piracy model that replicates the behaviors of the target GML model. Leveraging the piracy model and discovered vulnerabilities, the attacker can redistribute the proprietary model, undermining the model owners' competitive advantages in the market and their IP rights~\cite{chen2024scn_gnn}. In addition to model-level threats, data-level attacks also pose significant risks. Suppose a medical institution using a GML model deployed on GMLaaS to analyze patients' electronic healthcare records (EHR) and provide diagnostic reports~\cite{zhao2021exploiting,krall2020gradient}. The EHR data can be constructed as graphs, where the nodes correspond to patients, medical conditions, treatments, and healthcare providers, and the edges represent patient-diagnosis links or provider-treatment associations~\cite{murali2023towards,golmaei2021deepnote}. An attacker can masquerade as a legitimate healthcare provider and submits queries which involve falsified or manipulated patient EHR into the GMLaaS system~\cite{xu2023cover}. By analyzing the model responses to the queries, the attacker is able to infer sensitive information of re-identified patients, such as medical conditions, treatment history, or demographics~\cite{paul2024systematic}. This kind of attacks retrieve confidential medical details, compromising the privacy and IP of the healthcare institution.

\begin{figure*}[!htbp]
\centering
\includegraphics[width=0.90\textwidth]{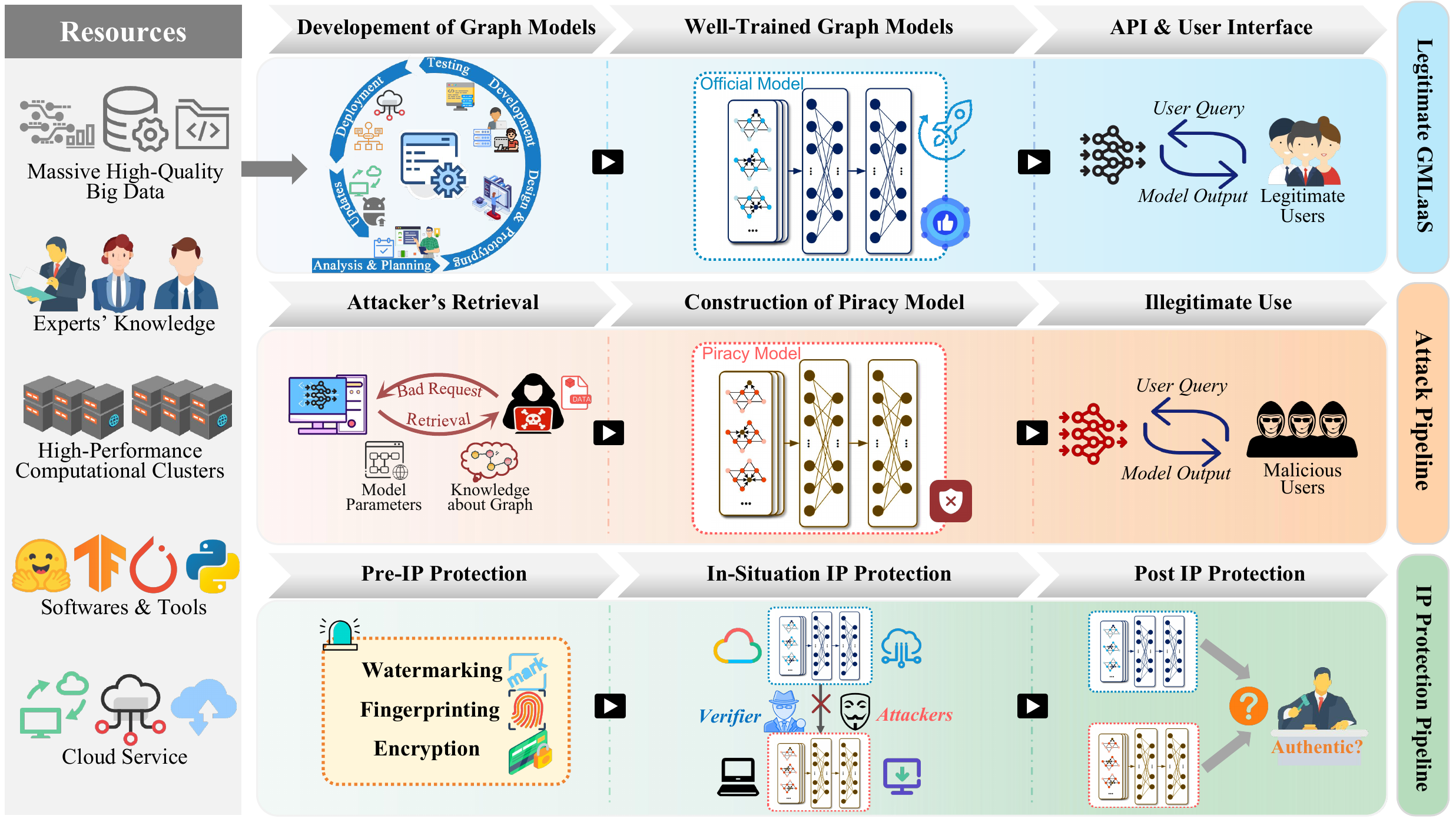}
\caption{The detailed illustration of GMLaaS ecosystem with the safeguard of intellectual property protection.}
\label{fig1}
\end{figure*}


The above-mentioned model-level and data-level threats have arisen in graph machine learning IP protection across various application fields~\cite{oliynyk2023know,gong2020model,kesarwani2018model,zhao2021exploiting}. Therefore, it is imperative to develop techniques for safeguarding the intellectual property of trained models and training data. Fortunately, researchers have contributed to the advancements in both understanding attack techniques and devising defensive approaches. We provide an overview in Fig.~\ref{fig1} to illustrate the basic GMLaaS paradigm and the dataflow of the corresponding attack and defense.

We begin by examining from the attack perspective. GML attacks can be categorized into two different forms, namely model extraction attacks (MEA)~\cite{sun2022adversarial} and model inversion attacks (MInvA)~\cite{fang2024privacy} based on the attackers' goals: MEA focus on retrieving model parameters and creating a piracy model that replicates the target model's performance. MInvA, on the other hand, seek to reconstruct sensitive attributes associated with the labels of the target deep learning model.~\cite{tramer2016stealing} introduced the first MEA on linear machine learning models, reconstructing target models by solving equations derived from queries and confidence values. Subsequent works extended MEA to attack more advanced GML models~\cite{milli2019model,orekondy2019knockoff,9833607,defazio2019adversarial,shen2022model}, with attack schemes tailored for different levels of knowledge assumption~\cite{wu2022model,chen2022knowledge,zhuang2024unveiling}. The concept of MInvA originated in linear regression models~\cite{yin2021comprehensive} and was later extended to neural networks as an optimization task using gradient descent~\cite{fredrikson2015model}. Recent studies have further developed MInvAs for GML~\cite{wang2022group,zhang2022model,zhang2021graphmi,liu2023model,aivodji2019gamin}, exploring black-box scenarios and leveraging generative methods. Meanwhile, it is equally important to explore the strategies designed to defend against these attacks. Inspired by copyright marking in classical digital multimedia~\cite{hartung2000digital}, watermarking and fingerprinting have emerged as key defense techniques. Watermarking~\cite{dai2024pregip,xu2023watermarking,zhao2024transferable} embeds unique identifiers into model parameters or outputs, enabling piracy detection through watermark extraction. In contrast, fingerprinting~\cite{you2024gnnfingers,you2024gnnguard} leverages inherent model properties, such as decision boundaries to generate "fingerprints" without altering the original model. Other defenses include adversarial training~\cite{chen2020smoothing,zheng2024improving}, differential privacy~\cite{mueller2022sok,hsieh2021netfense}, and graph structure perturbations~\cite{hsieh2021netfense,gnnguard}.

Building upon the practice of attacks and defenses, researchers have respectively reviewed privacy preservation methods, attack techniques, and defense mechanisms in GML. These studies provide insights for building robust and secure GML models. In particular,~\cite{dai2024comprehensive,wu2022trustworthy,zhang2022trustworthy} investigated the fairness, explainability, and privacy protection in graph neural networks. However, these works primarily discuss general principles and methodologies for trustworthy GNNs without sufficiently addressing the intellectual property protection issues specific to graph learning. Several studies~\cite{Wu2022_TrustGL,Sun2024_Adversarial} examined recent adversarial attack advancements on GML models. Despite the merits, these papers lack an in-depth analysis from the attack perspective and neglect investigation into corresponding defense methods. Authors in~\cite{zheng2graph,9878092} discussed standard benchmarks and evaluation protocols, but they overlooked threats specific to graph learning IP protection, such as model extraction and unauthorized redistribution, and did not adequately explore IP-targeted defensive strategies. Meanwhile, IP protection has also aroused considerable attention for securing convolutional neural networks (CNNs) and recurrent neural networks (RNNs)~\cite{xue2021intellectual, peng2023intellectual,sun2023deep}. Existing studies have primarily focused on individual tasks of IP protection in graph learning. Nevertheless, a comprehensive review considering multiple perspectives, if not all, is still missing.

To bridge the gap between the fragmented focus of existing studies and the need for a comprehensive understanding of GML intellectual property protection, in this survey, we aim to advance both research methodologies and practical implementations of this field. To start with, we introduce the first taxonomy for GML IP protection, organizing methods into distinct categories based on their mechanisms and scenarios. This taxonomy is accompanied by an in-depth examination of existing techniques, highlighting their strengths, weaknesses, and resistance to various levels of attacks. Next, this survey provides a systematic evaluation framework for analyzing IP protection performance, emphasizing critical aspects such as accuracy, effectiveness, robustness, and uniqueness. We also suggest potential improvements to enhance evaluation transparency and comparability. In addition, we provide a detailed exploration of benchmark datasets, offering guidance on their statistics, characteristics, and application scopes. The studied datasets include citation networks, social networks, chemical structures, protein interactions, e-commerce platforms, traffic networks, and collaborative environments. Finally, the survey identifies open problems in existing studies and suggests future research directions that advance towards more effective and reliable IP protection. To help a wider range of research communities, we develop a continuously updated online library that provides a comprehensive set of attack and defense examples for easily implementing benchmark methods and evaluating the results. Furthermore, we organized most of the existing literature according to the taxonomy, and this resource will be continuously updated once the latest research is released.

The main contributions of this research are as follows:
\begin{itemize}
\item \textbf{The First Comprehensive Taxonomy:} We introduce the first systematic taxonomy of graph machine learning IP protection according to their distinct mechanisms, goals, functions, and scenarios.
\item \textbf{Analysis of Threat Models and Defenses:} We investigate existing threat models and protective methods, emphasizing their advantages, drawbacks, and resilience against different levels of attacks.



\item \textbf{Establishment of GML IP Protection Criteria:} A comprehensive paradigm of IP protection evaluation metrics is presented, establishing criteria that focus on IP protection quality, fidelity, and efficiency.


\item \textbf{Systematic Characterization of Benchmark Datasets:} Existing benchmark datasets on GML IP protection are systematically categorized and introduced; we detail their statistics, features, and application scopes.

\item \textbf{Open-Source Library Tool:} We develop PyGIP\footnote{https://labrai.github.io/PyGIP}\footnote{https://github.com/LabRAI/PyGIP}, a comprehensive open-sourced library to help researchers easily implement and compare these state-of-the-art GML IP protection methods.

\item \textbf{Challenges \& Future Directions:} Discussions about existing challenges and potential solutions are detailed at the end of this work, offering executable insights for future research directions.
\end{itemize}

\noindent \textbf{Difference from Other Related Surveys.} Existing IP protection studies primarily focus on CNNs and RNNs that apply to Euclidean data~\cite{peng2023intellectual,xue2021intellectual}. A notable gap exists in systematic IP protection methodologies for graph machine learning models on non-Euclidean data. Current surveys in GML domain~\cite{dai2024comprehensive,wu2022trustworthy,zhang2022trustworthy,Wu2022_TrustGL,Sun2024_Adversarial,zheng2graph,9878092} have primarily focused on isolated aspects such as privacy preservation, attack techniques, or defense mechanisms. These works usually lack an in-depth analysis of specific threats and prortective approaches to GML IP protection~\cite{wu2022trustworthy,zhang2022trustworthy}. There is a lack of horizontal comparison among different techniques within model-level and data-level attacks, as well as their corresponding defense strategies. Moreover, existing evaluation protocols lack standardized metrics in certain aspects, failing to comprehensively assess the performance of attack and defense under real-world application scenarios~\cite{Wu2022_TrustGL,Sun2024_Adversarial}. Additionally, while some open-source GML datasets have been employed in IP Protection studies, none of the existing works provides a comprehensive and systematic summarization of benchmark datasets~\cite{dai2024comprehensive,ju2024survey}. Our survey uniquely integrates these perspectives by systematically addressing specific GML IP threats, including detailed analysis of model extraction/model inversion/membership inference attacks and targeted defensive strategies such as watermarking, fingerprinting, graph structure perturbation, etc. We introduce standardized evaluation frameworks equipped with tailored criteria that explicitly reflect the realistic attack and defense performance. Additionally, we summarize carefully selected benchmark datasets that align with practical applications, facilitating more applicable evaluations and comparisons of GML IP protection methods.

\noindent \textbf{Intended Audiences.} This survey is intended for a diverse audience including researchers, practitioners, and policymakers who are involved in the fields of deep learning on graphs, data security, intellectual property protection, and privacy preservation. Specifically, it targets academics and industry professionals who seek to understand the vulnerabilities and protection mechanisms associated with graph models and their intellectual property. Additionally, it informs policymakers and legal experts about the potential risks and protective measures for safeguarding intellectual property in the evolving landscape of MLaaS~\cite{guarino2021machine}. 


\noindent \textbf{Survey Structure.} 
The remainder of this article is organized as follows: Section~\ref{sec2} defines the related preliminaries and notations. Section~\ref{sec3} elaborates the taxonomy of graph machine learning IP protection approaches. In Section~\ref{sec4}, we introduce model-level and data-level threats, including the goals of threats, knowledge assumptions, attack methodologies, and detailed performance evaluation paradigms. Subsequently, Section~\ref{sec5} discusses the protective approaches in Graph Machine Learning, including model-level and data-level defense methods. For each category, we introduce representative techniques and their theoretical principles, then detail their performance evaluation metrics. In Section~\ref{sec6}, a wide range of benchmark datasets are summarized and discussed. Following that, we introduce real-world applications of graph machine learning IP protection. Discussions and suggestions about research outlooks are organized in Section~\ref{sec7}. Finally, we conclude this paper in Section~\ref{sec8}.


\section{Preliminaries}\label{sec2}
\subsection{Notations}
\begin{table}[htbp]
\small
\centering
\caption{Common notations and explanations.}
\label{tab1}
\begin{tabularx}{\columnwidth}{c|X}
\toprule
\textbf{\scalebox{1.2}{Notation}}              & \textbf{\scalebox{1.2}{Definitions or Descriptions}}           \\ \midrule
$\mathcal{G}=(\mathcal{V},\mathcal{A},X)$    &\begin{tabular}[c]{@{}l@{}} The whole set of graph data. Subgraphs\\ such as $\mathcal{G}_t$ and $\mathcal{G}_s$ are subsets of $\mathcal{G}$\end{tabular}    \\ \midrule
$\mathcal{G}_t = (\mathcal{V}_t, \mathcal{E}_t, \mathcal{A}_t)$   & \begin{tabular}[c]{@{}l@{}}The target graph data used for\\training the target GML model $f_{\phi}^{*}(\cdot)$\end{tabular} \\ \midrule
$\mathcal{G}_s = (\mathcal{V}_s, \mathcal{E}_s, \mathcal{A}_s)$  & The shadow graph data in Membership Inference Attacks \\ \midrule
$\mathcal{V} \in \mathbb{R}^n$           & The node set of $\mathcal{G}$                                                                                         \\ \midrule
$\mathcal{E}$                            & The edge set of $\mathcal{G}$                                                                                         \\ \midrule
$\mathcal{A} \in \{0,1\}^{n \times n}$   & The adjacency matrix of $\mathcal{G}$                                                                                 \\ \midrule
$\mathcal{N}(\mathcal{V})$          & \begin{tabular}[c]{@{}l@{}}The $k$-hop neighborhood of the node set $\mathcal{V}$\end{tabular}                 \\ \midrule
$f_{\phi}(\cdot)$              & \begin{tabular}[c]{@{}l@{}}A neural network-based graph machine\\ learning model with parameters $\phi$\end{tabular}  \\ \midrule
$X = \{x_i\}_{i=1}^{n}$                  & \begin{tabular}[c]{@{}l@{}}The attributes associated with node set $\mathcal{V}$\end{tabular}                    \\ \midrule
$Y=\{y_i\}_{i=1}^{n}$                & The label set corresponds to node set $\mathcal{V}$                                    \\ \midrule
$\hat{Y}=\{\hat{y}_i\}_{i=1}^{n}$                      & \begin{tabular}[c]{@{}l@{}}The prediction results set generated\\ from $f_{\phi}(v_i)$ for $v_i \in \mathcal{V}$\end{tabular} \\ \midrule
$f_{\phi}(\mathcal{G})$       & The original model (target model) to be attacked.                     \\ \midrule
$f_{\phi}^*(\mathcal{G})$       & The piracy model imitating $f_{\phi}(\mathcal{G})$               \\ \midrule
$\mathcal{S}$                 & \begin{tabular}[c]{@{}l@{}}Secret keys used in watermarking\\ protective methods\end{tabular}                      \\ \midrule
$\tilde{\mathcal{G}}$                       & Perturbed graph                                      \\ \midrule
$\tilde{\mathcal{A}}$                       & \begin{tabular}[c]{@{}l@{}}Reconstructed adjacency matrix \\from an attacker\end{tabular} \\ \midrule
$O_i^{k}$                          & \begin{tabular}[c]{@{}l@{}}The $k$-th fingerprint of node $i$,\\ used for node-level tasks\end{tabular}     \\ \midrule
$O_{ij}^{k}$           &\begin{tabular}[c]{@{}l@{}}The $k$-th fingerprint between node $i,j$,\\ used for edge-level tasks\end{tabular} \\ \midrule
$O_{G_i}^{k}$           &\begin{tabular}[c]{@{}l@{}}The $k$-th fingerprint of graph $\mathcal{G}_i$, used in\\ graph-level tasks\end{tabular}  \\ \midrule
$\mathcal{M}$             & An algorithm used for attack or defense      \\ \bottomrule
\end{tabularx}
\end{table}

The fundamental preliminaries and notations are presented for understanding the concepts in this survey. We begin by defining graph $\mathcal{G} = (\mathcal{V}, \mathcal{A}, X)$, where $\mathcal{V} \in \mathbb{R}^n$ represents the node set, $\mathcal{E}$ is the edge set, and $\mathcal{A} \in \{0,1\}^{n \times n}$ is the adjacency matrix. $\mathcal{N}(\mathcal{V})$ denotes the $k$-hop neighboring nodes of the full node set $\mathcal{V}$. For a graph learning model, $f_{\phi}(\cdot)$ denotes the neural network mapping function, with a learned parameter set $\phi$. We employ $f_{\phi}(\cdot)$ to process the attributes $X$, generating prediction results $\hat{Y}$. Additionally, $Y$ denotes the corresponding label set of $\mathcal{V}$. We further introduce the notations related to graph learning attack and defense. Specifically, $f_{\phi}(\mathcal{G})$ denotes the target model (the original model without modification), whereas $f_{\phi}^{*}(\mathcal{G})$ denotes the piracy model constructed by attackers. $\mathcal{S}$ stands for the secret keys used in watermarking protection. To represent the perturbations, we use $\tilde{\mathcal{G}}$ to indicate a perturbed graph by malicious attackers, and $\tilde{\mathcal{A}}$ to represent the reconstructed adjacency matrix by attackers. For different levels of tasks, $O_i^k$ denotes the $k$-th fingerprint of node $i$ within node-level tasks, whereas $O_{ij}^k$ signifies the $k$-th fingerprint between nodes $i$ and $j$ for edge-level tasks. In graph-level tasks, $O_{\mathcal{G}_i}^k$ refers to the $k$-th fingerprint of graph $\mathcal{G}_i$. Finally, $\mathcal{M}$ denotes a random algorithm employed for attack or defense purposes. We summarize the frequently appeared notations in Table~\ref{tab1} for the convenience of comprehension.


\subsection{Graph Machine Learning}


\noindent \textbf{From GML to Graph Neural Networks.}
Graph machine learning (GML) models are designed to process and analyze graph-structured data. These models aim to extract meaningful patterns by leveraging both the topology and the attributes of the graph. Among various GML models, graph neural networks (GNNs) have emerged as the most representative class that incorporates traditional graph learning approaches with neural network-based architectures. GNNs, including Graph Convolutional Networks (GCN)~\cite{wu2019simplifying}, GraphSAGE~\cite{hamilton2017inductive}, Graph Attention Networks (GAT)~\cite{velivckovic2017graph}, Graph Isomorphism Networks (GIN)~\cite{xu2018powerful}, etc, provide sophisticated mechanism for learning representations from non-Euclidean data. GNNs employ various \textit{information aggregation mechanisms} to capture and propagate structural information across nodes in a graph, enabling the model to learn decent node representations. GCNs use mean aggregation, where each node aggregates the average of its neighbors' feature vectors. GraphSAGE supports multiple options, including mean, LSTM-based, and max pooling aggregation, with more flexibility. GAT adopts an attention mechanism-based aggregation, where the model assigns learnable weights to neighbors based on their relevance. Finally, GINs use sum aggregation which sums up neighboring features to capture unique graph structural patterns. Most existing IP protection research for GMLaaS primarily focuses on graph neural networks (GNNs), with relatively few studies considering other types of GML models.

\noindent \textbf{GML Downstream Tasks.}
Existing graph learning tasks can be categorized into three types: node-level tasks, edge-level tasks, and graph-level tasks. Node-level tasks include node classification~\cite{grover2016node2vec,zhang2019heterogeneous,wang2025cega}, regression~\cite{park2019estimating,jia2020residual}, clustering~\cite{pmlr-v119-bianchi20a,tsitsulin2023graph}, and embedding~\cite{you2019position,chen2020iterative}, where the goal is to predict node labels, make continuous predictions, group nodes into clusters, or map node features into a lower-dimensional space, respectively. Edge-level tasks such as link prediction~\cite{zhang2018link,ying2018graph,wei2023dual} and edge classification~\cite{wu2020comprehensive,li2024drug} focus on predicting the existence of edges or classifying them. Graph-level tasks encompass graph classification~\cite{yang2023dgrec,zhao2021graphsmote}, matching~\cite{sarlin2020superglue,sun2022adversarial}, regression~\cite{hamilton2017representation,dai2024comprehensive}, and generation~\cite{liao2019efficient,you2018graphrnn}, aiming to predict graph labels, measure similarity between graphs, output continuous values, or generate new graph structures. Among these commonly seen graph learning tasks, Node Classification, Node Embedding, Link Prediction, and Graph Classification are the major concern of existing literature on IP Protection for GMLaaS.

\subsection{Overview: Model Extraction Attack \& Defense}
\noindent \textbf{Model Extraction Attacks.} Model extraction attacks (MEA) occur when an attacker attempts to replicate a target model $f_{\phi}(\cdot)$ by querying it and using the resulting output to train a piracy model $f_{\phi}^{*}(\cdot)$ that mimics the target's behavior. In classic MEA, the target model is deployed on a platform that provides predictions in response to user queries, most commonly in the form of prediction probabilities or confidence scores, sometimes as discrete output labels, and less frequently as embedding vectors~\cite{zhao2025survey}. The attacker, who lacks direct access to the model structure, weights, or training data, leverages query-response pairs to reconstruct a replica of the model. With carefully designed query strategies, the attacker issues multiple queries, observes the corresponding outputs, and minimizes the response difference between the piracy model and the target model. The success of MEA is indicated by similar performance between the piracy model and target model on test datasets.

In the context of graph machine learning, MEA involves an attacker aiming to replicate a private GML model $f_{\phi}(\cdot)$ deployed on GMLaaS framework. The GMLaaS server provides client-facing query interfaces that allow clients to input graphs or subgraphs $\mathcal{G}_j = (\mathcal{V}_j,\mathcal{A}_j,X_j)$ and receive the predicted outputs. To perform MEA, the attacker issues multiple queries and analyzes the GMLaaS server's responses to reconstruct a piracy model $f_{\phi}^{*}(\cdot)$ that approximates $f_{\phi}(\cdot)$. Given a limited query budget $Q$, the attacker's goal is to devise an efficient query strategy, collect query-response pairs, and train the piracy model to minimize the discrepancy between its outputs and the target model's responses. Likewise, the successful MEA under the GML setting is indicated by similar performance between the piracy model and the target model on test sets.

\noindent \textbf{Desiderata of Defense.}
Effective defense against model extraction attacks aims to minimize the risk of unauthorized replication while maintaining model utility. Core approaches include Watermarking and Fingerprinting, which embed unique identifiers to track unauthorized use, and Adversarial Training, which introduces resilience against extraction attempts. In Section~\ref{sec5-1}, we will explore these techniques in detail, focusing on how they contribute to the security and robustness of GML models under potential extraction threats.

\subsection{Overview: Model Inversion Attack \& Defense}

\noindent \textbf{Model Inversion Attacks.} Model inversion attacks (MInvA) seek to infer sensitive features of training data by using the outputs of a target model $f_{\phi}(\cdot)$. In classic MInvA scenarios, the attacker aims to reconstruct the feature attributes $X =\{x_1, x_2, \ldots, x_n\}$ that correspond to specific labels by querying the model and obtaining both the predicted labels and the associated confidence scores. The scores reflect the model's certainty on its predictions, thereby providing insight into which feature is likely associated with each label. Leveraging the information, the attacker can infer sensitive attributes by estimating the probability distribution $P(z_i \mid f_{\phi}(x_i))$, where $z_i$ is the sensitive attribute of interest.

Model Inversion Attacks target reconstructing node or graph attributes from the outputs of a trained GML model $f_{\theta}(\cdot)$. Here, graph attributes refer to sensitive information associated with the graph, which usually includes node features and edge features and, in a few cases, even structural properties such as node degrees or connectivity patterns. Given a target model $f_{\theta}(\cdot)$ and an input $x_i$, the attacker queries the model to obtain $f_{\theta}(x_i)$, and then aims to predict sensitive attributes $z_i$ associated with $x_i$. By analyzing multiple query-response pairs, the attacker refines the estimation of $P(z_i \mid f_{\theta}(x_i))$, reconstructing $\hat{z_i}$ to approximate the true attribute $z_i$. The success of the attack is assessed by comparing reconstructed attributes $\hat{z_i}$ against ground-truth attributes on test datasets.

\noindent \textbf{Desiderata of Defense.}
Defending against model inversion attacks encompasses methods that obscure sensitive data without significantly compromising model utility. Key strategies include Differential Privacy to limit exposure of sensitive features, Adversarial Training to increase model robustness, and Graph Structure Perturbation to protect data attributes. These defense approaches will be discussed in Section~\ref{sec5-2} and contribute to the security of GML models against inversion risks.

\subsection{Overview: Membership Inference Attack \& Defense}



\noindent \textbf{Membership Inference Attack.} Membership Inference Attacks (MInfA) aim to determine whether a specific data point (e.g., a node in a graph) was part of the training dataset used to build a target GML model $f_{\phi}(\cdot)$. Given a target graph $\mathcal{G}_t = (\mathcal{V}_t, \mathcal{E}_t, \mathcal{A}_t)$ and a query node $v \in \mathcal{V}$, MInfA seeks to infer whether $v \in \mathcal{V}_t$. The attacker leverages model outputs, such as class posterior probabilities $f^{*}_{\phi}(v)$, to predict the membership status of $v$. 

MInfA is typically formulated as a binary classification task. The attack involves three key phases: (i) \textit{Shadow Model Training}: The attacker constructs a shadow model using a shadow graph $\mathcal{G}_s = (\mathcal{V}_s, \mathcal{E}_s, \mathcal{A}_s)$, which mimics the distribution of the target graph $\mathcal{G}_t$. The shadow model is trained on a known subset $\mathcal{V}_s^{\text{train}} \subseteq \mathcal{V}_s$. (ii) \textit{Attack Model Training}: By querying the shadow model, the attacker collects posterior probabilities for both training nodes $\mathcal{V}_s^{\text{train}}$ and non-training nodes $\mathcal{V}_s^{\text{out}}=\mathcal{V}_s \setminus \mathcal{V}_s^{\text{train}}$. These features, paired with membership labels (1 for $\mathcal{V}_s^{\text{train}}$, 0 for $\mathcal{V}_s^{\text{out}}$), are used to train the attack model. (iii) \textit{Membership Inference}: For a query node $v$, the attacker queries the target model $f_{\phi}(\cdot)$ to obtain $f_{\phi}(v)$ and employs the attack model to predict whether $v \in \mathcal{V}_t$. The attacker's success is evaluated by comparing predictions with ground truth, highlighting the privacy risks of exposing GML model outputs.

\noindent \textbf{Desiderata of Defense.} 
Defending against membership inference attacks focuses on safeguarding the privacy of training data while preserving model utility. Key strategies include Output Perturbation, which introduces noise to model outputs to obscure membership signals, and Query Neighborhood Perturbation, which modifies the structure of queried neighborhoods to reduce information leakage. Additionally, techniques such as Regularization and Differential Privacy mitigate overfitting and limit exposure of sensitive data during training. These defense mechanisms, discussed in Section~\ref{sec5-2}, aim to enhance the privacy and robustness of GML models.

\section{A Taxonomy for IP Protection in GMLaaS}\label{sec3}

\begin{figure*}[ht]
\centering
\hspace*{-1.5cm}
\resizebox{0.8\textwidth}{!}{%
\begin{forest}
  for tree={
    forked edges,
    edge={-, draw=black, line width=1.0pt},
    edge path={
      \noexpand\path[\forestoption{edge}, rounded corners]
      (!u.parent anchor) -- +(-1pt,0pt) -| ([xshift=0pt].child anchor)\forestoption{edge label};
    },
    grow=east,
    reversed,
    anchor=base west,
    parent anchor=east,
    child anchor=west,
    base=middle,
    font=\scriptsize\bfseries,  
    rectangle,
    draw=black,
    line width=1pt,  
    rounded corners=2pt,    
    align=left,
    minimum width=2em,
    s sep=4pt,
    l sep=0.5cm,
    inner sep=1pt,
  },
  where level=1{text width=4.5em}{},
  where level=2{text width=5em}{},
  where level=3{}{},
  where level=4{edge path={\noexpand\path[\forestoption{edge}] (!u.parent anchor) -- (.child anchor)\forestoption{edge label};}}{},
  where level=5{}{},
  [\parbox{9.0em}{\centering\textbf{Graph Learning Intellectual Property Protection}}, fill=light-orange, anchor=north, edge=extraction,
    [\parbox{6.0em}{\centering\textbf{Model-Level Attack}}, fill=light-green, edge={draw=mid-green},
      [Goal of Attacks, fill=light-green, text width=8.4em, edge={draw=mid-green}
        [Replicate Model Functionality\fontsize{4pt}{4pt}{\selectfont{\cite{wu2022model}}}, fill=white, text width=12.5em, edge={draw=mid-green}]
      ]
      [Evaluation Metrics, fill=light-green, text width=8.4em, edge={draw=mid-green}
        [Attack Effectiveness\fontsize{4pt}{4pt}{\selectfont{\cite{sun2022adversarial,zhuang2024unveiling}}}, fill=white, text width=12.5em, edge={draw=mid-green}]
        [Attack Efficiency\fontsize{4pt}{4pt}{\selectfont{\cite{wu2022model,zhuang2024unveiling}}}, fill=white, text width=12.5em, edge={draw=mid-green}]
      ]
      [Representative Methods, fill=light-green, text width=8.4em, edge={draw=mid-green}
        [Random Querying Attack\fontsize{4pt}{4pt}{\selectfont{\cite{guan2024realistic, wu2022model}}}, fill=white, text width=12.5em, edge={draw=mid-green}]
        [Adaptive Querying Attack\fontsize{4pt}{4.2pt}{\selectfont \cite{cheng2025atom,jia2025sigfinger,li2022towards,zhuang2024unveiling}}, fill=white, text width=12.5em, edge={draw=mid-green}]
        [Generator-based Querying Attack\fontsize{4pt}{4.2pt}{\selectfont \cite{chen2025vgfl}}, fill=white, text width=12.5em, edge={draw=mid-green}]
      ]
    ]
    [\parbox{6.0em}{\centering\textbf{Model-Level Defense}}, fill=light-blue, edge={draw=mid-blue}, 
      [Goal of Defenses, fill=light-blue, text width=8.4em, edge={draw=mid-blue}
        [Detect Extraction Attempts\fontsize{4pt}{4pt}{\selectfont{\cite{juuti2019prada}}}, fill=white, text width=12.5em, edge={draw=mid-blue}]
        [${\Downarrow}
$Piracy Model Performance\fontsize{4pt}{4pt}{\selectfont{\cite{guan2024realistic}}}, fill=white, text width=12.5em, edge={draw=mid-blue}]
      ]
      [Representative Methods, fill=light-blue, text width=8.4em, edge={draw=mid-blue}
            [Watermarking\fontsize{4pt}{4pt}{\selectfont{\cite{wang2023making,zhang2024imperceptible,dai2024pregip,zhao2021watermarking,xu2023watermarking,bachina2024genie}}}, fill=white, text width=12.5em, edge={draw=mid-blue}]
            [Fingerprinting\fontsize{4pt}{4pt}{\selectfont{\cite{you2024gnnfingers,JIANG2022109309,you2024gnnguard,waheed2023using,zhang2024survey}}}, fill=white, text width=12.5em, edge={draw=mid-blue}]
            [Adversarial Training\fontsize{4pt}{4pt}{\selectfont{\cite{gosch2024adversarial,wang2019adversarial,chen2020smoothing,kumar2020adversary,liao2020graph,zheng2024improve}}}, fill=white, text width=12.5em, edge={draw=mid-blue}]
      ]
      [Evaluation Metrics, fill=light-blue, text width=8.4em, edge={draw=mid-blue}
        [Fidelity Degradation\fontsize{4pt}{4pt}{\selectfont{\cite{zhuang2024unveiling}}}, fill=white, text width=12.5em, edge={draw=mid-blue}]
        [Extraction Detection\fontsize{4pt}{4pt}{\selectfont{\cite{10646643}}}, fill=white, text width=12.5em, edge={draw=mid-blue}]
        [Discrimination \& Robustness\fontsize{4pt}{4pt}{\selectfont{\cite{ennadir2023unboundattack}}}, fill=white, text width=12.5em, edge={draw=mid-blue}]
      ]
    ]
    [\parbox{6.0em}{\centering\textbf{Data-Level Attack}}, fill=light-pink, edge=extraction, 
      [Goal of Attacks, fill=light-pink, text width=8.4em, edge={draw=mid-pink}
        [Reconstruct Sensitive Attribute\fontsize{4pt}{4pt}{\selectfont{\cite{zhang2022model}}}, fill=white, text width=12.5em, edge={draw=mid-pink}]
        [\scriptsize{Membership Identification}\fontsize{4pt}{4pt}{\selectfont{\cite{wu2021adapting}}}, fill=white, text width=12.5em, edge={draw=mid-pink}]        
      ]
      [Evaluation Metrics, fill=light-pink, text width=8.4em, edge={draw=mid-pink}
        [Attack Effectiveness\fontsize{4pt}{4pt}{\selectfont{\cite{zhang2022model,fang2024privacy}}}, fill=white, text width=12.5em, edge={draw=mid-pink}]
        [Attack Efficiency\fontsize{4pt}{4pt}{\selectfont{\cite{xu2024query,hu2022membership}}}, fill=white, text width=12.5em, edge={draw=mid-pink}]                
      ]
      [Representative Methods, fill=light-pink, text width=8.4em, edge={draw=mid-pink}
        [Node Attribute Attacks\fontsize{4pt}{4pt}{\selectfont \cite{wu2021adapting, he2021node, jnaini2022powerful, niuimproving, conti2022label, olatunji2021membership, yang2023membership, anand2024gradient, xiao2024fedgig, xu2024query}}, fill=white, text width=12.6em, edge={draw=mid-pink}]
        [Graph Structure Attacks\fontsize{4pt}{4.2pt}{\selectfont \cite{song2023gnnbleed, wang2023link, guan2025topology, shaikhelislamov2024study, wang2024subgraph, anand2024gradient, xiao2024fedgig, zhang2021graphmi, zhang2022model, lin2024stealing}}, fill=white, text width=12.5em, edge={draw=mid-pink}]   
      ]
    ]
    [\parbox{6.0em}{\centering\textbf{Data-Level Defense}}, fill=light-purple, edge={draw=mid-purple} 
      [Goal of Defenses, fill=light-purple, text width=8.4em, edge={draw=mid-purple}
        [Impede Attribute Reconstruction\fontsize{4pt}{4.2pt}{\selectfont \cite{zhang2021graphmi}}, fill=white, text width=12.5em, edge={draw=mid-purple}]
        [Prevent Membership Disclosure\fontsize{4pt}{4pt}{\selectfont{\cite{dai2025graph}}}, fill=white, text width=12.5em, edge={draw=mid-purple}]
      ]
      [Representative Methods, fill=light-purple, edge={draw=mid-purple}, text width=8.4em,
        [Differential Privacy\fontsize{4pt}{4pt}{\selectfont{\cite{olatunji2021releasing,xu2023mdp,tran2022heterogeneous,wei2024poincare,hu2024towards,mueller2022sok}}}, fill=white, text width=12.5em, edge={draw=mid-purple}]
        [Adversarial Training\fontsize{4pt}{4pt}{\selectfont{\cite{dai2019adversarial,zhang2021multi,zhou2023strengthening,zhang2021graphmi}}},fill=white, text width=12.5em,edge={draw=mid-purple}]
        [Topology Perturbation\fontsize{4pt}{4.2pt}{\selectfont \cite{jin2020graph,boratto2024robustness,wang2022group,he2021node,cai2024privacy,liu2024revisiting}}, fill=white, text width=12.5em, edge={draw=mid-purple}]
        [Regularization\fontsize{4pt}{4pt}{\selectfont{\cite{wu2021adapting,jnaini2022powerful,conti2022label,yang2023membership,guan2024topology}}},fill=white, text width=12.5em, edge={draw=mid-purple}]
        [Knowledge Distillation\fontsize{4pt}{4pt}{\selectfont{\cite{zheng2021resisting, mazzone2022repeated, tang2022mitigating,chen2024maskarmor}}}, fill=white, text width=12.5em, edge={draw=mid-purple}]
      ]
      [Evaluation Metrics, fill=light-purple, text width=8.4em, edge={draw=mid-purple}
        [Task Utility\fontsize{4pt}{4pt}{\selectfont{\cite{zhou2024model,chen2022graph}}}, fill=white, text width=12.5em, edge={draw=mid-purple}]
        [Privacy Protection Performance\fontsize{4pt}{4pt}{\selectfont{\cite{liu2022membership}}}, fill=white, text width=12.5em, edge={draw=mid-purple}]
        [Resource and Efficiency\fontsize{4pt}{4pt}{\selectfont{\cite{dai2023unified,dai2024comprehensive}}}, fill=white, text width=12.5em, edge={draw=mid-purple}]    
      ]
    ]
]
\end{forest}
}
\caption{Illustration of the Comprehensive Taxonomy of Graph Learning IP Protection Methodologies.}
\vspace{-3mm}
\label{fig:taxonomy}
\end{figure*}
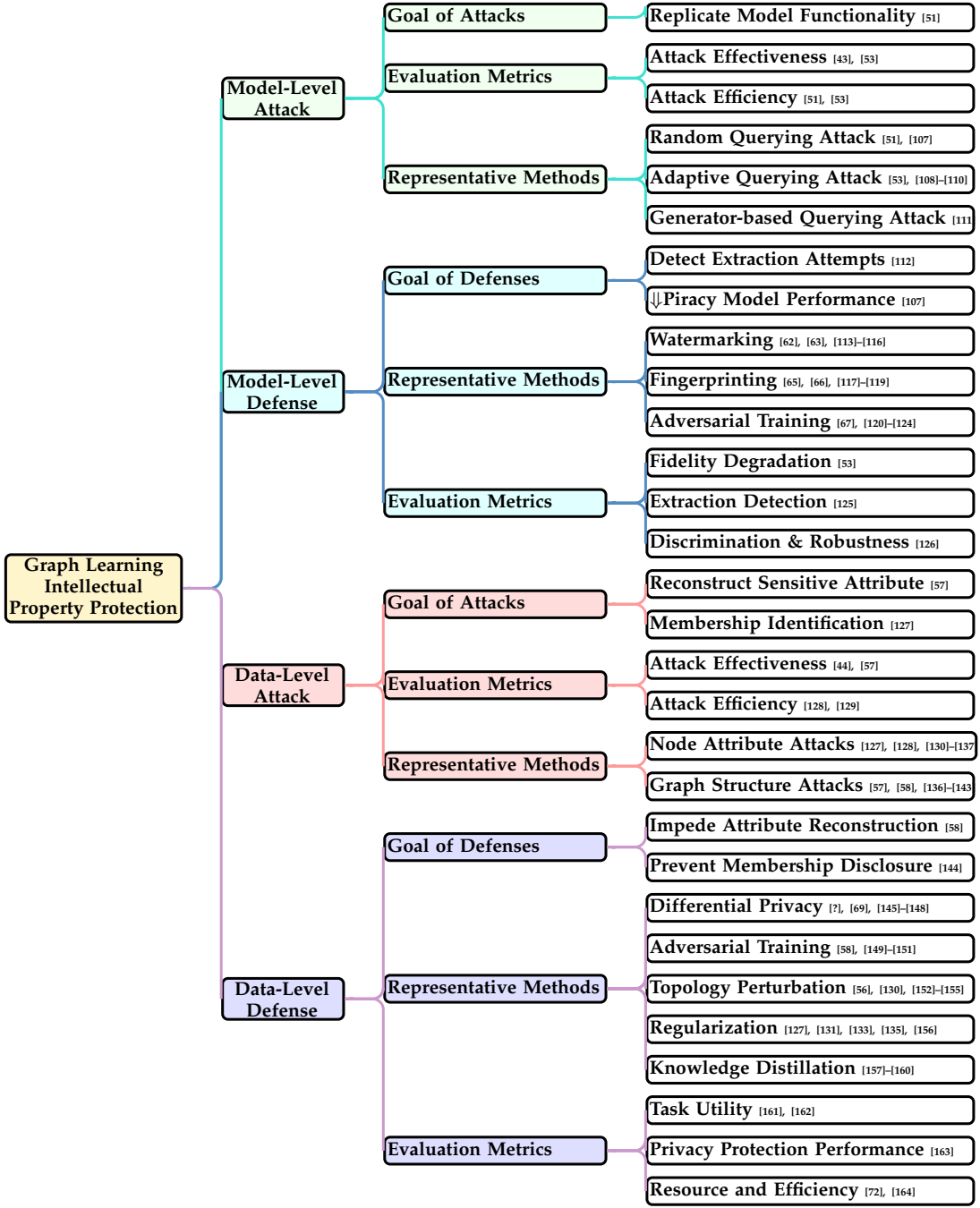

We for the first time present a comprehensive taxonomy for the intellectual property protection in graph learning, as illustrated in Fig.~\ref{fig:taxonomy}. The taxonomy is structured into four primary categories: Model-Level Attacks, Model-Level Defenses, Data-Level Attacks, and Data-Level Defenses, each encompassing various strategies and evaluation metrics.

\textit{Model-level attacks} aim to extract proprietary knowledge by querying the target model and reproducing its behavior. In this survey, we consider the widely adopted black-box setting, where attackers can only access model outputs given their chosen inputs. The primary goals of such attacks include building a piracy model that mimics the functionality and performance of the target model. Evaluation metrics typically include accuracy and fidelity, which measure how closely the stolen model replicates the target model's predictions. To counter these threats, \textit{model-level defenses} have been developed, focusing on tracing unauthorized usage and reducing the utility of pirated models. Representative defense methods include watermarking and fingerprinting, which embed identifiers into model parameters or predictions to enable ownership verification. Adversarial training is also employed to make the model more resilient to extraction attempts. These defenses are evaluated using metrics such as accuracy, attack success rate (ASR), robustness, uniqueness, AUC, and ROC, providing a comprehensive view of their effectiveness under black-box attack scenarios.

\textit{Data-level attacks} exploit the privacy vulnerabilities of the training data used in GML models. The primary objectives of such attacks are to infer sensitive graph attributes or determine whether specific nodes, edges, or subgraphs were part of the training dataset. In this work, we consider only the black-box setting, where the attacker can interact with the model through input-output queries but has no direct access to the training data or model internals. Evaluation metrics for data-level attacks include accuracy, AUC, average number of queries, and average inference time, which reflect both the effectiveness and efficiency of the attacks. Defending against data-level attacks (\textit{data-level defenses}) involves techniques designed to obscure the influence of training data on model outputs. Common strategies include differential privacy, adversarial training, graph topology perturbation, and representation regularization. These methods are evaluated using metrics such as accuracy, F1-score, AUC, attack success rate, and robustness, aiming to preserve model utility while minimizing privacy leakage.

The taxonomy provides a structured framework for understanding the landscape of graph learning IP protection. This systematic approach not only highlights the current state-of-the-art research but also identifies potential areas for future exploration and development.

\section{Threats to GMLaaS: Model-Level and Data-Level Attacks} \label{sec4}

This section thoroughly investigates threats on graph machine learning, focusing on both model-level and data-level attacks. \textit{Model-level attacks}, also known as model extraction attacks, aims to replicate the target model's functionality by extracting its parameters or architecture for unauthorized use. \textit{Data-level attacks} refers to model inversion attacks, seeking to infer sensitive input features by exploiting the model's learned representations. We begin by examining model-level attacks, outlining their goals, underlying knowledge assumptions, and key methodologies. Following that, we present a detailed discussion on data-level attacks, similarly covering their objectives, assumptions, and specific approaches. For both types of attacks, we include an evaluation framework, discussing metrics used to assess the effectiveness and impact of these attack methods.

\subsection{Model-Level Attacks}

\subsubsection{Threat Model and Objective of Model-Level Attacks} \label{sec3-2-1}
The most commonly adopted knowledge assumption of model-level attacks is \underline{black-box}~\cite{guan2024realistic}. Given its practical relevance and prevalence in real-world deployment, \textbf{we focus exclusively on this setting in our discussion}. Black-Box model-level attacks assume that the attacker can only observe the model outputs when providing input queries, with no direct knowledge of model parameters, gradients, or architectures~\cite{zhuang2024unveiling}. This more closely mirrors real-world scenarios, where a GML model is treated as a service deployed on GMLaaS environment, and attackers must infer model behaviors from its responses.

The model owner trains a target model $f_{\phi}(\cdot)$ on graph data $\mathcal{G} = (\mathcal{V}, \mathcal{A},X)$, with the GMLaaS provider offering an API for users. The local attacker queries the target model by submitting unlabeled data through the provided API. The model owner enforces query limits and restricts the response information. The responses to an attacker's query typically fall into two categories~\cite{gong2021model}. The first type is the predicted label, where the target model directly outputs the predicted class label~\cite{tang2024modelguard}. The second type is the prediction confidence, where the target model returns a softmax probability vector or a logit vector, representing the confidence scores associated with each class.~\cite{tang2024modelguard} Using these responses, the attacker trains a piracy model $f_{\phi}^{*}(\cdot)$. The piracy model replicates the target model's functionality, ensuring that $\forall v_i \in \mathcal{V}, f_{\phi}^{*}(x_i) \approx f_{\phi}(x_i)$, where $\mathcal{V}=\{v_i\}_{i=1}^{n}$ denotes the set of nodes, and $x_i$ is the node attributes.

As the most representative type of model-level attacks, MEA involves an attacker seeking to construct a piracy model that closely replicates the performance of the target model. An extraction is considered successful under either of the following conditions: The first is that the piracy model attains comparable accuracy on the test set, i.e., $f_{\phi}^{*}(x_i) = y_i$. The second is that its predictions closely match those of the target model, i.e., $f_{\phi}^{*}(x_i) \approx f_{\phi}(x_i)$. It is crucial to recognize that the architecture and parameters of the piracy model may differ from those of the target model, even if they perform similarly or yield similar outcomes.

\subsubsection{Model-Level Attacks on GML}

In black-box settings, the attacker has no internal access to the target model and can only interact with it via query-based techniques, aiming to reconstruct a piracy model that approximates the behavior and performance of the original one. This scenario reflects realistic threat models, where GMLs are deployed as remote services and their architecture and parameters remain confidential. Although some existing literature explores model-level attacks under gray-box or white-box settings, such assumptions are uncommon and often serve specialized purposes, such as worst-case security analysis~\cite{chen2025towards} or enabling white-box access in order to facilitate query crafting~\cite{yang2024black}. 

\noindent \textbf{Random Querying Attacks}:~\cite{wu2022model} is an early work proposing a systematic framework for model extraction attacks on GML models, where the attacker generates queries based on nodes they control and utilizes the returned predictions to train a surrogate model. The method does not rely on optimized query generation and instead assumes random or fixed selection of available nodes and structures, simulating realistic black-box attack settings. DeFazio et al.~\cite{defazio2019adversarial} similarly initiate queries over randomly selected nodes and exploit the model's output probabilities to iteratively reconstruct the target model. These approaches are simple and broadly applicable but require a large number of queries to achieve high fidelity due to their non-adaptive nature.

\noindent \textbf{Adaptive Querying Attacks}: To overcome the limitations of random query-based methods, researchers have developed adaptive querying approaches that dynamically adjust query generation based on feedback from the target model. GQA~\cite{li2022towards} introduces a query-efficient black-box attack that reduces query overhead by combining contrastive pretraining of piracy model features with a buffer-based reuse of informative queries. To further enhance adaptability, Jia et al.~\cite{jia2025sigfinger} proposed SIGFinger, an interactive query framework that incorporates subgraph sampling and feedback-aware surrogate learning to guide query selection inclined to more informative graph regions. Finally, ATOM~\cite{cheng2025atom} models the temporal dynamics of query sequences and integrates structural signals via $k$-core embeddings, enabling the detection and mitigation of sophisticated adaptive extraction attempts. These methods reflect an evolving trajectory from efficiency-oriented querying to behavior-aware and structure-guided adaptive attacks.

\noindent \textbf{Generator-based Querying Attacks}: Unlike random querying attacks which sample queries arbitrarily, and adaptive querying attacks which select queries based on previous model responses, generator-based querying attacks employ a dedicated generative model to synthesize realistic and diverse queries without relying on any real graph data. While most existing model-level attacks rely on real graph data sampled either randomly or adaptively, StealGNN~\cite{zhuang2024unveiling} introduces a generator-based querying approach, in which a graph generator is trained to synthesize informative synthetic graphs for querying the victim GNN, enabling model extraction in a fully data-free setting. However, StealGNN is primarily tailored for standard graph neural networks and relies on supervised objectives. To further advance this direction, VGFL-SA~\cite{chen2025vgfl} introduces a generator-based approach for vertical graph federated learning settings. This approach integrates a variational graph generator with a feedback-aligned learning mechanism, allowing the attacker to produce synthetic graphs that effectively stimulate the target model's decision boundaries. By jointly optimizing the generator and the piracy model, VGFL-SA ensures that generated queries are diverse and highly informative.~\cite{zhuang2024unveiling, chen2025vgfl} open up new directions for query-efficient black-box attacks in scenarios where no real data is available.

\subsubsection{Evaluation Metrics of Model-Level Attacks}\label{sec3-2-3}


Following the widely-recognized evaluation criteria employed in DNN model-level attacks~\cite{juuti2019prada,zhu2021hermes}, researchers in graph machine learning domain continue the fashion and evaluate attack performance based on two aspects: (I) \textit{Attack Effectiveness} and (II) \textit{Attack Efficiency}~\cite{zhao2025surveya,cheng2025atom}.

\noindent \textbf{(I) Attack Effectiveness Criteria.} In model extraction attacks, attack effectiveness is mainly reflected through the functional and performance similarity between the target model and the piracy model. These metrics are designed to evaluate the degree to which the extracted piracy model replicates the target model's behavior, regardless of the underlying model parameters or architectures. These metrics primarily focus on how well the piracy model can mimic the original model's decision-making on various tasks, thereby reflecting the core objective (functional replication) of model extraction attacks:
\begin{itemize}
\item \underline{Fidelity/Faithfulness}: Measures the proportion of inputs for which the extracted model and the target model produce identical outputs (e.g., classification labels). This metric directly quantifies the behavioral alignment between the two models and is considered a central indicator of extraction success.
\item \underline{Accuracy}: Assesses the extracted model's prediction accuracy on a ground-truth labeled test set. While fidelity captures similarity to the target model, accuracy measures the practical performance of the piracy model.
\item \underline{F1 Score}: Represents the harmonic mean of precision and recall, providing a balanced evaluation of the extracted model's classification performance, particularly in imbalanced class scenarios.

\end{itemize}

\noindent \textbf{(II) Attack Efficiency Criteria.} The evaluation of attack efficiency in model extraction attacks can be further divided into two principal subcategories: \textit{query and computation cost metrics}, which capture the resource expenditure required for the attack, and \textit{robustness \& generalization metrics}, which assess the stability and adaptability of the attack methods under varying conditions. 

\noindent \textit{Query \& Computation Cost Metrics}: Query and computation cost metrics are intended to measure the overall resource consumption incurred by the attacker during the model extraction process. These metrics are particularly important in practical scenarios, where query limitations, financial costs, and computational constraints may significantly affect the implementation and scalability of an attack. The specific query \& computation cost metrics are:

\begin{itemize}
\item \underline{Query Budget}: The total number of queries or API calls made during the attack. This metric directly reflects the economic viability and efficiency of the attack, especially when the model is only accessible through a pay-per-query interface.
\item \underline{Time Cost \& Computational Overhead}: The total time required and computational resources consumed to complete the extraction process. This metric provides insight into the engineering practicality and operational burden associated with different attack techniques.
\end{itemize}

\noindent \textit{Robustness \& Generalization Metrics}: Robustness and generalization metrics are used to evaluate the stability, resilience, and adaptability of model extraction attacks across various conditions and in the presence of defense mechanisms. These metrics help to determine whether an attack method remains effective when the target model employs specific countermeasures or when the attack scenario deviates from the idealized assumptions. The specific metrics of this subcategory include:

\begin{itemize}
\item \underline{Robustness to Defense:} Assesses how well the attack's efficacy is maintained when different defense strategies are used on the target model. This metric is essential for assessing the true threat level posed by model extraction attacks in adversarial settings.
\item \underline{Generalization Ability:} Assesses whether the attack remains effective across different datasets, model architectures, or application scenarios. This metric reflects the adaptability and broader applicability of the extraction strategy beyond a specific experimental setup.
\end{itemize}



\subsection{Data-Level Attacks}


\subsubsection{Threat Model and Objective of Data-Level Attacks} \label{sec3-4-1}

Following the most common practices, we consider data-level attacks under \underline{black-box} settings~\cite{zhang2022model,olatunji2021membership}. Within this setting, we focus on two representative attack types: \textbf{model inversion attacks (MInvA)} and \textbf{membership inference attacks (MInfA)}. Out of the two, membership inference attacks have received more attention than model inversion attacks in graph learning IP protection. Despite the lack of access to internal model parameters or data samples, black-box attackers can still mount effective inference by exploiting statistical patterns in model responses.

In the context of Model Inversion Attacks (MInvA), the attacker aims to reconstruct sensitive properties of the input graph data, such as node attributes and graph structures~\cite{zhang2022model}. Given only the model outputs (e.g., predicted labels or confidence scores), the attacker seeks to reverse-engineer the underlying data distribution that the model used for training. Assuming the targeted graph learning model $f_{\phi}(\cdot)$ is trained on $\mathcal{G}$, the objective is to maximize the similarity between the attacker's reconstruction and the original data. For instance, if the attacker focuses on reconstructing the graph structure represented by $\mathcal{A}$, the attack can be formulated as: 
\begin{equation}
\max \text{Sim}(\tilde{\mathcal{A}}, \mathcal{A}), \quad s.t. \quad \tilde{\mathcal{A}} = \mathcal{K}(X, Y, f_{\phi}(\cdot))
\label{eq:eq6}
\end{equation} 
\noindent where $Y$ denotes the set of labels, $\text{Sim}(\cdot,\cdot)$ is a similarity function, $\tilde{\mathcal{A}}$ is the reconstructed adjacency matrix.

In contrast, Membership Inference Attacks (MInfA) aim to determine whether a specific node, edge, or subgraph in $\mathcal{G} = (\mathcal{V}, \mathcal{A}, X)$ is part of the model's training dataset $\mathcal{G}_{train}$~\cite{wu2021adapting}. The attacker queries the target graph learning model $f_{\phi}(\cdot)$ with a target query $x_i \in \mathcal{V}$ or $e_{ij} \in \mathcal{A}$ and leverages the model outputs $f_{\phi}(x_i)$ or $f_{\phi}(e_{ij})$ to infer the membership status of the queried node, edge, or subgraph in the training dataset. This is formulated as a binary classification task: $f_{cls}(x_i, f_{\phi}(x_i)) \rightarrow \{0, 1\}$, where $1$ indicates $x_i \in \mathcal{V}$ (or $e_{ij} \in \mathcal{A}$), and $0$ otherwise.

With MInvA and MInfA as the mainstream approach, data-level attacks present a multifaceted threat to GML models, exposing sensitive structural and attribute information, as well as training data membership.

\subsubsection{Data-Level Attacks on GML} \label{data-level-attack}
Data-level attacks in graph learning aim to infer sensitive properties or identify memberships of the underlying graph data $\mathcal{G}_s = (\mathcal{V}_s, \mathcal{A}_s, X_s)$, by interacting with the target model. In this work, we consider these attacks exclusively under the \textbf{black-box} setting, where the attacker has no access to the internal parameters or training data of the model. Instead, the attacker can only issue queries to the model to request input instances (e.g., node features or subgraph structures) and observe the corresponding output predictions. Despite this limited access, recent works have demonstrated impressive success in attacking MInvA and MInfA.


\noindent \textbf{Node-Level Membership Inference Attack:} A growing body of work has investigated membership inference attacks that target node attributes. Olatunji et al.~\cite{olatunji2021membership} first formalized node-level MInfA in GML models by adapting the shadow model framework from the classical machine learning paradigm. This approach relies on posterior prediction features (i.e., entropy and confidence scores) obtained from querying the target model to train a binary classifier that predicts membership status. Building on this foundation,~\cite{he2021node} proposed a multi-view attack framework that extracts both attribute-based and prediction-based distances to reveal discrepancies in model behavior between members and non-members. Wu et al.~\cite{jnaini2022powerful} further advanced the attack fidelity by evaluating multiple graph neural network architectures and demonstrated that membership leakage strongly correlates with node attribute uniqueness.~\cite{olatunji2023does} introduced a defense-aware analysis showing that differential training practices can affect the attack success rate and proposed more resilient attack strategies even in noisy or defense-hardened settings. Complementarily, Conti et al.~\cite{conti2022label} explored the label-only setting, where the attacker can only access predicted class labels rather than full probability vectors, and still achieved effective membership inference via decision-boundary sensitivity.

\noindent \textbf{Edge-Level Membership Inference Attack:} Edge-level (Link-level) membership inference attack aims to determine whether a specific edge, representing a link relationship between two nodes, is presented as the training data of a graph machine learning model.~\cite{wang2023link} studied edge-level membership inference attacks in unsupervised graph representation learning. By leveraging only the node embeddings generated by the proposed UGRL model, this work investigated two attack strategies to infer whether a pair of nodes is connected in the original graph. Shaikhelislamov et al.~\cite{shaikhelislamov2024study} conducted a comprehensive empirical analysis of link prediction tasks, revealing graph neural networks trained for structural prediction are inherently susceptible to link-based inference threats. Lately, Wang et al.~\cite{wang2024gcl} examined edge-level MInfA under contrastive learning settings and proposed \textit{GCL-Leak}, which leverages multi-view embeddings generated during training to determine edge membership, showcasing the risks of the widely adopted self-supervised pretraining schemes.

\noindent \textbf{Subgraph-Level Membership Inference Attack:}
This type of membership inference attack aims to determine whether a specific subgraph or structural pattern exists in the training data of a graph machine learning model.~\cite{wang2024subgraph} proposed a novel subgraph-level membership inference attack by designing tailored shadow models and subgraph extraction techniques, uncovering how structural motifs and local connectivity patterns may leak sensitive information in GML models. ~\cite{guan2025topology} focused on topology-driven membership information leakage, demonstrating that certain structural roles or positions in the graph topology can amplify the success of node-level membership inference. Most recently,~\cite{niuimproving} revisited subgraph-structure MInfA. The authors proposed a robust evaluation framework and a novel two-stage defense mechanism that exposes the limitations of existing defense approaches, including differential privacy. Their findings demonstrate that even with advanced protections, sensitive information about subgraph structures remains vulnerable to inference under realistic graph machine learning scenarios.

\noindent \textbf{Graph Structure/Node Attribute Model Inversion Attack:}
GraphMI~\cite{zhang2021graphmi} pioneers model inversion attacks on graph learning models by focusing on reconstructing graph structure, specifically recovering private edges, and reveals the limitations of differential privacy in providing structure-level protection. Building on these findings, Zhang et al.~\cite{zhang2022inference} systematically investigated inference attacks that target graph structure recovery from embeddings, emphasizing the risk of structural information leakage and highlighting the need for more robust defenses. To address the challenge of diverse graph types, Liu et al.~\cite{liu2023model} extended graph structure inversion attacks to both homogeneous and heterogeneous graph neural networks by developing HomoGMI and HeteGMI, thereby enhancing the accuracy of structure reconstruction across a variety of real-world graphs. As research shifted toward federated settings,~\cite{sinha2024gradient} demonstrated that both node features and graph structure can be inverted from shared gradients, exposing the privacy vulnerabilities of federated GML environments and motivating the development of more advanced inversion techniques. In response to the challenges of recovering sparse and discrete graph structures, Xiao et al. introduced FedGIG~\cite{xiao2024fedgig}, a method that focuses on structure inversion in federated learning settings. FedGIG leverages graph-specific constraints to significantly improve the recovery of private edges. More recently, TrendAttack~\cite{zhang2025unlearning} shows that even after graph unlearning methods are applied and edges are intended to be erased, structure-level inversion attacks still present a significant threat. TrendAttack demonstrates that deleted links can still be inferred from black-box graph machine learning APIs, exposing the limitations of current machine unlearning approaches.

\subsubsection{Evaluation Metrics of Data-Level Attacks} \label{sec3-4-3}


Evaluation metrics for data-level attacks can be organized into (I) \textit{Attack Effectiveness}, which quantifies how successfully the attack achieves its privacy-compromising objectives; and (II) \textit{Attack Efficiency}, which measures the actual resource cost and practicality of the attack~\cite{struppek2022, zhang2022model}.

\noindent \textbf{(I) Attack Effectiveness Criteria.} These criteria evaluate how accurately and reliably the attack can infer sensitive information from the target model. Common metrics include Attack Success Rate (ASR)~\cite{wu2021adapting}, which measures the proportion of correctly inferred samples among all queries, and Area Under the ROC Curve (AUC)~\cite{olatunji2021membership}, reflecting the attacker's ability to distinguish members from non-members (in MInfA) or reconstruct true attributes (in MInvA). For regression-style attribute inference, Mean Squared Error (MSE)~\cite{zhou2024model} or Attribute Reconstruction Accuracy are often used~\cite{olatunji2023does}. High values in these metrics indicate a more powerful and successful attack.

\noindent \textbf{(II) Attack Efficiency Criteria.} These criteria assess the practical cost and speed of carrying out the attack, especially under black-box access constraints. Representative criteria include Average Number of Queries (AQ)~\cite{dai2025graph}, which quantifies how many queries the attacker must submit to achieve their objective, and Average Attack Time (AT)~\cite{zhang2022model}, measuring the total runtime or latency of the attack. More efficient attacks require fewer queries and less time, making them more threatening in real-world scenarios.




\section{Protection Approaches in GMLaaS: Model-Level and Data-Level Defenses} \label{sec5}

This section explores protective methods designed to safeguard graph machine learning against various attack vectors, focusing on both model-level and data-level defenses. Model-level defenses are designed to protect against model extraction attacks, aiming to prevent unauthorized replication of the target model’s functionality. Data-level defenses, on the other hand, focus on mitigating model inversion attacks that attempt to infer sensitive input data from the model. We begin by discussing model-level defense approaches, including watermarking, fingerprinting, and adversarial training. Following that, we delve into data-level defense approaches, including differential privacy, adversarial training, graph structure perturbations, regularization techniques, and knowledge distillation. For each defense type, we also introduce relevant evaluation metrics to assess its effectiveness in mitigating threats.


\subsection{Model-Level Protective Approaches} \label{sec5-1}

\subsubsection{Defense with Watermarking}



\noindent \textbf{Goals of Watermarking.} Watermarking is an effective technique for protecting the intellectual property (IP) of general graph learning models. This method involves embedding distinctive identifiers, referred to as ``watermarks'', into the model during its training phase~\cite{zhang2024imperceptible}. These identifiers serve as proof of ownership and can be used to identify unauthorized copies or modifications of the model~\cite{dai2024pregip}. Specifically, a piracy model, $\mathcal{M}_p$, is an unauthorized copy or fine-tuned version of a protected model $\mathcal{M}$, exploited by adversaries for their specific purposes. The goal of watermarking is defined as ensuring that for a protected model $\mathcal{M}$, given a secret key $\mathcal{S}$ and IP message $\mathcal{I}$, the piracy model $\mathcal{M}_p$ should reproduce $\mathcal{I}$ with high probability, while an independent model $\mathcal{M}'$ can not achieve this:
\begin{equation}
\left\{
\begin{aligned}
P(\mathcal{I} = \text{ExtractIP}(\mathcal{M}_p, \mathcal{S})) &\approx 1 \\
P(\mathcal{I} = \text{ExtractIP}(\mathcal{M}', \mathcal{S})) &\approx 0
\end{aligned}
\right.
\label{eq-GW}
\end{equation}

\begin{figure}[!htbp]
\centering
\includegraphics[width=0.99\columnwidth]{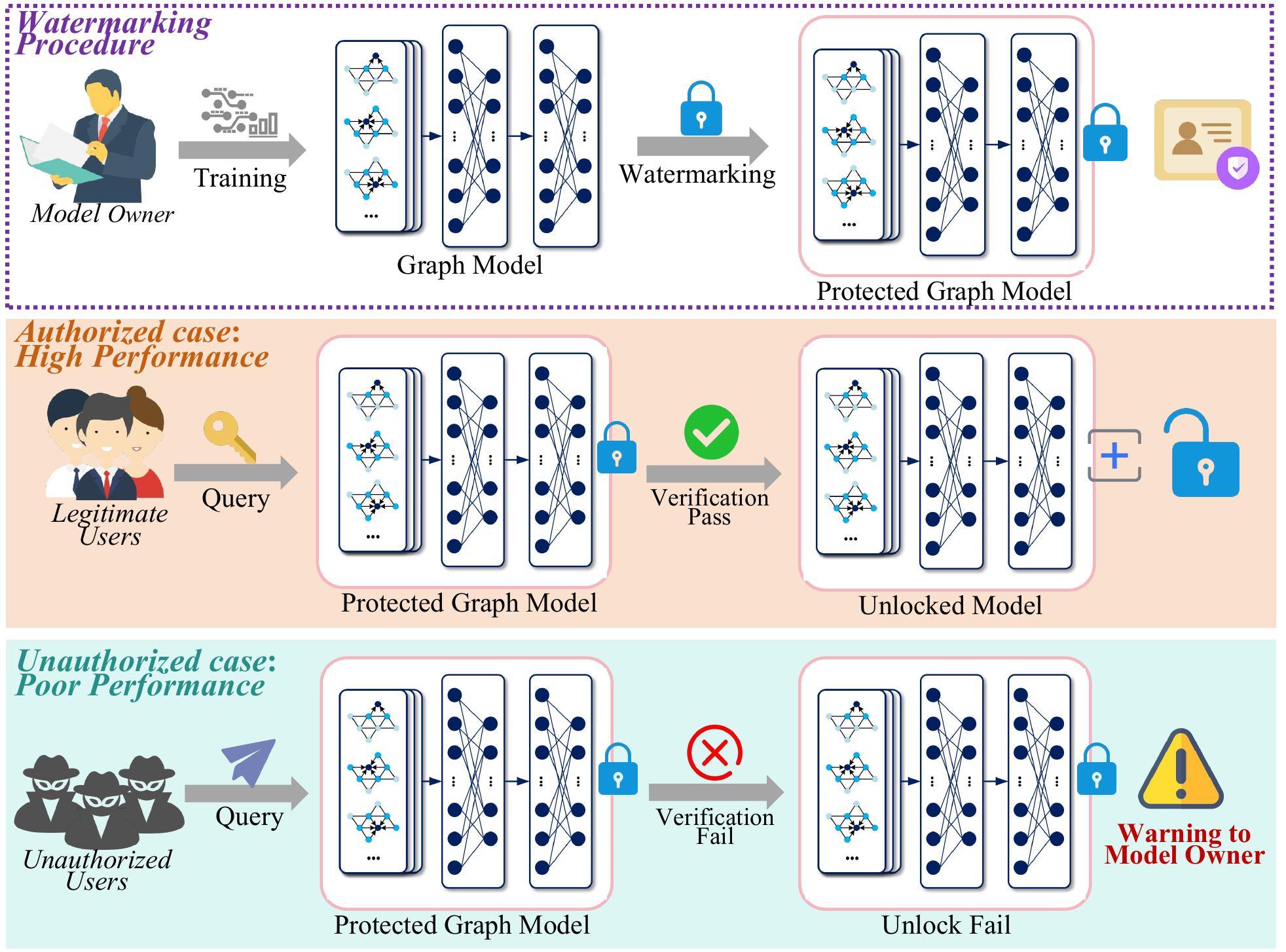}
\caption{Overview of watermarking verification workflow.}
\label{fig2}
\end{figure}

\noindent \textbf{Procedure of Watermarking.} As shown in Fig.~\ref{fig2}, the process of watermarking in GML models typically consists of two stages: watermark embedding and watermark verification~\cite{bachina2024genie}. In the first stage, secret keys $\mathcal{S}$ and corresponding IP messages $\mathcal{I}$ are generated and injected into the model $f_{\phi}(\cdot)$. These secret keys and IP messages can take the form of special input samples ${X}_w$ and their corresponding model predictions ${Y}_w$. Specifically, during the watermark embedding phase, the objective is to minimize a joint loss that combines the model's main task loss $\mathcal{L}_{\text{task}}$ and the watermark embedding loss $\mathcal{L}_{\text{wm}}$~\cite{wang2023making}:
\begin{equation}
\min_{\phi} \mathcal{L}_{\text{total}} = \mathcal{L}_{\text{task}}(f_{\phi},{X},{Y}) + \lambda \mathcal{L}_{\text{wm}}(f_{\phi},{X}_w,{Y}_w)
\label{eq-PW1}
\end{equation}
\noindent where $\lambda$ balances the trade-off between the task performance and the watermark embedding. 

In the second stage of watermarking verification, the model owner extracts the IP message $\mathcal{I}$ using the secret keys through the following process~\cite{bachina2024genie}:
\begin{equation}
\mathcal{I} = \text{ExtractIP}(f_{\phi}(\cdot), \mathcal{S})
\label{eq-PW2}
\end{equation}
If the extracted IP message matches the original, it confirms the ownership of the model.

\noindent \textbf{Representative Methods.} Watermarking is a popular defensive approach in graph learning IP protection, particularly against model-level attacks. Based on the unique characteristics of these techniques, watermarking can be categorized into Static/Dynamic watermarking. \textit{Static watermarking} embeds the watermark into the model in a way that remains unchanged over time, while \textit{Dynamic watermarking} embeds watermarks that are only revealed when triggered by specific inputs, making the watermark dependent on the model behavior. 

\textit{Static Watermarking Methods}~\cite{wang2023making,zhang2024imperceptible,dai2024pregip} embed fixed watermarks that remain unchanged throughout the model's lifecycle.~\cite{wang2023making} introduces a static watermarking technique. The method embeds watermarks into graph learning models that are resistant to extraction by merging the distributions of the normal tasks and watermarks using a soft nearest neighbor loss. This technique ensures that the watermark cannot be easily removed through querying, providing robustness against extraction attacks while maintaining model performance. Zhang et al.~\cite{zhang2024imperceptible} further proposed an imperceptible and owner-unique static watermarking method that improves resilience against a wider range of attacks, including fine-tuning, pruning, and evasion attacks. By utilizing a bi-level optimization framework,~\cite{zhang2024imperceptible} enhances the imperceptibility and uniqueness of the watermark, effectively verifying model ownership with minimal performance degradation. This approach improves on \cite{wang2023making} by targeting additional vulnerabilities, such as evasion attacks, that were not fully addressed in the previous work. Lastly, PreGIP~\cite{dai2024pregip} enhanced the robustness of watermarking by targeting on pre-trained GNNs rather than watermarking on classifiers. While previous methods~\cite{wang2023making,zhang2024imperceptible} primarily focus on task-specific watermarking, PreGIP embeds watermarks in the embedding space of pre-trained GNN encoders, making it applicable to self-supervised pretraining. The method broadens the scope of watermarking to protect a wider range of models, particularly those used in multiple downstream tasks, ensuring robustness not only against model extraction but also against ownership piracy.

In contrast to static approaches, \textit{Dynamic Watermarking}~\cite{zhao2021watermarking,xu2023watermarking,bachina2024genie} techniques embed watermarks that are triggered by specific inputs.~\cite{zhao2021watermarking} introduces a black-box dynamic watermarking method using random Erdos-Renyi (ER) graphs~\cite{gilbert1959random} as triggers, where the watermark is reconstructed based on the output of the GNN when presented with a random graph. This method is robust against model compression and fine-tuning and does not affect the model’s original task performance. However, it is vulnerable to sophisticated adaptive attacks, as the randomness of the ER graph can sometimes allow for indirect watermark removal. Xu et al.~\cite{xu2023watermarking} introduce a more sophisticate dynamic watermarking based on backdoor attacks, enabling watermark embedding for both graph and node classification tasks. Unlike~\cite{zhao2021watermarking}, Xu et al. introduce adaptive defenses that make the watermarking more resilient against attacks where the adversary has partial knowledge of the watermark. This ensures higher ownership verification probabilities while maintaining robustness against model extraction techniques. In terms of more task-specific watermarking, GENIE~\cite{bachina2024genie} extends dynamic watermarking to link prediction task that is untouched by previous works. GENIE integrates a backdoor attack-based dynamic watermarking scheme that secures graph learning models for link prediction. GENIE shows enhanced resilience against a variety of state-of-the-art watermark removal techniques and attacks, outperforming previous works in terms of robustness against ownership piracy attacks.

\subsubsection{Defense with Fingerprinting}

\begin{figure*}[!htbp]
\centering
\includegraphics[width=0.90\textwidth]{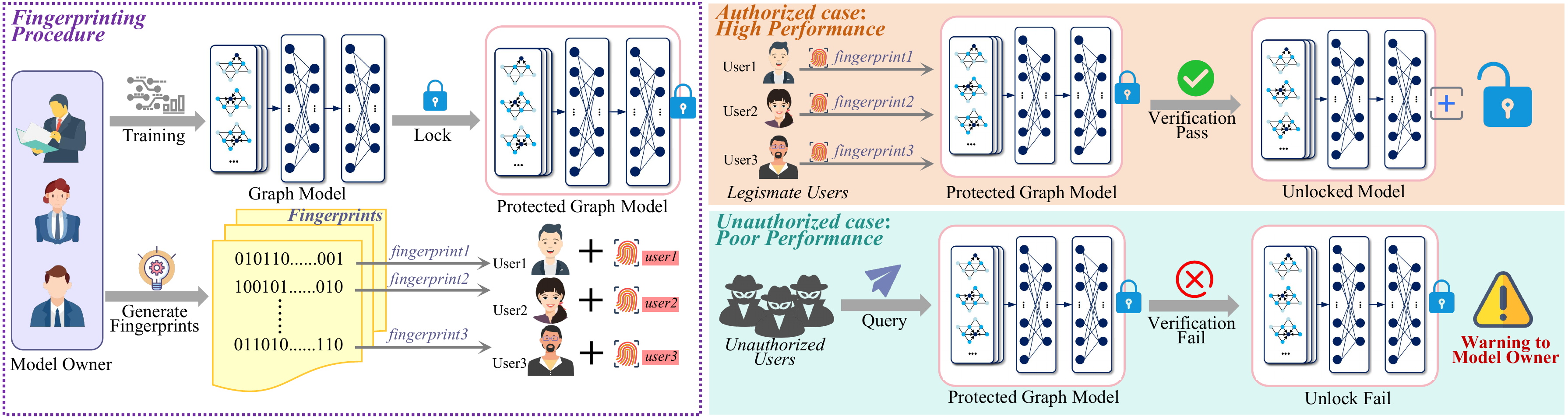}
\caption{The workflow of fingerprinting, including active authorization control and users' identities verification. Each legitimate user is assigned a distinct fingerprint that will be verified in the query step.}
\label{fig3}
\end{figure*}

\noindent \textbf{Goals of Fingerprinting.}
Fingerprinting is operated by embedding unique, verifiable patterns into the model without affecting the performance of the target model~\cite{you2024gnnguard}. It involves several key steps and components, as shown in Fig.~\ref{fig3}. The primary goal of Fingerprinting is to guarantee that a GML model $f_{\phi}(\cdot)$ produces distinct output patterns (or ``fingerprints'') when queried with specific inputs, which can then be used for ownership verification~\cite{jia2025sigfinger}. Formally, for a set of input-output pairs $\{(x_i, y_i)\}$ used for creating a fingerprint, the objective is to ensure that a suspect model $\hat{f_{\phi}}$ produces the same outputs if and only if it is a derived or copied version of the original model $f_{\phi}$, which can be formulated as:
\vspace{-1mm}
\begin{equation}
\hspace{-3mm}
P(\text{Verify}(f_{\phi}, \hat{f_{\phi}}, \{(x_i, y_i)\})) \approx 
\begin{cases} 
1, & \text{if} \ \hat{f_{\phi}} = f_{\phi} \\
0, & \text{otherwise}
\end{cases}
\label{eq-Fg1}
\end{equation}
\noindent The equation make sure that only the rightful model reproduces the correct fingerprinted outputs.

\noindent \textbf{Procedure of Fingerprinting.}
Fingerprinting consists of two stages: fingerprint generation and fingerprint verification~\cite{10646643}. In fingerprint generation stage, a set of unique input-output pairs, denoted as $\{(x_i, y_i)\}_{i=1}^n$, is embedded into the model $f_{\phi}$. These pairs are chosen such that for specific inputs $x_i$, the model produces unique outputs $y_i$.


In verification stage, ownership is verified by querying the model $f_{\phi}$ with the fingerprinted inputs $\{x_i\}$ and checking whether the outputs match the corresponding expected values $\{y_i\}$. The verification process can be expressed as:
\vspace{-1mm}
\begin{equation}
\hspace{-3mm}
\text{Verify}(f_{\phi}) = 
\begin{cases} 
\text{True}, & \text{if} \ f_{\phi}(x_i) = y_i, \forall(x_i, y_i) \\
\text{False}, & \text{otherwise}
\end{cases}
\label{eq-Fg2}
\end{equation}

\noindent This ensures that only models that have embedded the specific fingerprint during training can reproduce the correct outputs during verification.

\noindent \textbf{Representative Methods.} Similar to Watermarking, Fingerprinting can also be characterized as static/dynamic approaches. Static Fingerprinting embeds the fingerprint in a way that remains unchanged over time, ensuring its persistence even through common model modifications. Conversely, Dynamic Fingerprinting embeds fingerprints that are only revealed under specific circumstances or inputs, making the fingerprint more resilient to tampering by adversaries who try to avoid detection.

Waheed et al. introduces GrOVe~\cite{10646643}, a \textit{white-box static fingerprinting} method for GML model ownership verification. GrOVe extracts fingerprints from the embeddings learned by the GNN’s internal structure and verifies ownership by comparing the fingerprints of the target and suspect models. This method is robust in distinguishing independently trained models from surrogate models, even when both models use the same architecture and training dataset. However, GrOVe relies on white-box access to the model, which may not be feasible in scenarios where internal access to the model parameters is restricted. GrOVe provides certain-level of protection against extraction attacks, but it does not address dynamic fingerprinting, where model behavior under specific inputs could be exploited. 

To overcome the limitations of white-box methods, Wu et al.~\cite{10646777} proposes a \textit{black-box dynamic fingerprinting} approach in the context of GMLaaS. It includes a query-based integrity verification scheme where random nodes are fingerprinted and queried to detect potential model-centric attacks. This method has the advantage of working in scenarios where only black-box access to the model is available, making it more practical for real-world deployments. Additionally, the randomized nature of the fingerprint mechanism makes it resilient against attackers who are aware of the fingerprinting scheme. Compared to GrOVe~\cite{10646643}, this approach introduces dynamic fingerprinting, where ownership verification is based on the model response to specific queries rather than static internal features. The characteristic makes it more adaptive and robust against various adversarial attacks. 

GNNFingers~\cite{you2024gnnfingers} further present a \textit{dynamic black-box fingerprinting} scheme that enhances both the fingerprinting construction and verification phases. The fingerprints of GNNFingers are constructed based on a variety of graph-related tasks, such as graph classification, graph matching, and node classification. GNNFingers enjoys robust verification module that allows ownership verification across multiple graph neural network architectures and tasks. It achieves 100\% true positives and true negatives on the test of 200 suspect GNNs. GNNFingers extends query-based dynamic fingerprinting to various GNN models, and ensures the fingerprints remain effective even after post-processing or fine-tuning of the suspect models. The method offers stronger resilience against model post-processing and a broader application scope across graph learning.

\subsubsection{Defense with Adversarial Training}

Aside from watermarking and fingerprinting, adversarial training scheme~\cite{gosch2024adversarial,wang2019adversarial,chen2020smoothing,kumar2020adversary,liao2020graph,zheng2024improve} is another significant approach in graph learning IP protection. 

\begin{figure}[htbp]
\centering
\includegraphics[width=0.98\columnwidth]{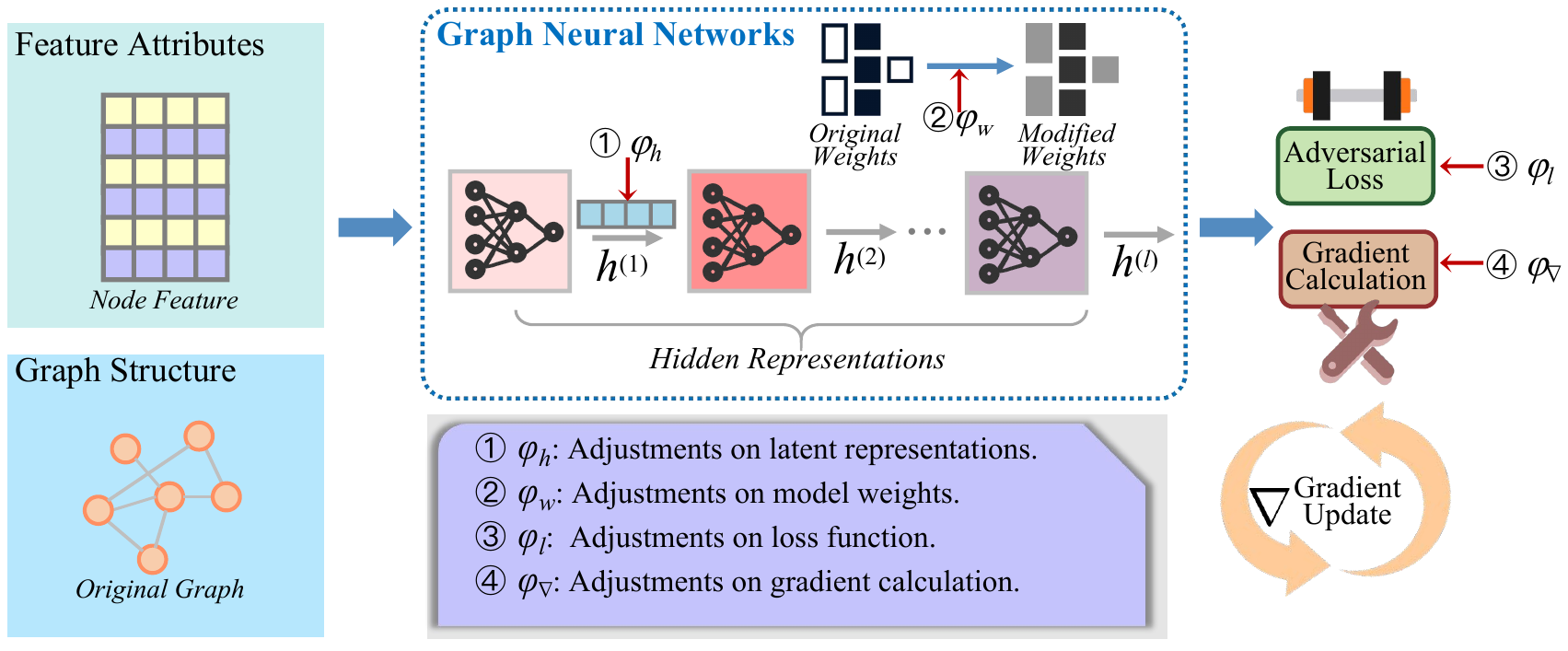}
\caption{The paradigm of model-level adversarial training defensive method.}
\label{fig6-1}
\end{figure}

\noindent \textbf{Goals of Model-Level Adversarial Training.}
This schema aims to enhance the robustness of graph learning models against various model-level attacks, where attackers attempt to replicate the model through extensive querying. As illustrated in Fig.~\ref{fig6-1}, model-level adversarial training involves strategies such as adjusting the latent representations~\cite{Xue2021CAP}, modifying model weights~\cite{Xue2022Adversarial}, constructing adversarial loss functions~\cite{Gosch2023Adversarial}, and refining gradient calculations~\cite{chen2020smoothing}. These elaborately designed adversarial adjustments during training phase make the model resilient to such malicious efforts, thereby safeguarding proprietary algorithms and data~\cite{gosch2024adversarial}. Moreover, adversarial training strives to maintain accuracy and performance in legitimate applications while complicating the reverse engineering process for potential attackers. 

\noindent \textbf{Procedure of Model-Level Adversarial Training.}
Assume we are given a graph $\mathcal{G}=(\mathcal{V},\mathcal{A},X)$ that consists of a total number of $n$ nodes, of which $c$ are labeled. Model-Level Adversarial Training executes by modifying the GML model parameters, gradients, latent representations, etc; and the objective function can be defined as:
\begin{equation}
\arg \underset{\phi}{\min} \max_{\delta_{\phi} \in \mathcal{P}(\delta)} \sum_{i=1}^{c} \mathcal{L}(f_{\phi + \delta_{\phi}}(\mathcal{G})_i, y_i)
\label{eq:eq3-3-3}
\end{equation}
\noindent where $\phi$ denotes model parameters, which are being optimized during training. $\delta_{\phi}$ denotes the adversarial perturbation on model parameters. $\delta_{\phi}$ simulates malicious modifications to model parameters. $\mathcal{P}(\delta)$ refers to the set of allowed perturbations on model parameters, defining constraints on the size and scope of $\delta_{\phi}$ to ensure perturbations remain within certain limits. $f_{\phi +\delta_{\phi}}(\mathcal{G})_i$ indicates model prediction for node $i$ using the perturbed parameters $\phi+\delta_{\phi}$ on original graph $\mathcal{G}$. $\mathcal{L}$ is the loss function, $y_i$ is the ground truth label for node $i$, and $c$ is the total number of labeled nodes.

This approach ensures GML models maintain high performance and robustness, even in the presence of adversarial attacks targeting model training process or gradients. 

\noindent \textbf{Representative Methods.} Adversarial training-based defenses against model-level attacks can be categorized into (a) model weight modification-based, (b) model gradient modification-based, and (c) objective function modification-based methods, depending on their characteristics.

The first set of methods \textit{modifies model weights} to defend against attacks. Wang et al.~\cite{wang2019adversarial} proposed DefNet, an adversarial defense framework that investigates vulnerabilities in different layers of GNNs. DefNet focuses on dual-stage aggregation and bottleneck perceptron layers to strengthen the robustness against various model-level attacks. It also introduces adversarial contrastive learning to improve performance under limited training data. Although showing satisfactory performance, DefNet primarily tackles vulnerabilities in model architecture without addressing more sophisticated attack types like targeted node attacks or structure perturbations. MO-GAA~\cite{zheng2024improve} extends the idea of weight modification in DefNet with a two-step framework for GNN anomaly detection. The method employs multiple-objective generative adversarial attack to disrupt both node features and graph structures, simulating an adversary's behavior. Then in the second stage, Purification-Based Adversarial Attack Defense (PB-AAD) cleanses the graph by correcting adversarial disruptions. The method enhances the robustness of GNN-based anomaly detectors, offering an improvement over DefNet by focusing on both features and structure to safeguard GML models against more severe poisoning attacks.

\textit{Model gradient modification-based} methods addresses the weakness of weight-based modification techniques to enhance the robustness of adversarial training. Smoothing Adversarial Training (SAT)~\cite{chen2020smoothing} tackles model-level attacks by smoothing the gradients of GML models. SAT proposes smoothing distillation and smoothing cross-entropy loss, which work to reduce the magnitude of adversarial gradients, helping to mitigate the effects of both global and targeted node attacks. While SAT specifically addresses the targeted attacks at the gradient level, it fails to directly address malicious attacks targeting on the graph structure itself. Gosch et al.~\cite{gosch2024adversarial} develop a more advanced adversarial training method that not only smooths gradients but also addresses graph structure perturbations. The method reveals how learnable graph diffusion can adjust to adversarial attacks by modifying the gradient flow and adapting the message-passing scheme. By introducing an attack that handles both global and local (node-level) constraints, this method outperforms earlier gradient-based defenses, especially in handling attacks on graph structure.

Finally, methods based on \textit{modifying model training objective functions} often incorporate game-theoretic or MinMax approaches to counter adversaries. Liao et al.~\cite{liao2020graph} present Graph Adversarial Networks, where a MinMax game is played between the GNN encoder and a worst-case attacker. This approach ensures the GNN encoder can protect sensitive node-level information even in face of adversarial challenges. Graph Adversarial Networks outperform previous methods that only focus on gradients or weights, as they incorporate a more sophisticated MinMax objective to handle adversaries, providing a more comprehensive defense against information leakage. Kumar et al.~\cite{kumar2020adversary} expand the concept of adversarial training for privacy protection with their framework that aims to protect familial privacy on social networks. The method introduces an adversarial objective that perturbs both node features and edges in social networks to prevent attackers from identifying family relationships. This framework emphasizes ethical considerations by using adversarial training not just for defense but also for privacy protection in social settings.

\subsubsection{Evaluation Metrics of Model-Level Defenses} \label{sec3-3-4}


The most commonly adopted evaluation criteria for model-level defenses can be categorized into three types: (I) \textit{Fidelity Degradation}, (II) \textit{Extraction Detection}, (III) \textit{Discrimination and Robustness}~\cite{zhao2025systematic,cheng2025misleader}.

\noindent \textbf{(I) Fidelity Degradation Metrics.} These metrics assess how effectively a defense mechanism reduces the ability of an extracted (piracy) model to replicate the target model's predictions or performance~\cite{bachina2024genie}. Typical indicators include the decrease in test accuracy, drop in fidelity scores (e.g., output similarity), or increased prediction divergence between the target and extracted models~\cite{podhajski2024efficient}. Greater reductions shown in these metrics of the piracy model indicate more successful defenses in preventing model functionality duplication.

\noindent \textbf{(II) Extraction Detection Metrics.} This category of metrics evaluates the defense mechanism's ability to detect and flag unauthorized extraction activities~\cite{cheng2025atom}. Common measures include True Positive Rate (i.e., correct identification of extraction attempts) and False Positive Rate (i.e., misidentification of benign users as attackers), as well as overall detection accuracy~\cite{xu2025adage}. An effective defense mechanism should maximize the detection of extraction attempts while minimizing false alarms.

\noindent \textbf{(III) Discrimination and Robustness Metrics.} This type of metric measures the defense mechanism's ability to distinguish between legitimate and pirated models and to remain robust under adversarial conditions~\cite{10646643}. Representative metrics include Area Under the ROC Curve (AUC) and the Receiver Operating Characteristic (ROC), which summarize the trade-off between detection sensitivity and specificity across thresholds. High AUC and robustness scores reflect reliable identification of piracy attempts and strong resistance against adaptive extraction strategies.

\subsection{Data-Level Protective Approaches} \label{sec5-2}

\subsubsection{Differential Privacy-based Defense}

\noindent \textbf{Goals of Differential Privacy.} To protect graph learning models from data-level attacks, Differential Privacy (DP) is employed as an effective defense mechanism. DP aims to ensure that the inclusion or exclusion of a single individual's data does not significantly affect the outcome of any analysis, thereby protecting sensitive information~\cite{olatunji2021releasing}. In essence, DP-based defenses in graph learning provide a robust framework to prevent adversaries from accurately reconstructing the original data from the model, thus safeguarding intellectual property against attacks~\cite{bhaila2024local}. DP can be applied at different levels, such as node-level or edge-level, to obscure the contributions of individual nodes or edges while preserving the overall utility of the data. This is particularly important in GML models where sensitive information can be inferred from the graph's structure and node relationships.

\begin{figure}[htbp]
\centering
\includegraphics[width=0.90\columnwidth]{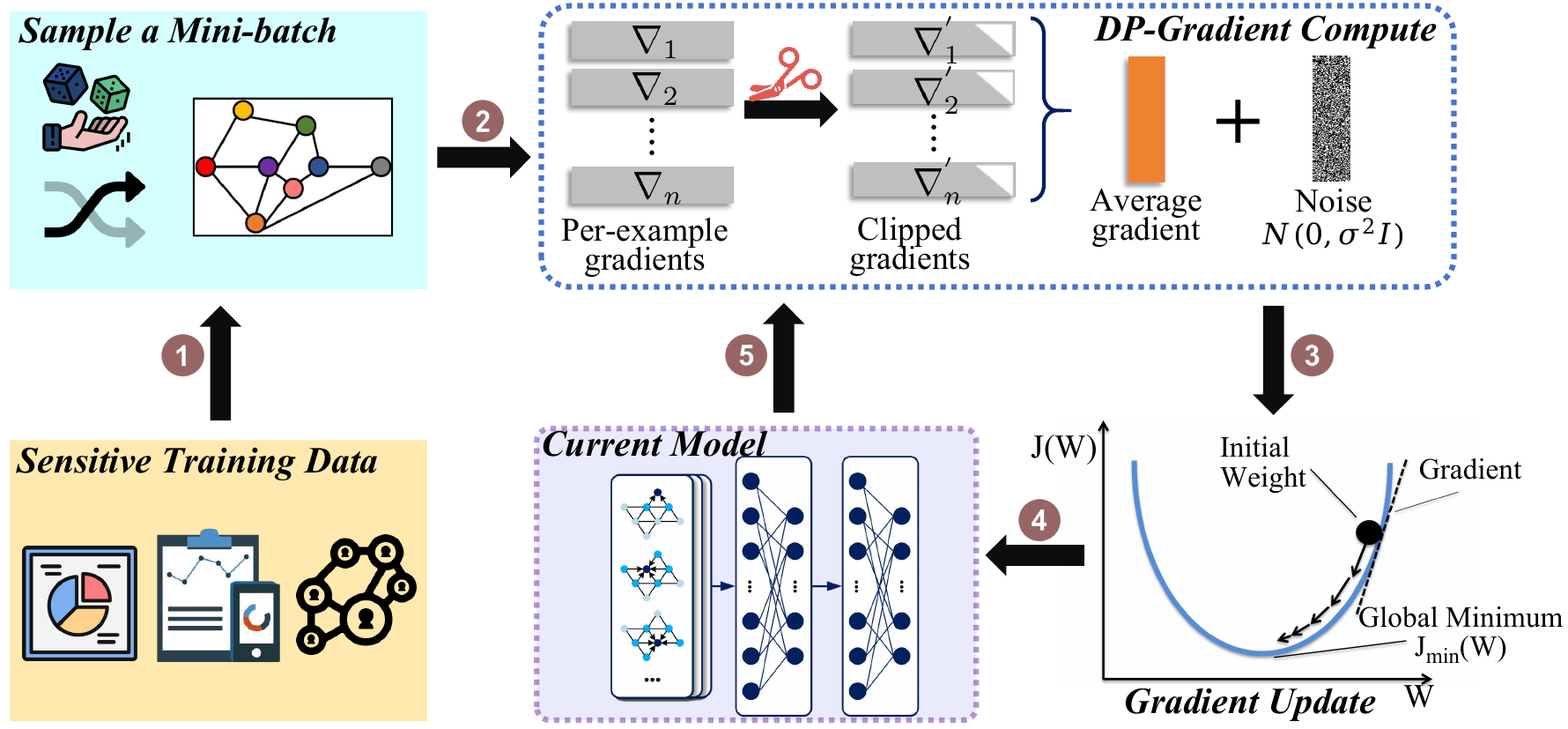}
\caption{The general procedure paradigm of differential privacy protection.}
\label{fig4}
\end{figure}

\noindent \textbf{Procedure of Differential Privacy.} 
The application of Differential Privacy in graph learning IP protection involves a structured process to ensure that sensitive information is adequately protected from potential attacks. As illustrated in Fig.~\ref{fig4}, the general procedure begins with the sampling of training data, where a subset of sensitive graph data is selected to form a mini-batch for the training process~\cite{tran2022heterogeneous}. Next, during the gradient computation phase, gradients are calculated for each example in the mini-batch. To preserve privacy, these gradients undergo a clipping process to limit their influence, ensuring that any single data point does not disproportionately affect the model's learning~\cite{bhaila2024local}. After clipping, an average gradient is computed, and a calibrated amount of noise is added to this gradient. The noise is carefully designed to adhere to the principles of DP, balancing the trade-off between privacy and utility by controlling the privacy loss parameter and the failure probability~\cite{bhaila2024local}. Finally, model parameters are updated based on the perturbed gradients, thus maintaining the model's utility while protecting the sensitive training data from potential privacy breaches~\cite{mueller2022sok}. 

Differential privacy can be formulated as follows~\cite{dwork2014algorithmic,chen2020vertically}: A randomized algorithm $\mathcal{M}$, which takes a dataset consisting of individuals as input, is said to be $(\eta, \varphi)$-differentially private if, for any pair of neighboring samples $X$ and $X'$ that differ in a single entry and for any result $y$ of the randomized algorithm, the following condition holds: if $\varphi = 0$, then $\mathcal{M}$ satisfies $\epsilon$-differential privacy, where $\epsilon$ is a measure of privacy loss, with smaller values of $\epsilon$ corresponding to stronger privacy guarantees.

\begin{equation}
P[\mathcal{M}(X)=y]\leq e^{\eta}P[\mathcal{M}(X')=y]+\varphi
\label{eq:eq3-5-1}
\end{equation}
\noindent where $\eta$ is the privacy loss parameter in DP, often referred to as the privacy budget, $\varphi$ is another parameter that represents the allowable probability of the algorithm $\mathcal{M}$ failing to meet $\eta$-differential privacy; $P$ denotes the probability. In summary, Eq.~\ref{eq:eq3-5-1} describes a probabilistic bound guaranteed by DP, balancing the trade-off between privacy protection and data utility by controlling $\eta$ and $\varphi$.

\noindent \textbf{Representative Methods.} 
Differential Privacy-based defense methods can be intuitively categorized based on their focus on specific privacy protection object, including (1) task-level privacy preservation, (2) privacy protection for node attributes and edges, and (3) hierarchical differential privacy protection.

The first category of DP-based methods focus on ensuring \textit{task-level privacy protection}. PRIVGNN~\cite{olatunji2021releasing} operates in three phases: private data selection, noisy pseudo-label retrieval, and student model training. It uses Poisson sampling to generate private subgraphs from public datasets and adds noise to node features via the Laplace mechanism to maintain differential privacy. These noisy features are then used to generate pseudo-labels through a teacher GNN, which trains the student GNN model in a transductive setting. Although PRIVGNN integrates privacy mechanisms into GNN training, its over-reliance on noisy pseudo-labels could result in accuracy degradation due to inappropriate injection of noise, especially in complex graph structures. To address the over-reliance on noisy pseudo-labels, MDP~\cite{xu2023mdp} offers an alternative approach by leveraging matrix decomposition and topological secret sharing (TSS) to protect GML models. MDP starts with eigenvalue decomposition of the graph adjacency and attribute matrices, splitting them into sub-matrices that preserve privacy while allowing accurate model training. These sub-matrices are processed using differential privacy mechanisms, and the GML model is trained using Laplacian-normalized adjacency matrices, which enhances privacy while maintaining message passing efficiency. Compared to PRIVGNN, MDP reduces the dependency on noisy labels by introducing matrix decomposition, but it still faces challenges in balancing computational complexity and privacy guarantees.

Moving beyond task-level DP, researchers further propose more specific \textit{privacy protection for node attributes and edges}. \cite{tran2022heterogeneous} addresses the challenge of balancing privacy and computational complexity for both node attributes and edge structures. It introduces two components: Feature-aware Randomized Response (FEATURERR) and Edge-aware Randomized Response (EDGERRR). FEATURERR optimizes privacy by randomizing node features based on their sensitivity and importance, ensuring that critical features remain intact while sensitive ones are perturbed. EDGERRR applies a similar approach to edges, utilizing a hierarchical random GML model and Markov Chain Monte Carlo sampling to preserve graph structure. \cite{tran2022heterogeneous} offers a more fine-grained approach, balancing privacy at both the feature and structural levels, but it may still suffer from reduced model accuracy due to the randomized process.

The third category centers on the latest \textit{hierarchical differential privacy protection}. PoinDP~\cite{wei2024poincare} takes DP a step further by focusing on hierarchical sensitivity in the embedding process. PoinDP leverages Poincaré embeddings, which is particularly suited for hierarchical graphs to compute personalized hierarchy-aware sensitivity (PHS) for nodes. It uses the Hyperbolic Gaussian Mechanism to add noise that aligns with the graph's curvature, ensuring that privacy is preserved without disrupting the hierarchical structure. This approach addresses a major limitation in previous methods, where sensitivity with respect to hierarchical structure was not explicitly accounted for, leading to reduced privacy or distorted embeddings. The optimization process of PoinDP further refines node embeddings, striking a balance between privacy and model accuracy.

\subsubsection{Adversarial Training Defense}


\noindent \textbf{Goals of Data-Level Adversarial Training.}
Adversarial Training is also employed to help graph learning models defend against data-level attacks, particularly model inversion attacks. This approach integrates adversarial perturbations into the training process to improve the model's generalization and resistance to adversarial manipulation~\cite{gosch2024adversarial}. Specifically, data-level adversarial training introduces two types of perturbations: disturbances to the graph structure (e.g., edge manipulations) or to the node feature attributes~\cite{kumarasinghe2022}. These adversarial perturbations are designed to maximize the model's prediction error, forcing the model to learn more robust representations that can withstand adversarial attempts to reverse-engineer or manipulate the data~\cite{dai2019adversarial}.
\begin{figure}[H]
\centering
\includegraphics[width=0.98\columnwidth]{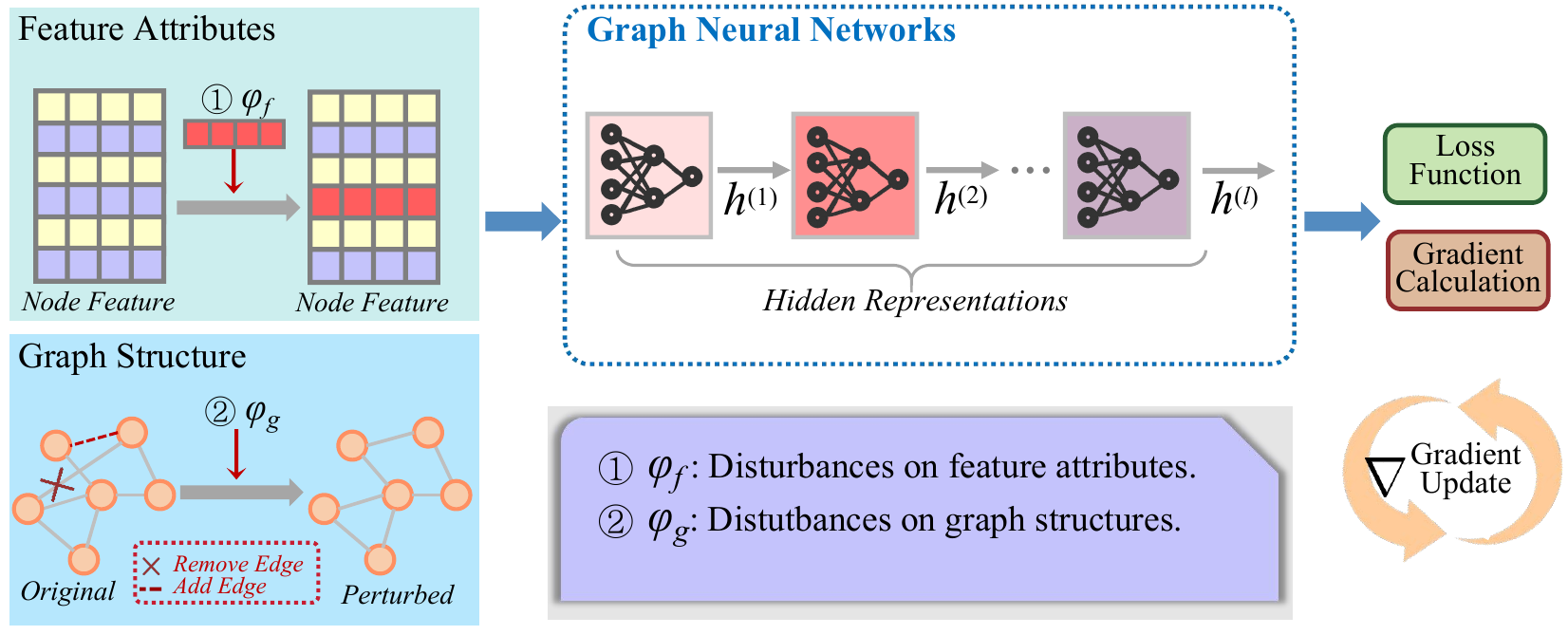}
\caption{The paradigm of data-level adversarial training defensive method.}
\label{fig6-2}
\end{figure}

\noindent \textbf{Procedures of Data-Level Adversarial Training.}
Data-Level Adversarial Training is a critical defense mechanism designed for data-level attacks, particularly those that target the input data, such as node features or graph structure~\cite{Gosch2023Adversarial}. The procedure integrates adversarial techniques by introducing adversarial noise into node embeddings or altering the graph topology. The main goal of this process is to adjust the model training objective to include an adversarial component that strengthens model resilience. In most common case, the overall objective function is modified to include both the original task loss and an additional adversarial loss that accounts for the perturbations of the graph data~\cite{song2024two}. The design ensures that the model performs well on its primary tasks while learning to defend against adversarial manipulations. By systematically exposing our model to elaborately crafted adversarial examples during training, our model becomes more robust to potential inversion attacks, effectively safeguarding sensitive information and enhancing its overall security~\cite{zheng2024improving}.

The general objective function for data-level adversarial training can be formulated as: 
\begin{equation} 
\arg \underset{\phi}{\min} \max_{\delta_{\mathcal{G}} \in \mathcal{P}(\delta)} \sum_{i=1}^{c} \mathcal{L}(f_{\phi}(\mathcal{G} + \delta_{\mathcal{G}})_i, y_i) 
\label{eq3-5-2} 
\end{equation} 

\noindent where $\phi$ is the model parameters, $\delta_{\mathcal{G}}$ denotes the adversarial perturbations applied on graph data. $\mathcal{P}(\delta)$ defines the set of allowed perturbations, ensuring that the perturbations remain within certain constraints. $f_{\phi}(\mathcal{G} + \delta_{\mathcal{G}})_i$ indicates the model prediction of node $i$ using perturbed data $\mathcal{G}+\delta_{\mathcal{G}}$. Finally, $y_i$ is the ground truth label for node $i$, and $c$ is the total number of labeled nodes used in the training process.

\noindent \textbf{Representative Methods.} Data-Level Adversarial Training methods can be classified into three types based on the specific perturbations they involve: (1) Feature Perturbation-based, (2) Graph Structural Perturbation-based, and (3) Hybrid Perturbation-based Adversarial Training, each focusing on different aspects.

Dai et al.~\cite{dai2019adversarial} proposed one of the earliest node \textit{feature perturbation-based adversarial training} frameworks.~\cite{dai2019adversarial} is built upon the DeepWalk algorithm and introduces adversarial perturbations by modifying node embeddings to maximize the loss function. Specifically, the framework perturbs the node embedding vectors $\mathbf{v}_i$ in order to introduce adversarial noise during training. By combining the original embedding loss with an adversarial regularization term, the model is trained to withstand a certain level of perturbations in the input features. However, this approach primarily focuses on node embeddings and does not fully account for the structural relationships between nodes, which still leave the model vulnerable to structural attacks.

To address the limitations of feature-only perturbations, graph \textit{structural perturbation-based adversarial training} is introduced. MC-GRA~\cite{zhou2023strengthening} improves upon feature perturbation methods by incorporating structural information into the adversarial training process. The method strengthens GML models against structural attacks by introducing a Markov Chain-based defense strategy. MC-GRA optimizes an objective function that balances the trade-off between graph representation accuracy and privacy preservation. It utilizes mutual information between the graph representation and its reconstructed version to extract valuable information while mitigating privacy risks. This approach effectively captures the structural properties of the graph, helping to defend against attacks targeting both the node level and the graph structure level. Nevertheless, while MC-GRA enhances robustness against structural perturbations, it still faces challenges in integrating the two kinds of perturbations into its defense strategy.

Hybrid approaches aim to overcome the weaknesses of methods focusing solely on either feature attributes or graph structural perturbations by innovatively combining both types of adversarial attacks. MCCNIFTY~\cite{zhang2021multi} introduces a \textit{hybrid perturbation adversarial training} framework that integrates both feature and structural perturbations to provide comprehensive robustness in graph neural networks. MCCNIFTY consists of multiple components, including a Multi-View Extractor (MVE), Multi-View Confidence-calibrated GNN Encoder (MCGC), and Embedding Consistency Regularization Module (ECRM). MVE generates subgraph-level augmented views, capturing variations in both node features and graph structures. Meanwhile, MCGC module handles uncertainty-aware regularization, further improving robustness against adversarial attacks by adjusting embeddings based on the uncertainty derived from multiple views. The hybrid approach addresses the limitations of existing methods that solely focus on one aspect. The ECRM module of MCCNIFTY ensures consistency across different views of the graph, promoting both robustness and fairness in the presence of adversarial perturbations.

\subsubsection{Graph Structure Perturbation Defense}

\begin{figure}[htbp]
\centering
\includegraphics[width=0.70\columnwidth]{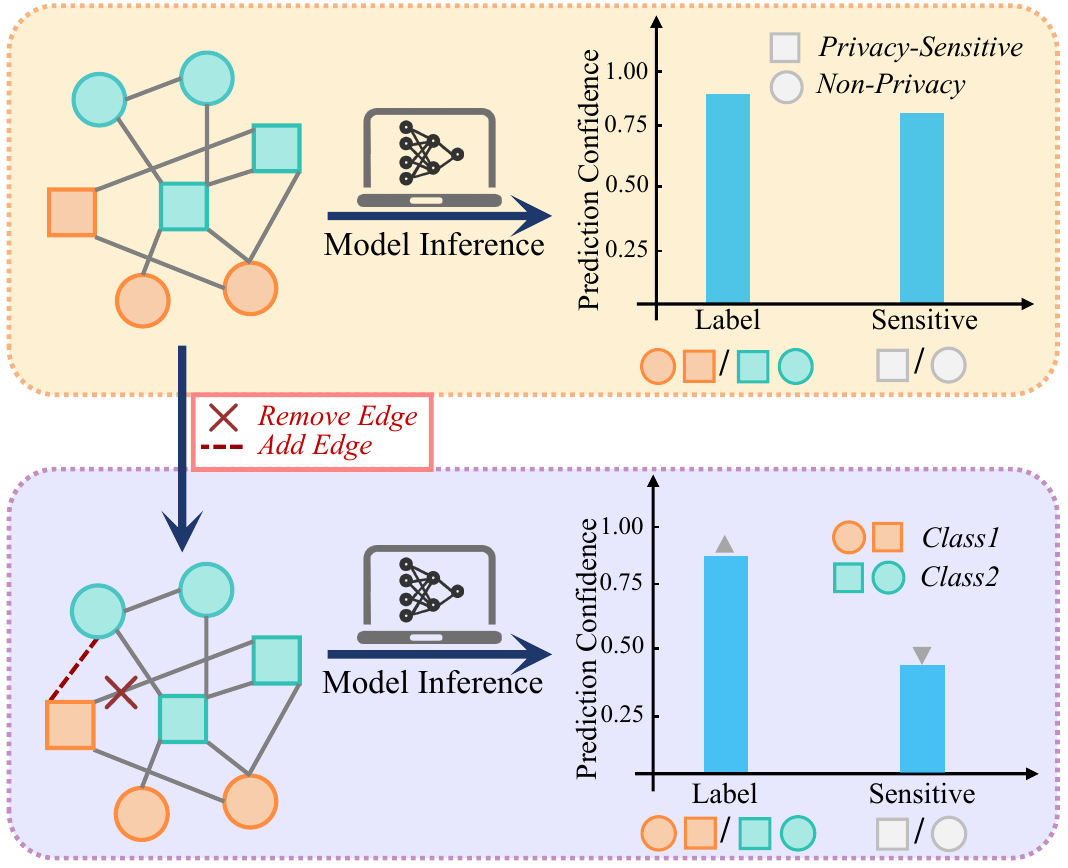}
\caption{The graph is perturbed by deleting an edge and adding a new one, ensuring two conditions are met: (1) There is a decrease of prediction confidence on private labels (square and circle), and (2) the prediction confidence on target categories (orange and blue) remains stable or even increased, keeping the data utility.}
\label{fig5}
\end{figure}

\noindent \textbf{Goals of Graph Structure Perturbation.} Graph Structure Perturbations (GSP) works by adding or removing edges/nodes and modifying edge weights in such a way that the overall utility of the graph for the target task is maintained, while significantly reducing the attack success rate~\cite{Xue2021CAP}. As shown in Fig.~\ref{fig5}, GSP aims to perturb the graph structure so that the perturbed graph $\mathcal{G}^{'}$ remains highly similar to the original graph $\mathcal{G}$ in  functionality, yet makes it much harder for attackers to accurately infer sensitive information, such as private attributes~\cite{liu2024revisiting}. By carefully designing perturbation strategies, GSP ensures the local neighborhoods of nodes retain high utility for tasks like node classification and link prediction, while the structural changes obscure patterns that attackers typically exploit.

\noindent \textbf{Procedures of Graph Structure Perturbations.}
The process of GSP typically involves several key steps to ensure effective defense against data-level attacks such as model inversion. First, the original graph structure is analyzed to identify vulnerable edges and nodes that could potentially reveal sensitive information~\cite{jin2020graph}. Next, a perturbation strategy is designed, which may involve adding, removing, or rewiring edges and nodes, or adjusting edge weights~\cite{hsieh2021netfense}. This strategy must balance the need to obscure sensitive information with the requirement to preserve the graph's overall structural properties and utility for the target task. Third, the perturbations are applied systematically, often using stochastic methods to introduce uncertainty and prevent attackers from reversing the modifications~\cite{wang2022group}. In the final step, the perturbed graph is validated to ensure that it still performs well on the intended tasks while significantly hindering the attack success rate.

GSP can be formalized as a process where the original graph $\mathcal{G} = (\mathcal{V}, \mathcal{A}, X)$ is perturbed to produce a new graph $\mathcal{G}^{'} = (\mathcal{V}, \mathcal{A}^{'}, X^{'})$, with $\mathcal{A}^{'}$ representing the modified adjacency matrix and $X^{'}$ being the perturbed feature matrix if applicable. The primary objective is to preserve the utility of the graph for tasks such as node classification or link prediction while reducing the success rate of adversarial attacks. The objective function for GSP is formulated as:
\begin{equation}
\arg \underset{\mathcal{G}^{'}}{\min} \left[ \mathcal{L}_{\text{task}}(f_{\phi}(\mathcal{G}^{'}), Y) - \lambda \cdot \mathcal{L}_{\text{adv}}(f_{\phi}(\mathcal{G}^{'}), \mathcal{A}, X) \right]
\label{eq:gsp-1}
\end{equation}

\noindent where $\mathcal{L}_{\text{task}}(f_{\phi}(\mathcal{G}^{'}), Y)$ is the task loss measuring the performance of the graph on the primary task, $Y$ is the ground truth labels, and $\mathcal{L}_{\text{adv}}(f_{\phi}(\mathcal{G}^{'}), \mathcal{A}, X)$ is the adversarial loss that reflects the difficulty faced by attackers when trying to infer sensitive information from the graph. $\lambda$ is a balancing factor that determines the trade-off between task performance and attack resistance.

\noindent \textbf{Representative Methods.} 
Graph structure perturbation methods can be categorized into three types based on the specific modifications they applied to the graph structure: (1) Edge Insertion/Deletion-based Perturbation, (2) Node Injection/Removal-based Perturbation, and (3) Subgraph Perturbation, each aiming to disrupt adversarial attacks while preserving graph utility for target tasks.

For \textit{Edge Insertion/Deletion-based perturbation} methods, Pro-GNN~\cite{jin2020graph} is one of earliest work that introduces a bi-level optimization framework to defend against attacks by perturbing the graph structure. Pro-GNN focuses on modifying the adjacency matrix to ensure robustness while preserving key graph properties like low-rankness and sparsity. Pro-GNN introduces feature smoothness regularization, encouraging connected nodes to have similar features, which helps mitigate the impact of adversarial edge manipulations. However, this method focuses primarily on edge-level perturbations, and its regularization terms do not fully account for more sophisticated attacks on the graph's global structure. Built upon Pro-GNN, NetFense~\cite{hsieh2021netfense} introduces a more comprehensive combinatorial optimization framework. It refines the process of edge insertion and deletion by incorporating personalized PageRank (PPR) to select edge candidates that maximize both utility and unnoticeability. This approach addresses the limitations of Pro-GNN by balancing privacy and accuracy, while ensuring that the selected edges for perturbation are less detectable by attackers. 

The second type of perturbation is \textit{Node Injection/Removal-based}. Wang et al.~\cite{wang2022group} extend beyond edge perturbations by introducing mechanisms that focus on node-level perturbations, specifically noisy embedding and embedding truncation. By adding Laplace noise to node embeddings and reducing their dimensionality, this method protects sensitive node information from being inferred by attackers. In contrast to edge-based methods like Pro-GNN and NetFense, this method targets the node-level representations, making it effective in scenarios where edge manipulations alone are insufficient. However, the over-reliance on noise can degrade the performance of~\cite{wang2022group} if not carefully calibrated, leaving room for improvement in handling more sophisticated attacks.

Finally, \textit{subgraph perturbation-based methods} address the limitations of edge and node perturbations by focusing on subgraph-level modifications.~\cite{he2021node} propose to schematically add random edges to create unpredictable subgraph structures, which significantly lowers the attacker's inference accuracy.~\cite{he2021node} shows that large-scale subgraph perturbations can effectively obscure node relationships and protect sensitive information. However, this approach might overly disturb the graph utility, as random edge addition can lead to suboptimal performance on various tasks. More recently,~\cite{boratto2024robustness} further refines subgraph perturbation strategies by focusing on improving both robustness and fairness. Their approach adjusts the adjacency matrix through controlled edge additions and deletions, with a specific focus on maintaining fairness metrics like demographic parity. The proposed method tackles the trade-off between robustness and fairness, which previous methods fail to address.

\subsubsection{Regularization Techniques for MInfA}\label{sec3-5-4}

\noindent \textbf{Goals of Regularization Defense.} Regularization techniques serve as fundamental mechanisms to mitigate \textit{\textbf{Membership Inference Attacks}} by enhancing the generalization ability of graph learning models and reducing the unintended memorization of training data~\cite{wu2021adapting}. Formally, regularization in the defense context refers to a set of strategies that constrain the model's capacity to overfit specific training instances, thereby decreasing the distinguishability between members and non-members in the dataset~\cite{conti2022label}. Given a model $f_{\phi}(\cdot)$ trained on private graph dataset $\mathcal{G}_{\text{train}}$, regularization modifies the Empirical Risk Minimization (ERM) objective function by incorporating an additional penalty term $\Omega_{\phi}(\cdot)$, leading to the optimization equation:
\begin{equation}
\min_{\phi} \frac{1}{N} \sum_{i=1}^{N} \mathcal{L}(y_i, f_{\phi}(x_i)) + \lambda \Omega({\phi})
\label{eq:regularization_defense}
\end{equation}

\noindent where $\mathcal{L}(\cdot)$ denotes the loss function, $\lambda$ is a tunable regularization coefficient, and $\Omega(\phi)$ denotes the constraint on model complexity, such as $L_2$ regularization, dropout, or adversarial regularization. $N$ is the total number of training samples. The goal is to increase the behavioral similarity between training and testing samples such that an attacker attempting to conduct membership inference attacks cannot effectively differentiate whether a given data point is included in $\mathcal{G}_{\text{train}}$. By enforcing smoother decision boundaries and reducing over-confidence in predictions, regularization plays a critical role in GML IP protection paradigms.

\noindent \textbf{Procedures of Regularization Defense.} Regularization-based defense against membership inference attacks follows a structured procedure. Step one is \textit{Risk Assessment and Attack Analysis}, which involves evaluating the model's vulnerability to MInfA by assessing its reliance on training data. Techniques such as shadow model training and white-box gradient inspection help measure the risk~\cite{yang2023membership}. The second step——\textit{Regularization Strategy Selection} focuses on selecting appropriate regularization techniques to mitigate overfitting and reduce membership distinguishability~\cite{yang2023membership}. Common strategies include $L_2$ regularization to constrain model complexity, Dropout to randomly deactivate neurons during training, Label Smoothing to reduce overconfident predictions, and Adversarial Regularization to enhance robustness against inference attacks. The third step is \textit{Model Training with Regularization}, which incorporates the chosen regularization techniques into model training process and optimizes the objective function (Eq.~\ref{eq:regularization_defense}).

\noindent where $\mathcal{L}(\cdot)$ is the loss function, and $\Omega(\phi)$ is the selected regularization constraints. The final step is \textit{Evaluation and Fine-tuning}. This step assesses the defense effectiveness, followed by hyperparameter tuning (e.g., adjust $\lambda$), to balance model accuracy and privacy protection trade-offs.

\noindent \textbf{Representative Methods.}
Regularization-based techniques are a major class of defense mechanisms against membership inference attacks in graph learning settings. The categories include, but are not limited to, the following~\cite{hu2022membership}: (1) L2-norm regularization, (2) Dropout, (3) Data augmentation, (4) Label smoothing, and (5) Adversarial regularization.

As early as 2021, Wu et al.~\cite{wu2021adapting} identified overfitting as a primary vulnerability of GML models to membership inference attacks and proposed $L_2$ regularization to mitigate this issue. While effective in reducing attack success rate, this approach only constrains parameter norms without decreasing the model's reliance on individual training samples, limiting its effectiveness on datasets with high information leakage risks. To further reduce dependency on individual samples, Jnaini et al.~\cite{jnaini2022powerful} introduced Dropout techniques, which randomly deactivate neurons during training to enhance generalization. However, excessive Dropout led to performance degradation and the failure of mitigating the confidence score imbalance. To address the confidence imbalance issue, Conti et al.~\cite{conti2022label} proposed a Label Smoothing method, ensuring that the model predictions remain less confident, thereby reducing the likelihood of membership inference. However, this method only modifies the output distribution without fundamentally altering the latent representations and model learning dynamics, making it vulnerable to more advanced attacks. To enhance defense robustness against MInfA,~\cite{hasegawa2023membership} introduced Representation Self-Challenging (RSC) as a domain generalization method to mitigate model overfitting. RSC works by masking large gradient updates in the final layer, preventing a GML model from making drastic parameter changes that can lead to overfitting on training samples. Yet, RSC primarily focuses on imbalanced datasets, making it less effective in balanced datasets. Moreover, it does not alter the model learning process, leaving it vulnerable to being attacked by advanced membership inference schemes. Building upon these efforts, Yang et al.~\cite{yang2023membership} employed a hybrid defense strategy. On the one hand, it combined regularization and dropout to model structure; on the other hand, it added perturbations to increase model robustness. The method shows superior defense capabilities across various benchmark datasets including ENZYMES, PROTEINS, and NCI1. However, this method did not leverage the structural properties of GML models and merely stacked existing regularization techniques, making it susceptible to more sophisticated structural attacks. Lately, Guan et al.~\cite{guan2024topology} introduced Topology Modification, which strategically alters edge connections during training. The strategy ensures that GML models observe slightly different graph structures across training epochs, thereby disrupting attackers' ability to infer membership information. Unlike prior regularization-based methods that primarily target parameters or confidence scores, this approach modifies the data structure itself, enhancing privacy protection. Nevertheless, it still requires a careful design of trade-offs between graph structure modifications and model performance.

\subsubsection{Knowledge Distillation for MInfA}\label{sec3-5-5}

\noindent \textbf{Goals of Knowledge Distillation Defense.} Knowledge Distillation (KD) aims to mitigate \textit{\textbf{Membership Inference Attacks}} by transferring knowledge from a pre-trained teacher GML model $f_{\text{teacher}}$ to a student GML model $f_{\text{student}}$ while minimizing information leakage related to the private training graph data $\mathcal{G}_{\text{train}}=(\mathcal{V}_{\text{train}},\mathcal{E}_{\text{train}},X_{\text{train}})$~\cite{zheng2021resisting}. The primary objective of KD in GML defense is to reduce the sensitivity of the student model to specific graph components in $\mathcal{G}_{\text{train}}$, thereby obscuring the membership information~\cite{chen2024maskarmor}.

Formally, given a teacher GML model $f_{\text{teacher}}$ trained on private graph data ${\mathcal{G}}_{\text{train}}$, KD optimizes the student GML model by minimizing the loss:
\begin{equation}
\mathcal{L}_{\text{KD}} = \alpha \cdot \mathcal{L}_{\text{soft}} + (1-\alpha) \cdot \mathcal{L}_{\text{hard}}
\label{eq:KD-MInfA}
\end{equation}

\noindent where $\mathcal{L}_{\text{soft}}$ represents the Kullback-Leibler divergence between the softened outputs (logits) of the teacher and student models, and $\mathcal{L}_{\text{soft}} = \text{KL}(\sigma(Z_{\text{teacher}}/T), \sigma(Z_{\text{student}}/T))$, where $T$ is the temperature parameter that controls the granularity of knowledge transfer. $\mathcal{L}_{\text{hard}}$ is the standard cross-entropy loss computed on true labels. By carefully tuning $T$ and the balance factor $\alpha$, KD ensures that the student GML model retains high task utility while effectively obfuscating sensitive membership signals in ${\mathcal{G}}_{\text{train}}$.

\begin{figure*}[!htbp]
\centering
\includegraphics[width=0.90\textwidth]{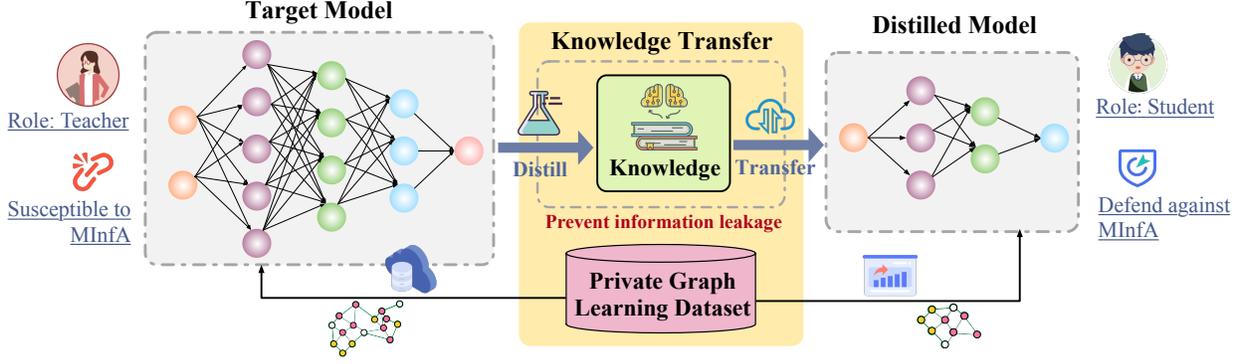}
\caption{The general procedure of knowledge distillation defense for graph machine learning models.}
\label{knowledge_distillation}
\end{figure*}

\noindent \textbf{Procedures of Knowledge Distillation Defense.} Fig.~\ref{knowledge_distillation} demonstrates the general procedure of Knowledge Distillation Defense. Specifically, the procedure of Knowledge Distillation defense consists of several carefully designed steps. First, a pre-trained teacher GML model $f_{\text{teacher}}$ is used to generate soft labels for a training graph dataset $\mathcal{G}_{\text{train}}$. These soft labels are represented as the logits output by the teacher GML model, which encode rich inter-class relationships and obfuscate sensitive membership information~\cite{tian2025knowledge}. Next, a student model $f_{\text{student}}$ is trained using these soft labels as guidance, alongside the original hard labels~\cite{zheng2021resisting}. The optimization objective is $\mathcal{L}_{\text{KD}} = \alpha \cdot \mathcal{L}_{\text{soft}} + (1-\alpha) \cdot \mathcal{L}_{\text{hard}}$. Subsequently, the trained student model $f_{\text{student}}$ is deployed as the final model for inference. This process ensures the student model mimics the teacher model's functionality while reducing the likelihood of leaking sensitive membership information about the private training data.

\noindent \textbf{Representative Methods.}
Knowledge Distillation defensive methods are characterized into four groups based on the employed specific distillation technology: (1) Complementary Knowledge Distillation (CKD); (2) Pseudo Complementary Knowledge Distillation (PCKD); (3) Self-Distillation; (4) Masked Confidence Distillation.

The evolution of Knowledge Distillation defenses for mitigating Membership Inference Attacks has led to the development of several innovative approaches. Among them, CKD and PCKD are the most foundational and widely recognized approaches in KD-based defenses.~\cite{zheng2021resisting} is an early work in GML domain which proposes to employ CKD and PCKD for data-level defense. CKD relies on training teacher models on complementary subsets of the private graph dataset, ensuring that soft labels generated for the student model $f_{\text{student}}$ do not directly expose sensitive information from the private graph training dataset $\mathcal{G}_{\text{train}}$. PCKD refines this approach by reducing computational costs through model averaging and pre-training, enabling efficient knowledge transfer while maintaining high task utility. Despite their effectiveness, these methods require careful construction of complementary datasets, which may cause unaffordable computational expenses. To address the limitations of CKD and PCKD, RepKD~\cite{mazzone2022repeated} proposed an iterative \textit{masked confidence distillation} method. RepKD introduced a confidence masking function during each distillation round to perturb the teacher model's predictions. This iterative approach creates a chain of models where the deployed student model is increasingly distanced from the private whole training dataset $\mathcal{G}_{\text{train}}$. RepKD allows fine-grained control over the privacy-utility tradeoff, demonstrating notable reductions in attack success rates. Building on these achievements, SELENA~\cite{tang2022mitigating} introduced a self-distillation method with a novel ensemble architecture. SELENA splits the training graph dataset $\mathcal{G}_{\text{train}}$ into smaller subsets and trains individual models on these subsets. The outputs of these models are then used to train a final student model through self-distillation, without relying on external or synthetic datasets. This ensemble strategy leverages diversity across subsets to obscure membership signals while ensuring robust model performance.

\subsubsection{Evaluation Metrics of Data-Level Defenses} \label{sec3-5-6}



We categorize the evaluation metrics for assessing defense mechanisms against data-level attacks into three types, based on insights synthesized from existing literature~\cite{dibbo2023sok,zhang2022model,zhou2024model,guan2025topology,wang2024gcl}: (I) \textit{Task Utility Metrics}, (II) \textit{Privacy Protection Metrics}, and (III) \textit{Resource and Efficiency Metrics}. These metrics comprehensively reflect the defense effectiveness in application scenarios.

\noindent \textbf{(I) Task Utility Metrics.} These metrics measure how well the defense mechanism preserves the original model's performance on intended downstream tasks (e.g., node classification, link prediction, or graph classification) after defensive interventions~\cite{zhang2022model}. Common indicators include Accuracy, F1 Score, and AUC, reflecting the model's ability to correctly classify or rank instances while under protection~\cite{zhang2022model}. High task utility implies that the defense does not sacrifice essential predictive capability.

\noindent \textbf{(II) Privacy Protection Metrics.} This type of metrics evaluates the ability of the defense to prevent information leakage or unauthorized inference. For model inversion attacks, metrics such as Attribute Reconstruction Accuracy (or reconstruction error) and Mean Squared Error (MSE) are used to quantify how well sensitive attributes are protected~\cite{dibbo2023sok}. For membership inference attacks, Attack Success Rate (ASR), AUC, and True/False Positive Rates are used to measure the reduction of attack effectiveness caused by the defense~\cite{wang2024gcl}.
    
\noindent \textbf{(III) Resource and Efficiency Metrics.} These metrics assess the computational and practical cost of the defense method. Relevant indicators include Inference Overhead (i.e., the additional computation or time required by the defense)~\cite{zhou2024model}, Model Complexity Increase~\cite{guan2025topology}, and, where applicable, Query Budget Impact (i.e., how the defense affects the attacker’s required number of queries)~\cite{dibbo2023sok}. Efficient defenses provide strong privacy protection with minimal performance and resource penalties.

\section{Benchmark Datasets and Applications}\label{sec6}
This section broadly investigates existing open-sourced datasets that were employed for developing graph learning intellectual property protection methods and empirical studies. The benchmark datasets can be divided according to their application domains, including Citation, Social Network, Chemical, Protein, E-Commerce Platform, Traffic Network, Collaboration, etc. A detailed summary of dataset characteristics and statistics is provided in Table~\ref{tab3}. Because of the space in this article, we provide the links to our online GitHub library, where the implementations of benchmark graph learning IP protection methods are available and ready-to-use for interested readers. We hope this work can arouse attention from the research community towards developing more robust and comprehensive benchmarks for graph learning IP protection.

\subsection{Comprehensive Guide to Benchmark Datasets}
\noindent \textbf{(i) Citation.} The commonly adopted datasets in citation domain include Cora\cite{duddu2020quantifying}, Citeseer\cite{hu2022membership}, DBLP\cite{cong2022sslguard}, ogbn-arxiv\cite{ding2022data}, PubMed\cite{li2022graph}, and Wiki-CS\cite{mernyei2020wiki}. In citation networks, nodes represent papers, and edges depict the citation relationships among them. Although citation graphs typically do not contain sensitive private information, making the impact of intellectual property theft less severe, their high popularity and widespread availability in graph learning domain make them a popular choice for practice.

\noindent \textbf{(ii) Social Network.} The domain of social network analysis can identify sub-communities, assist in fraud detection, conduct personalized social recommendation, and others. In a social network, the nodes usually represent social media users, while the edges denote the connections between these users. User features in social networks often encompass attributes like gender, age, location, personal preferences, relationship status, etc. The open-source benchmark datasets of Social Networks include Facebook\cite{liu2022federated}, Flickr\cite{tao2022revisiting}, PolBlogs\cite{adamic2005political}, LastFM\cite{xu2023openp5}, Reddit\cite{chen2022understanding}, and Twitter\cite{yang2021consisrec}. 

\noindent \textbf{(iii) Molecular \& Protein Networks.} Graph learning methodologies have revolutionized molecular and protein research. Under this context, molecules and proteins are represented as graphs, where nodes correspond to atoms or amino acids, and edges represent chemical bonds or interactions. Unlike traditional datasets that form a single graph, each data point in molecular and protein datasets is an individual graph, capturing the unique structure and properties of distinct molecules or proteins. This framework facilitates a deeper understanding of molecular behavior, interactions, and functions, enabling advancements in areas such as drug discovery, toxicity prediction, and enzyme activity analysis. In this emerging field, open-source datasets like MUTAG\cite{debnath1991structure}, PTC\cite{toivonen2003statistical}, NCT1 \& NCI109\cite{wale2008comparison}, AIDS\cite{riesen2008iam}, ENZYMES\cite{zhang2022model}, PROTEINS\cite{borgwardt2005protein}, and Disease Pathway (DP)\cite{agrawal2018large} have aroused great attention from acadeMInvA. 

\noindent \textbf{(iv) E-Commerce Platform.} Computers\cite{kumar2018rev2}, Photo\cite{kumar2018rev2}, Yelp\cite{asghar2016yelp}, ML-1M\cite{chin2022datasets}, and Bitcoin-Alpha\cite{you2022roland} are among the most popular open-source E-Commerce Platform datasets in graph learning domain. Graph learning for E-commerce involves using complicated networks that capture interactions between users and products. The interactions can be represented as bipartite graphs, where nodes correspond to users and products, and edges represent user-product interactions. Leveraging these graphs enables advanced recommendation systems, enhancing user experience by providing personalized suggestions. However, privacy concerns arise as these datasets can reveal sensitive information about user preferences and product attributes. Thus, it is crucial to develop and evaluate graph-based recommendation models, ensuring both effectiveness and privacy protection.

\noindent \textbf{(v) Other Domain Applications.} Aside from common applications such as recommendation, social networks, and bioinformatics, graph learning has been extended into a diverse range of emerging domains. Here, we explore datasets that demonstrate the versatility of graph-based models across various fields, including transportation networks, scientific collaboration networks, and the movie industry. {Brazil}\cite{ribeiro2017struc2vec}, {USA}\cite{ribeiro2017struc2vec}, {COLLAB}\cite{xu2023watermarking}, and {IMDB-BINARY}\cite{wei2021pooling} are the representative benchmarks in these fields.

\begin{table*}[htb]
\footnotesize
\caption{Detailed Statistics and Comparison of existing benchmark datasets been used for practicing graph learning intellectual property protection.}
\label{tab3}
\begin{tabular}{c|c|c|c|c|c|c}
\toprule
\textbf{Dataset}     & \textbf{Application Domain} & \textbf{(Avg.) \#Nodes} & \textbf{(Avg.) \#Edges} & \textbf{\#Graphs} & \textbf{Feature Dim} & \textbf{Num of Labels/Classes} \\ \midrule
Cora \cite{duddu2020quantifying}   & Citation                    & 2,708                 & 5,429                 & 1               & 1,433                      & 7                                 \\ 
Citeseer \cite{hu2022membership}  & Citation                    & 2,110                 & 3,668                 & 1               & 3,703                      & 6                                 \\ 
DBLP \cite{oliynyk2023know,cong2022sslguard}      & Citation                    & 17,716                & 105,734               & 1               & 1,639                      & 4                                 \\ 
ogbn-arxiv \cite{wu2022nodeformer,ding2022data} & Citation                    & 169,343               & 1,166,243             & 1               & 128                        & 40                                \\ 
PubMed \cite{li2022graph,reau2023deeprank}   & Citation                    & 19,717                & 44,338                & 1               & 500                        & 3                                 \\ 
WiKi-CS \cite{mernyei2020wiki}  & Citation                    & 11,701                & 216,123               & 1               & 300                        & 10                                \\ \hline
Facebook \cite{liu2022federated} & Social Network              & 22,470                & 170,912               & 1               & 4,714                      & 4                                 \\ 
Flickr \cite{tao2022revisiting}  & Social Network              & 8,252                 & 327,815               & 8,358           & -                         & 81                                \\ 
Polblogs \cite{adamic2005political}  & Social Network              & 1,490                 & 19,090                & -              & -                         & 2                                 \\ 
LastFM \cite{xu2023openp5}   & Social Network              & 7,083                 & 25,814                & 1               & 7,842                      & 10                                \\ 
Reddit \cite{chen2022understanding}   & Social Network              & 232,965               & 57,307,946            & 2,000           & 602                        & 41                                \\ 
Twitter \cite{yang2021consisrec,guo2020deep}   & Social Network              & 81,306                & 1,768,149             & 1               & -                         & 4                                 \\ \hline
MUTAG \cite{debnath1991structure} & Molecular                   & 17.93                 & 19.79                 & 188             & 7                          & 2                                 \\ 
PTC \cite{toivonen2003statistical}  & Molecular                   & 25.5                  & -                    & 344             & 19                         & 2                                 \\ 
NCI1 \cite{wale2008comparison}  & Molecular                    & 29.68                 & 32.13                 & 4110            & 37                         & 2                                 \\ 
NCI109 \cite{wale2008comparison} & Molecular                  & 29.6                  & -                    & 4,127           & 38                         & 2                                 \\ 
AIDS \cite{riesen2008iam}    & Molecular                  & 31,385                & 64,780                & -              & 4                          & 38                                \\ \hline
ENZYMES \cite{zhang2022model}  & Protein                     & 19,580                & 74,564                & -              & 18                         & 3                                 \\ 
PROTEINS\cite{borgwardt2005protein}  & Protein                     & 39.06                 & 72.82                 & 1,113           & 3                          & 2                                 \\ 
Disease Pathway (DP) \cite{agrawal2018large} & Protein                     & 22,552                & 342,353               &                 & 73                         & 519                               \\ \hline
COLLAB \cite{xu2023watermarking}   & Collaboration               & 74.49                 & 2,457.78             & 5,000          & -                         & 3                                 \\ 
IMDB-BINARY \cite{wei2021pooling}  & Collaboration               & 19.8                  & 96.5                  & 1,000           & -                         & 2                                 \\ \hline
Computers \cite{kumar2018rev2} & E-Commerce Platform         & 13,752                & 245,861               & -              & -                         & 10                                \\ 
Photo \cite{kumar2018rev2}  & E-Commerce Platform         & 7,487                 & 119,043               & -              & -                         & 8                                 \\ 
Tmall \cite{hu2023towards}  & E-Commerce Platform         & 132,724               & 3,338,788             & -              & 2                          & -                                \\ 
Yelp\cite{asghar2016yelp}   & E-Commerce Platform         & 99,011                & 1,842,418             & -              & 6                          & -                                \\ 
ML-1M \cite{chin2022datasets} & E-Commerce Platform         & 5,945                 & 365,535               & -              & 3                          & -                                \\
Bitcoin-Alpha \cite{you2022roland}  & E-Commerce Platform         & 3,783                 & 24,186                & -              & -                         & -                                \\ \hline
USA \cite{ribeiro2017struc2vec}  & Traffic Network             & 1,190                 & 13,599                & -              & -                         & 4                                 \\ 
Brazil \cite{ribeiro2017struc2vec} & Traffic Network             & 131                   & 1,038                 & -              & -                         & 4                                 \\ \bottomrule
\end{tabular}
\end{table*}

\subsection{IP Protection Applications in Real-World Scenarios}

\noindent \textbf{Application in Citation Networks.} Citation networks represent the structure of academic literature, where nodes denote research papers and edges represent citations between them. Common datasets such as Cora~\cite{duddu2020quantifying}, Citeseer~\cite{hu2022membership},DBLP~\cite{oliynyk2023know}, ogbn-arxiv \cite{ding2022data}, PubMed~\cite{reau2023deeprank}, and WiKi-CS~\cite{mernyei2020wiki} are widely used to analyze trends and relationships in scientific research. In the context of Intellectual Property protection, GNNs and other advanced graph learning models can be employed to predict potential collaborations, discover influential works, and detect patterns of knowledge flow, which are critical for safeguarding the proprietary methodologies and innovations documented in these networks. However, the use of various GML models in citation networks also poses risks, including model inversion and extraction attacks. These attacks can lead to unauthorized access to sensitive information, such as unpublished research insights or proprietary algorithms, potentially jeopardizing intellectual property rights. Consequently, robust defenses against such attacks are essential to maintain the integrity and confidentiality of data in citation networks, ensuring that intellectual property remains protected against unauthorized exploitation.

\noindent \textbf{Application in Social Networks.} Social networks encompass a vast array of platforms, including major internet sites like Facebook and IoT systems exemplified by Vehicular Social Networks (VSN). GML models have extensive applications within these networks, notably in link prediction and in security and fraud detection. For instance, GNNs can forecast the potential formation of new social connections between users, as demonstrated by~\cite{islam2020comparative}. Furthermore,~\cite{guo2022mixed} introduces an innovative method, FRD-VSN, designed to detect fake news within VSNs. Despite these advancements, link prediction models are vulnerable to model inversion attacks, which can lead to the illicit acquisition of users' personal data. Therefore, developing robust defenses against model inversion is imperative. Similarly, graph learning models used in fake news detection are susceptible to model extraction attacks, where attackers, leveraging publicly available data, can create a functionally similar model to the original model. These threats underscore the need for model extraction defenses to safeguard intellectual property and ensure the integrity of proprietary algorithms.

\noindent \textbf{Application in Molecular \& Protein Networks.} GML models have transformative applications in the fields of biology and medicine, including but not limited to protein-protein interfaces (PPI) prediction, drug side-effect forecasting, molecular graph generation, and disease prediction. The employment of graph learning models in PPI prediction is pivotal for elucidating protein-protein interactions and predicting the quaternary structures of protein complexes, as evidenced by~\cite{lane2013milliseconds}. Decagon~\cite{zitnik2018modeling} exemplifies the use of multimodal graphs, incorporating protein-protein, drug-protein, and drug-drug interactions, to predict side effects through graph convolutional networks (GCNs). The generation of molecular graphs is a critical step in drug discovery, facilitating the design of numerous potential drug candidates. Molecular Graph Network (MGN)~\cite{bongini2022biognn} employs a sequential generator to create molecular structures atom by atom, guided by a node expansion algorithm. Furthermore, by integrating BERT and GNNs, Basaad~\cite{basaad2024bert} seeks to advance cancer metastasis detection through a non-invasive breast cancer classification model. In biomedical domain, the integrity of GML models is of paramount importance, necessitating robust defenses against model extraction and inversion attacks. These models often encapsulate proprietary methods and utilize sensitive patient data, making their protection essential to prevent significant intellectual property losses and breaches of patient privacy.

\noindent \textbf{Application in E-Commerce Platforms.} E-Commerce Platforms underpinned by deep and large graph learning models have become ubiquitous in contemporary digital ecosystems. By leveraging the intricate web of interactions and connections among users, alongside the relationships between users and content, graph learning models can effectively personalize recommendations, including friend suggestions~\cite{li2023survey}, content curation~\cite{guo2020deep}, and product recommendations~\cite{wu2022graph}. Despite their utility, these systems are susceptible to various attacks, which pose significant privacy risks by potentially exposing users' private data. According to~\cite{zhang2022model}, if an attacker accesses a trained GML model through a compromised client and supplements it with auxiliary information from the internet, the attacker can reconstruct sensitive relationships, such as user friendships. Consequently, the development of robust defense mechanisms is imperative. Moreover, the proprietary nature of well-crafted recommendation algorithms confers substantial commercial advantages, necessitating the implementation of effective defense methodologies. Besides their functionality in safeguarding intellectual property, IP protection methods are also significant for preserving the unique competitiveness that the algorithms provide.

\noindent \textbf{Application in Fintech Technology.} In the dynamic landscape of e-Financial Technology, effective regulation is indispensable for maintaining market stability. GML models have emerged as a pivotal tool in this sector, offering advanced capabilities in default risk prediction, fraud detection, and event nowcasting. For instance, MotifGNN~\cite{wang2023financial} is developed to enhance the accuracy of financial default predictions, demonstrating significant improvements in predictive performance. GEM~\cite{liu2018heterogeneous}, on the other hand, focuses on identifying fraudulent accounts on Alipay, a major cashless payment platform, by aggregating device and activity data. Yang et al.~\cite{yang2019using} further explore the integration of graph neural networks with external knowledge sources (e.g.,financial news), to forecast financial events. Despite their utility, fintech-related models are vulnerable to malicious attacks, which pose a serious threat to user privacy by potentially exposing the sensitive data of clients and companies. The confidential nature of the data used in financial event nowcasting models highlights the critical need for robust IP protections. The unauthorized extraction of these models not only threatens the competitive advantage and unique insights they provide but also risks significant economic losses, allowing the attackers to unscrupulously manipulate e-financial markets. 

\section{Discussions and Outlooks} \label{sec7}

Despite significant efforts in this field, accurate and efficient graph learning IP protection for real-world applications remains at an early stage~\cite{xu2024idea,peng2023intellectual,hubscher2022graph}, posing considerable challenges. Therefore, this section outlines several promising future research directions, emphasizing the need for innovative strategies to enhance the robustness and applicability of graph learning IP protection methods in practical and commercial settings.

\subsection{Efficient Graph Learning IP Protection Frameworks}
IP protection frameworks must be efficiently adapted to continuously growing deep and large-scale GML models, as constructing and verifying IP identifiers often requires extensive computations and optimization. Key research directions to enhance efficiency include but not limited to: \textit{(a) Robust IP identifiers~\cite{ma2023graphnei}:} Developing methods to create robust IP identifiers that are sensitive during construction but resistant to tampering or removal, potentially eliminating the need for retraining or extensive tuning of the target GML model. \textit{(b) Efficient verification~\cite{you2024gnnfingers}:} Reducing verification costs by leveraging surrogate models to optimize query efficiency and minimizing the number of necessary queries for remote verification processes. \textit{(c) Scalable IP protection~\cite{10646643}:} Improving the scalability of IP protection across different model architectures and tasks, ensuring that a single IP protection framework can be effectively applied to various applications and use cases in the graph domain.

\subsection{Secured IP Protection Pipelines}
A comprehensive graph learning IP protection pipeline encompasses various stages, from design and deployment to forensic analysis. This pipeline includes not only the construction and verification of IP identifiers but also additional components such as infringement detection, credible notarization mechanisms, and secure IP management platforms. The following are some possible research directions for enhancing the security of graph learning IP protection pipeline: \textit{(a) Active Defense Techniques~\cite{10.14778/3659437.3659457}:} Graph learning IP protection can prevent malicious attackers by identifying specific patterns in queries or graph data access that indicate malicious intent, leveraging anomaly detection in network states and query behaviors. \textit{(b) Automated Infringement Detection~\cite{yasaei2021gnn4ip}:} IP protection systems should effectively and economically identify model misuse instances, particularly within the GMLaaS service framework. This involves balancing the cost of verification with its accuracy, especially given the prevalence of pay-per-request models among service providers. In scenarios where no direct access to the model is available, mechanisms must be in place to track suspect data usage and its source model. \textit{(c) Credible Notarization Mechanisms and Platforms~\cite{sun2023deep}:} To prevent disputes over IP claims, it is essential to have a reliable platform for verifying IP identifiers. This platform can be operated by a trusted third party or a decentralized authority. In addition, the platform should maintain a comprehensive database of IP identifiers and support detecting the chronological order of their creation, thereby enabling the tracing of model distribution and usage history. 

\subsection{Unified and Modular Evaluation Benchmarks} 
Although various approaches have been developed for graph learning IP protection recently, these methods employ disparate criteria for evaluating essential attributes in graph learning IP protection~\cite{wu2022survey,ganz2021explaining}. The evaluation inconsistency hampers a fair comparison between different methods, complicates the choice and deployment of these methods. Consequently, establishing a unified, systematic, and empirical evaluation benchmark is crucial for advancing graph learning IP protection~\cite{platonov2023critical,xue2023turn}. The new benchmark should encompass a comprehensive spectrum of evaluation criteria, including a series of standardized datasets, quantitative criteria, and fundamental qualitative metrics like fidelity, robustness, capacity, and efficiency. Additionally, it should incorporate common attack scenarios and diverse application contexts. This standardization will facilitate more rigorous comparisons and encourage the development of more robust and applicable graph learning IP protection methods.

\subsection{IP Protection for Graph Learning-based LLMs} 
The rise of large foundation models in computer vision and natural language processing has also sparked a surge in developing large models in the field of graph learning~\cite{galkin2023towards,ye2023natural}. Graph learning in combination with Large Language Models (LLM) has permeated numerous domains, including social networks~\cite{deng2024advances}, e-commerce platforms~\cite{yu2024cosmo}, biomedical~\cite{xiao2024fuselinker}, and navigation~\cite{yuan2024gnnavi}. Developing high-performance graph learning-based LLMs in these fields requires substantial investment in infrastructure, data acquisition, and model training, highlighting the urgent need for robust IP protection tools. Effective IP protection methods must balance strong attack resistance, minimal performance overhead, and imperceptible impact on the model's functionality. Several challenges are associated with the IP protection for large-scale GML models, which waiting for future investigation: (a) The inherent complexity and expansive nature of large-scale GML models make them difficult to fully explore and interpret~\cite{wang2023empower}, complicating the integration of protective measures like watermarks and fingerprints, which may degrade performance or introduce new security vulnerabilities. (b) graph learning-based LLMs often support multimodal data inputs and leverage advanced learning techniques~\cite{liu2024git}, such as graph transformers and probabilistic models, which are not yet fully addressed by existing IP protection frameworks. (c) The scalability of these models across various applications poses additional risks, as their capabilities can be misused for malicious purposes~\cite{zhang2023adversarial}. Therefore, IP protection strategies for large-scale graph machine learning models should focus not only on preventing unauthorized use but also on detecting and mitigating model misuse.

\section{Conclusion} \label{sec8} 
This survey provides a pioneering and comprehensive exploration of Intellectual Property (IP) protection in general graph machine learning (GML) domain. We begin by introducing the first systematic taxonomy of GML IP protection methodologies, categorizing them based on key attributes including their mechanism, goal, function, and scenario. Our deep investigation into existing methods revealed how these techniques conduct proactive protection and their resistance to different level of attacks, with a particular focus on defending against model-level and data-level attacks. Furthermore, we develop a comprehensive evaluation framework to assess the quality, fidelity, and efficiency of GML IP protection, ensuring a robust comparison across different approaches and application scenarios. To support a wider range of research community, we elaborately compile existing benchmark datasets in GML IP protection, providing insights into data statistics, characteristics, and application scope. Additionally, we introduce PyGIP, a continuously updated online resource repository that facilitates the easy implementation and comparison of various advanced GML IP protection methods. Finally, we discuss current challenges and highlight the promising directions for future research. We believe this survey will serve as a foundational resource for both academic research and industrial practitioners, offering valuable guidance for the development of secure and robust IP protection in graph machine learning.


\bibliographystyle{IEEEtran}
\normalem
\bibliography{TKDE_main_file}

\begin{thebibliography}{100}
\providecommand{\url}[1]{#1}
\csname url@samestyle\endcsname
\providecommand{\newblock}{\relax}
\providecommand{\bibinfo}[2]{#2}
\providecommand{\BIBentrySTDinterwordspacing}{\spaceskip=0pt\relax}
\providecommand{\BIBentryALTinterwordstretchfactor}{4}
\providecommand{\BIBentryALTinterwordspacing}{\spaceskip=\fontdimen2\font plus
\BIBentryALTinterwordstretchfactor\fontdimen3\font minus \fontdimen4\font\relax}
\providecommand{\BIBforeignlanguage}[2]{{%
\expandafter\ifx\csname l@#1\endcsname\relax
\typeout{** WARNING: IEEEtran.bst: No hyphenation pattern has been}%
\typeout{** loaded for the language `#1'. Using the pattern for}%
\typeout{** the default language instead.}%
\else
\language=\csname l@#1\endcsname
\fi
#2}}
\providecommand{\BIBdecl}{\relax}
\BIBdecl

\bibitem{luo2024cross}
Z.~Luo, H.~Huang, J.~Lian, X.~Song, X.~Xie, and H.~Jin, ``Cross-links matter for link prediction: rethinking the debiased gnn from a data perspective,'' \emph{Advances in Neural Information Processing Systems}, vol.~36, 2024.

\bibitem{wang2023topological}
Y.~Wang, T.~Zhao, Y.~Zhao, Y.~Liu, X.~Cheng, N.~Shah, and T.~Derr, ``A topological perspective on demystifying gnn-based link prediction performance,'' \emph{arXiv preprint arXiv:2310.04612}, 2023.

\bibitem{zhang2023page}
S.~Zhang, J.~Zhang, X.~Song, S.~Adeshina, D.~Zheng, C.~Faloutsos, and Y.~Sun, ``Page-link: Path-based graph neural network explanation for heterogeneous link prediction,'' in \emph{Proceedings of the ACM Web Conference 2023}, 2023, pp. 3784--3793.

\bibitem{xu2022contrastive}
Z.~Xu, X.~Huang, Y.~Zhao, Y.~Dong, and J.~Li, ``Contrastive attributed network anomaly detection with data augmentation,'' in \emph{Pacific-Asia conference on knowledge discovery and data mining}.\hskip 1em plus 0.5em minus 0.4em\relax Springer, 2022, pp. 444--457.

\bibitem{han2022adbench}
S.~Han, X.~Hu, H.~Huang, M.~Jiang, and Y.~Zhao, ``Adbench: Anomaly detection benchmark,'' \emph{Advances in Neural Information Processing Systems}, vol.~35, pp. 32\,142--32\,159, 2022.

\bibitem{ding2021few}
K.~Ding, Q.~Zhou, H.~Tong, and H.~Liu, ``Few-shot network anomaly detection via cross-network meta-learning,'' in \emph{Proceedings of the Web Conference 2021}, 2021, pp. 2448--2456.

\bibitem{wu2019graph}
Z.~Wu, S.~Pan, G.~Long, J.~Jiang, and C.~Zhang, ``Graph wavenet for deep spatial-temporal graph modeling,'' \emph{arXiv preprint arXiv:1906.00121}, 2019.

\bibitem{shao2022pre}
Z.~Shao, Z.~Zhang, F.~Wang, and Y.~Xu, ``Pre-training enhanced spatial-temporal graph neural network for multivariate time series forecasting,'' in \emph{Proceedings of the 28th ACM SIGKDD conference on knowledge discovery and data mining}, 2022, pp. 1567--1577.

\bibitem{bai2020adaptive}
L.~Bai, L.~Yao, C.~Li, X.~Wang, and C.~Wang, ``Adaptive graph convolutional recurrent network for traffic forecasting,'' \emph{Advances in neural information processing systems}, vol.~33, pp. 17\,804--17\,815, 2020.

\bibitem{yang2023dgrec}
L.~Yang, S.~Wang, Y.~Tao, J.~Sun, X.~Liu, P.~S. Yu, and T.~Wang, ``Dgrec: Graph neural network for recommendation with diversified embedding generation,'' in \emph{Proceedings of the sixteenth ACM international conference on web search and data mining}, 2023, pp. 661--669.

\bibitem{wang2024distribution}
B.~Wang, J.~Chen, C.~Li, S.~Zhou, Q.~Shi, Y.~Gao, Y.~Feng, C.~Chen, and C.~Wang, ``Distributionally robust graph-based recommendation system,'' in \emph{Proceedings of the ACM on Web Conference 2024}, 2024, pp. 3777--3788.

\bibitem{sun2020disease}
Z.~Sun, H.~Yin, H.~Chen, T.~Chen, L.~Cui, and F.~Yang, ``Disease prediction via graph neural networks,'' \emph{IEEE Journal of Biomedical and Health Informatics}, vol.~25, no.~3, pp. 818--826, 2020.

\bibitem{choi2020learning}
E.~Choi, Z.~Xu, Y.~Li, M.~Dusenberry, G.~Flores, E.~Xue, and A.~Dai, ``Learning the graphical structure of electronic health records with graph convolutional transformer,'' in \emph{Proceedings of the AAAI conference on artificial intelligence}, vol.~34, no.~01, 2020, pp. 606--613.

\bibitem{mo2023map}
X.~Mo, Y.~Xing, H.~Liu, and C.~Lv, ``Map-adaptive multimodal trajectory prediction using hierarchical graph neural networks,'' \emph{IEEE Robotics and Automation Letters}, vol.~8, no.~6, pp. 3685--3692, 2023.

\bibitem{liu2023graph}
Z.~Liu, Y.~Zhai, J.~Li, G.~Wang, Y.~Miao, and H.~Wang, ``Graph relational reinforcement learning for mobile robot navigation in large-scale crowded environments,'' \emph{IEEE Transactions on Intelligent Transportation Systems}, vol.~24, no.~8, pp. 8776--8787, 2023.

\bibitem{zeng2022gnn}
L.~Zeng, C.~Yang, P.~Huang, Z.~Zhou, S.~Yu, and X.~Chen, ``Gnn at the edge: Cost-efficient graph neural network processing over distributed edge servers,'' \emph{IEEE Journal on Selected Areas in Communications}, vol.~41, no.~3, pp. 720--739, 2022.

\bibitem{philipp2020machine}
R.~Philipp, A.~Mladenow, C.~Strauss, and A.~V{\"o}lz, ``Machine learning as a service: Challenges in research and applications,'' in \emph{Proceedings of the 22nd International Conference on Information Integration and Web-based Applications \& Services}, 2020, pp. 396--406.

\bibitem{peng2023intellectual}
S.~Peng, Y.~Chen, J.~Xu, Z.~Chen, C.~Wang, and X.~Jia, ``Intellectual property protection of dnn models,'' \emph{World Wide Web}, vol.~26, no.~4, pp. 1877--1911, 2023.

\bibitem{he2022protecting}
X.~He, Q.~Xu, L.~Lyu, F.~Wu, and C.~Wang, ``Protecting intellectual property of language generation apis with lexical watermark,'' in \emph{Proceedings of the AAAI Conference on Artificial Intelligence}, vol.~36, no.~10, 2022, pp. 10\,758--10\,766.

\bibitem{chakraborty2020hardware}
A.~Chakraborty, A.~Mondai, and A.~Srivastava, ``Hardware-assisted intellectual property protection of deep learning models,'' in \emph{2020 57th ACM/IEEE Design Automation Conference (DAC)}, 2020, pp. 1--6.

\bibitem{10035510}
S.~Wang, Y.~Zheng, and X.~Jia, ``Secgnn: Privacy-preserving graph neural network training and inference as a cloud service,'' \emph{IEEE Transactions on Services Computing}, vol.~16, no.~4, pp. 2923--2938, 2023.

\bibitem{hunt2018chiron}
T.~Hunt, C.~Song, R.~Shokri, V.~Shmatikov, and E.~Witchel, ``Chiron: Privacy-preserving machine learning as a service,'' \emph{arXiv preprint arXiv:1803.05961}, 2018.

\bibitem{zhao2021veriml}
L.~Zhao, Q.~Wang, C.~Wang, Q.~Li, C.~Shen, and B.~Feng, ``Veriml: Enabling integrity assurances and fair payments for machine learning as a service,'' \emph{IEEE Transactions on Parallel and Distributed Systems}, vol.~32, no.~10, pp. 2524--2540, 2021.

\bibitem{tanuwidjaja2020privacy}
H.~C. Tanuwidjaja, R.~Choi, S.~Baek, and K.~Kim, ``Privacy-preserving deep learning on machine learning as a service—a comprehensive survey,'' \emph{IEEE Access}, vol.~8, pp. 167\,425--167\,447, 2020.

\bibitem{liang2024model}
J.~Liang, R.~Pang, C.~Li, and T.~Wang, ``Model extraction attacks revisited,'' in \emph{Proceedings of the 19th ACM Asia Conference on Computer and Communications Security}, 2024, pp. 1231--1245.

\bibitem{kesarwani2018model}
M.~Kesarwani, B.~Mukhoty, V.~Arya, and S.~Mehta, ``Model extraction warning in mlaas paradigm,'' in \emph{Proceedings of the 34th Annual Computer Security Applications Conference}, 2018, pp. 371--380.

\bibitem{hu2024learn}
H.~Hu, S.~Wang, T.~Dong, and M.~Xue, ``Learn what you want to unlearn: Unlearning inversion attacks against machine unlearning,'' \emph{arXiv preprint arXiv:2404.03233}, 2024.

\bibitem{ravindranathan2024cloud}
M.~K. Ravindranathan, D.~S. Vadivu, and N.~Rajagopalan, ``Cloud-driven machine learning with aws: A comprehensive review of services,'' in \emph{2024 International Conference on Intelligent and Innovative Technologies in Computing, Electrical and Electronics (IITCEE)}, 2024, pp. 1--8.

\bibitem{bhol2024machine}
S.~G. Bhol, S.~Mohanty, and P.~K. Pattnaik, ``Machine learning as a service cloud selection: An mcdm approach for optimal decision making,'' \emph{Procedia Computer Science}, vol. 233, pp. 909--918, 2024.

\bibitem{duan2024cat}
Y.~Duan, G.~Zhang, S.~Wang, X.~Peng, W.~Ziqi, J.~Mao, H.~Wu, X.~Jiang, and K.~Wang, ``Cat-gnn: Enhancing credit card fraud detection via causal temporal graph neural networks,'' \emph{arXiv preprint arXiv:2402.14708}, 2024.

\bibitem{li2022internet}
R.~Li, Z.~Liu, Y.~Ma, D.~Yang, and S.~Sun, ``Internet financial fraud detection based on graph learning,'' \emph{IEEE Transactions on Computational Social Systems}, vol.~10, no.~3, pp. 1394--1401, 2022.

\bibitem{an2024finsformer}
H.~An, R.~Ma, Y.~Yan, T.~Chen, Y.~Zhao, P.~Li, J.~Li, X.~Wang, D.~Fan, and C.~Lv, ``Finsformer: A novel approach to detecting financial attacks using transformer and cluster-attention,'' \emph{Applied Sciences}, vol.~14, no.~1, p. 460, 2024.

\bibitem{zhao2025surveya}
K.~Zhao, L.~Li, K.~Ding, N.~Z. Gong, Y.~Zhao, and Y.~Dong, ``A survey on model extraction attacks and defenses for large language models,'' \emph{arXiv preprint arXiv:2506.22521}, 2025.

\bibitem{chen2024scn_gnn}
J.~Chen, Q.~Chen, F.~Jiang, X.~Guo, K.~Sha, and Y.~Wang, ``Scn\_gnn: A gnn-based fraud detection algorithm combining strong node and graph topology information,'' \emph{Expert Systems with Applications}, vol. 237, p. 121643, 2024.

\bibitem{zhao2021exploiting}
X.~Zhao, W.~Zhang, X.~Xiao, and B.~Lim, ``Exploiting explanations for model inversion attacks,'' in \emph{Proceedings of the IEEE/CVF international conference on computer vision}, 2021, pp. 682--692.

\bibitem{krall2020gradient}
A.~Krall, D.~Finke, and H.~Yang, ``Gradient mechanism to preserve differential privacy and deter against model inversion attacks in healthcare analytics,'' in \emph{2020 42nd Annual International Conference of the IEEE Engineering in Medicine \& Biology Society (EMBC)}, 2020, pp. 5714--5717.

\bibitem{murali2023towards}
L.~Murali, G.~Gopakumar, D.~M. Viswanathan, and P.~Nedungadi, ``Towards electronic health record-based medical knowledge graph construction, completion, and applications: A literature study,'' \emph{Journal of biomedical informatics}, vol. 143, p. 104403, 2023.

\bibitem{golmaei2021deepnote}
S.~N. Golmaei and X.~Luo, ``Deepnote-gnn: predicting hospital readmission using clinical notes and patient network,'' in \emph{Proceedings of the 12th ACM International Conference on Bioinformatics, Computational Biology, and Health Informatics}, 2021, pp. 1--9.

\bibitem{xu2023cover}
X.~Xu, Q.~Hao, Z.~Yang, B.~Li, D.~Liebovitz, G.~Wang, and C.~A. Gunter, ``How to cover up anomalous accesses to electronic health records,'' in \emph{32nd USENIX Security Symposium (USENIX Security 23)}, 2023, pp. 229--246.

\bibitem{paul2024systematic}
S.~G. Paul, A.~Saha, M.~Z. Hasan, S.~R.~H. Noori, and A.~Moustafa, ``A systematic review of graph neural network in healthcare-based applications: Recent advances, trends, and future directions,'' \emph{IEEE Access}, vol.~12, pp. 15\,145--15\,170, 2024.

\bibitem{oliynyk2023know}
D.~Oliynyk, R.~Mayer, and A.~Rauber, ``I know what you trained last summer: A survey on stealing machine learning models and defences,'' \emph{ACM Computing Surveys}, vol.~55, no. 14s, pp. 1--41, 2023.

\bibitem{gong2020model}
X.~Gong, Q.~Wang, Y.~Chen, W.~Yang, and X.~Jiang, ``Model extraction attacks and defenses on cloud-based machine learning models,'' \emph{IEEE Communications Magazine}, vol.~58, no.~12, pp. 83--89, 2020.

\bibitem{sun2022adversarial}
L.~Sun, Y.~Dou, C.~Yang, K.~Zhang, J.~Wang, S.~Y. Philip, L.~He, and B.~Li, ``Adversarial attack and defense on graph data: A survey,'' \emph{IEEE Transactions on Knowledge and Data Engineering}, vol.~35, no.~8, pp. 7693--7711, 2022.

\bibitem{fang2024privacy}
H.~Fang, Y.~Qiu, H.~Yu, W.~Yu, J.~Kong, B.~Chong, B.~Chen, X.~Wang, and S.-T. Xia, ``Privacy leakage on dnns: A survey of model inversion attacks and defenses,'' \emph{arXiv preprint arXiv:2402.04013}, 2024.

\bibitem{tramer2016stealing}
F.~Tram{\`e}r, F.~Zhang, A.~Juels, M.~K. Reiter, and T.~Ristenpart, ``Stealing machine learning models via prediction $\{$APIs$\}$,'' in \emph{25th USENIX security symposium (USENIX Security 16)}, 2016, pp. 601--618.

\bibitem{milli2019model}
S.~Milli, L.~Schmidt, A.~D. Dragan, and M.~Hardt, ``Model reconstruction from model explanations,'' in \emph{Proceedings of the Conference on Fairness, Accountability, and Transparency}, 2019, pp. 1--9.

\bibitem{orekondy2019knockoff}
T.~Orekondy, B.~Schiele, and M.~Fritz, ``Knockoff nets: Stealing functionality of black-box models,'' in \emph{Proceedings of the IEEE/CVF conference on computer vision and pattern recognition}, 2019, pp. 4954--4963.

\bibitem{9833607}
Y.~Shen, X.~He, Y.~Han, and Y.~Zhang, ``Model stealing attacks against inductive graph neural networks,'' in \emph{2022 IEEE Symposium on Security and Privacy (SP)}, 2022, pp. 1175--1192.

\bibitem{defazio2019adversarial}
D.~DeFazio and A.~Ramesh, ``Adversarial model extraction on graph neural networks,'' \emph{arXiv preprint arXiv:1912.07721}, 2019.

\bibitem{shen2022model}
Y.~Shen, X.~He, Y.~Han, and Y.~Zhang, ``Model stealing attacks against inductive graph neural networks,'' in \emph{2022 IEEE Symposium on Security and Privacy (SP)}, 2022, pp. 1175--1192.

\bibitem{wu2022model}
B.~Wu, X.~Yang, S.~Pan, and X.~Yuan, ``Model extraction attacks on graph neural networks: Taxonomy and realisation,'' in \emph{Proceedings of the 2022 ACM on Asia conference on computer and communications security}, 2022, pp. 337--350.

\bibitem{chen2022knowledge}
J.~Chen, W.~Fan, G.~Zhu, X.~Zhao, C.~Yuan, Q.~Li, and Y.~Huang, ``Knowledge-enhanced black-box attacks for recommendations,'' in \emph{Proceedings of the 28th ACM SIGKDD Conference on Knowledge Discovery and Data Mining}, 2022, pp. 108--117.

\bibitem{zhuang2024unveiling}
Y.~Zhuang, C.~Shi, M.~Zhang, J.~Chen, L.~Lyu, P.~Zhou, and L.~Sun, ``Unveiling the secrets without data: Can graph neural networks be exploited through $\{$Data-Free$\}$ model extraction attacks?'' in \emph{33rd USENIX Security Symposium (USENIX Security 24)}, 2024, pp. 5251--5268.

\bibitem{yin2021comprehensive}
X.~Yin, Y.~Zhu, and J.~Hu, ``A comprehensive survey of privacy-preserving federated learning: A taxonomy, review, and future directions,'' \emph{ACM Computing Surveys (CSUR)}, vol.~54, no.~6, pp. 1--36, 2021.

\bibitem{fredrikson2015model}
M.~Fredrikson, S.~Jha, and T.~Ristenpart, ``Model inversion attacks that exploit confidence information and basic countermeasures,'' in \emph{Proceedings of the 22nd ACM SIGSAC conference on computer and communications security}, 2015, pp. 1322--1333.

\bibitem{wang2022group}
X.~Wang and W.~H. Wang, ``Group property inference attacks against graph neural networks,'' in \emph{Proceedings of the 2022 ACM SIGSAC Conference on Computer and Communications Security}, 2022, pp. 2871--2884.

\bibitem{zhang2022model}
Z.~Zhang, Q.~Liu, Z.~Huang, H.~Wang, C.-K. Lee, and E.~Chen, ``Model inversion attacks against graph neural networks,'' \emph{IEEE Transactions on Knowledge and Data Engineering}, 2022.

\bibitem{zhang2021graphmi}
Z.~Zhang, Q.~Liu, Z.~Huang, H.~Wang, C.~Lu, C.~Liu, and E.~Chen, ``Graphmi: Extracting private graph data from graph neural networks,'' \emph{arXiv preprint arXiv:2106.02820}, 2021.

\bibitem{liu2023model}
R.~Liu, W.~Zhou, J.~Zhang, X.~Liu, P.~Si, and H.~Li, ``Model inversion attacks on homogeneous and heterogeneous graph neural networks,'' \emph{arXiv preprint arXiv:2310.09800}, 2023.

\bibitem{aivodji2019gamin}
U.~A{\"\i}vodji, S.~Gambs, and T.~Ther, ``Gamin: An adversarial approach to black-box model inversion,'' \emph{arXiv preprint arXiv:1909.11835}, 2019.

\bibitem{hartung2000digital}
F.~Hartung and F.~Ramme, ``Digital rights management and watermarking of multimedia content for m-commerce applications,'' \emph{IEEE communications magazine}, vol.~38, no.~11, pp. 78--84, 2000.

\bibitem{dai2024pregip}
E.~Dai, M.~Lin, and S.~Wang, ``Pregip: Watermarking the pretraining of graph neural networks for deep intellectual property protection,'' \emph{arXiv preprint arXiv:2402.04435}, 2024.

\bibitem{xu2023watermarking}
J.~Xu, S.~Koffas, O.~Ersoy, and S.~Picek, ``Watermarking graph neural networks based on backdoor attacks,'' in \emph{2023 IEEE 8th European Symposium on Security and Privacy (EuroS\&P)}, 2023, pp. 1179--1197.

\bibitem{zhao2024transferable}
X.~Zhao, H.~Wu, and X.~Zhang, ``Transferable watermarking to self-supervised pre-trained graph encoders by trigger embeddings,'' \emph{arXiv preprint arXiv:2406.13177}, 2024.

\bibitem{you2024gnnfingers}
X.~You, Y.~Jiang, J.~Xu, M.~Zhang, and M.~Yang, ``Gnnfingers: A fingerprinting framework for verifying ownerships of graph neural networks,'' in \emph{Proceedings of the ACM on Web Conference 2024}, 2024, pp. 652--663.

\bibitem{you2024gnnguard}
X.~You, Y.~Jiang, J.~Xu, and M.~Zhang, ``Gnnguard: A fingerprinting framework for verifying ownerships of graph neural networks,'' in \emph{The Web Conference 2024}, 2024.

\bibitem{chen2020smoothing}
J.~Chen, X.~Lin, H.~Xiong, Y.~Wu, H.~Zheng, and Q.~Xuan, ``Smoothing adversarial training for gnn,'' \emph{IEEE Transactions on Computational Social Systems}, vol.~8, no.~3, pp. 618--629, 2020.

\bibitem{zheng2024improving}
X.~Zheng, B.~Wu, A.~X. Zhang, and W.~Li, ``Improving robustness of gnn-based anomaly detection by graph adversarial training,'' in \emph{Proceedings of the 2024 Joint International Conference on Computational Linguistics, Language Resources and Evaluation (LREC-COLING 2024)}, 2024, pp. 8902--8912.

\bibitem{mueller2022sok}
T.~T. Mueller, D.~Usynin, J.~C. Paetzold, D.~Rueckert, and G.~Kaissis, ``Sok: Differential privacy on graph-structured data,'' \emph{arXiv preprint arXiv:2203.09205}, 2022.

\bibitem{hsieh2021netfense}
I.-C. Hsieh and C.-T. Li, ``Netfense: Adversarial defenses against privacy attacks on neural networks for graph data,'' \emph{IEEE Transactions on Knowledge and Data Engineering}, vol.~35, no.~1, pp. 796--809, 2021.

\bibitem{gnnguard}
X.~Zhang and M.~Zitnik, ``Gnnguard: Defending graph neural networks against adversarial attacks,'' \emph{Advances in neural information processing systems}, vol.~33, pp. 9263--9275, 2020.

\bibitem{dai2024comprehensive}
E.~Dai, T.~Zhao, H.~Zhu, J.~Xu, Z.~Guo, H.~Liu, J.~Tang, and S.~Wang, ``A comprehensive survey on trustworthy graph neural networks: Privacy, robustness, fairness, and explainability,'' \emph{Machine Intelligence Research}, pp. 1--51, 2024.

\bibitem{wu2022trustworthy}
B.~Wu, Y.~Bian, H.~Zhang, J.~Li, J.~Yu, L.~Chen, C.~Chen, and J.~Huang, ``Trustworthy graph learning: Reliability, explainability, and privacy protection,'' in \emph{Proceedings of the 28th ACM SIGKDD Conference on Knowledge Discovery and Data Mining}, 2022, pp. 4838--4839.

\bibitem{zhang2022trustworthy}
H.~Zhang, B.~Wu, X.~Yuan, S.~Pan, H.~Tong, and J.~Pei, ``Trustworthy graph neural networks: Aspects, methods and trends,'' \emph{arXiv preprint arXiv:2205.07424}, 2022.

\bibitem{Wu2022_TrustGL}
B.~Wu, Y.~Bian, H.~Zhang, J.~Li, J.~Yu, L.~Chen, C.~Chen, and J.~Huang, ``Trustworthy graph learning: Reliability, explainability, and privacy protection,'' in \emph{Proceedings of the 28th ACM SIGKDD Conference on Knowledge Discovery and Data Mining}, 2022, p. 4838–4839.

\bibitem{Sun2024_Adversarial}
H.~Sun, W.~Yang, and Y.~Xiao, ``A review of adversarial attacks and defenses on graphs,'' in \emph{Proceedings of the 4th International Conference on Artificial Intelligence and Computer Engineering}, 2024, p. 416–421.

\bibitem{zheng2graph}
Q.~Zheng, X.~Zou, Y.~Dong, Y.~Cen, D.~Yin, J.~Xu, Y.~Yang, and J.~Tang, ``Graph robustness benchmark: Benchmarking the adversarial robustness of graph machine learning,'' in \emph{Thirty-fifth Conference on Neural Information Processing Systems Datasets and Benchmarks Track}, 2021.

\bibitem{9878092}
L.~Sun, Y.~Dou, C.~Yang, K.~Zhang, J.~Wang, P.~S. Yu, L.~He, and B.~Li, ``Adversarial attack and defense on graph data: A survey,'' \emph{IEEE Transactions on Knowledge and Data Engineering}, vol.~35, no.~8, pp. 7693--7711, 2023.

\bibitem{xue2021intellectual}
M.~Xue, Y.~Zhang, J.~Wang, and W.~Liu, ``Intellectual property protection for deep learning models: Taxonomy, methods, attacks, and evaluations,'' \emph{IEEE Transactions on Artificial Intelligence}, vol.~3, no.~6, pp. 908--923, 2021.

\bibitem{sun2023deep}
Y.~Sun, T.~Liu, P.~Hu, Q.~Liao, S.~Fu, N.~Yu, D.~Guo, Y.~Liu, and L.~Liu, ``Deep intellectual property protection: A survey,'' \emph{arXiv preprint arXiv:2304.14613}, 2023.

\bibitem{ju2024survey}
W.~Ju, S.~Yi, Y.~Wang, Z.~Xiao, Z.~Mao, H.~Li, Y.~Gu, Y.~Qin, N.~Yin, S.~Wang \emph{et~al.}, ``A survey of graph neural networks in real world: Imbalance, noise, privacy and ood challenges,'' \emph{arXiv preprint arXiv:2403.04468}, 2024.

\bibitem{guarino2021machine}
A.~Guarino, N.~Lettieri, D.~Malandrino, and R.~Zaccagnino, ``A machine learning-based approach to identify unlawful practices in online terms of service: analysis, implementation and evaluation,'' \emph{Neural Computing and Applications}, vol.~33, pp. 17\,569--17\,587, 2021.

\bibitem{wu2019simplifying}
F.~Wu, A.~Souza, T.~Zhang, C.~Fifty, T.~Yu, and K.~Weinberger, ``Simplifying graph convolutional networks,'' in \emph{International conference on machine learning}, 2019, pp. 6861--6871.

\bibitem{hamilton2017inductive}
W.~Hamilton, Z.~Ying, and J.~Leskovec, ``Inductive representation learning on large graphs,'' \emph{Advances in neural information processing systems}, vol.~30, 2017.

\bibitem{velivckovic2017graph}
P.~Veli{\v{c}}kovi{\'c}, G.~Cucurull, A.~Casanova, A.~Romero, P.~Lio, and Y.~Bengio, ``Graph attention networks,'' \emph{arXiv preprint arXiv:1710.10903}, 2017.

\bibitem{xu2018powerful}
K.~Xu, W.~Hu, J.~Leskovec, and S.~Jegelka, ``How powerful are graph neural networks?'' \emph{arXiv preprint arXiv:1810.00826}, 2018.

\bibitem{grover2016node2vec}
A.~Grover and J.~Leskovec, ``node2vec: Scalable feature learning for networks,'' in \emph{Proceedings of the 22nd ACM SIGKDD international conference on Knowledge discovery and data mining}, 2016, pp. 855--864.

\bibitem{zhang2019heterogeneous}
C.~Zhang, D.~Song, C.~Huang, A.~Swami, and N.~V. Chawla, ``Heterogeneous graph neural network,'' in \emph{Proceedings of the 25th ACM SIGKDD international conference on knowledge discovery \& data mining}, 2019, pp. 793--803.

\bibitem{wang2025cega}
Z.~Wang, M.~Lin, B.~Shen, K.~Anderson, M.~Liu, T.~Cai, and Y.~Dong, ``Cega: A cost-effective approach for graph-based model extraction and acquisition,'' \emph{arXiv preprint arXiv:2506.17709}, 2025.

\bibitem{park2019estimating}
N.~Park, A.~Kan, X.~L. Dong, T.~Zhao, and C.~Faloutsos, ``Estimating node importance in knowledge graphs using graph neural networks,'' in \emph{Proceedings of the 25th ACM SIGKDD international conference on knowledge discovery \& data mining}, 2019, pp. 596--606.

\bibitem{jia2020residual}
J.~Jia and A.~R. Benson, ``Residual correlation in graph neural network regression,'' in \emph{Proceedings of the 26th ACM SIGKDD international conference on knowledge discovery \& data mining}, 2020, pp. 588--598.

\bibitem{pmlr-v119-bianchi20a}
F.~M. Bianchi, D.~Grattarola, and C.~Alippi, ``Spectral clustering with graph neural networks for graph pooling,'' in \emph{Proceedings of the 37th International Conference on Machine Learning}, 2020, pp. 874--883.

\bibitem{tsitsulin2023graph}
A.~Tsitsulin, J.~Palowitch, B.~Perozzi, and E.~M{\"u}ller, ``Graph clustering with graph neural networks,'' \emph{Journal of Machine Learning Research}, vol.~24, no. 127, pp. 1--21, 2023.

\bibitem{you2019position}
J.~You, R.~Ying, and J.~Leskovec, ``Position-aware graph neural networks,'' in \emph{International conference on machine learning}.\hskip 1em plus 0.5em minus 0.4em\relax PMLR, 2019, pp. 7134--7143.

\bibitem{chen2020iterative}
Y.~Chen, L.~Wu, and M.~Zaki, ``Iterative deep graph learning for graph neural networks: Better and robust node embeddings,'' \emph{Advances in neural information processing systems}, vol.~33, pp. 19\,314--19\,326, 2020.

\bibitem{zhang2018link}
M.~Zhang and Y.~Chen, ``Link prediction based on graph neural networks,'' \emph{Advances in neural information processing systems}, vol.~31, 2018.

\bibitem{ying2018graph}
R.~Ying, R.~He, K.~Chen, P.~Eksombatchai, W.~L. Hamilton, and J.~Leskovec, ``Graph convolutional neural networks for web-scale recommender systems,'' in \emph{Proceedings of the 24th ACM SIGKDD international conference on knowledge discovery \& data mining}, 2018, pp. 974--983.

\bibitem{wei2023dual}
X.~Wei, Y.~Liu, J.~Sun, Y.~Jiang, Q.~Tang, and K.~Yuan, ``Dual subgraph-based graph neural network for friendship prediction in location-based social networks,'' \emph{ACM Transactions on Knowledge Discovery from Data}, vol.~17, no.~3, pp. 1--28, 2023.

\bibitem{wu2020comprehensive}
Z.~Wu, S.~Pan, F.~Chen, G.~Long, C.~Zhang, and S.~Y. Philip, ``A comprehensive survey on graph neural networks,'' \emph{IEEE transactions on neural networks and learning systems}, vol.~32, no.~1, pp. 4--24, 2020.

\bibitem{li2024drug}
W.~Li, W.~Ma, M.~Yang, and X.~Tang, ``Drug repurposing based on the dtd-gnn graph neural network: revealing the relationships among drugs, targets and diseases,'' \emph{BMC genomics}, vol.~25, 2024.

\bibitem{zhao2021graphsmote}
T.~Zhao, X.~Zhang, and S.~Wang, ``Graphsmote: Imbalanced node classification on graphs with graph neural networks,'' in \emph{Proceedings of the 14th ACM international conference on web search and data mining}, 2021, pp. 833--841.

\bibitem{sarlin2020superglue}
P.-E. Sarlin, D.~DeTone, T.~Malisiewicz, and A.~Rabinovich, ``Superglue: Learning feature matching with graph neural networks,'' in \emph{Proceedings of the IEEE/CVF conference on computer vision and pattern recognition}, 2020, pp. 4938--4947.

\bibitem{hamilton2017representation}
W.~L. Hamilton, R.~Ying, and J.~Leskovec, ``Representation learning on graphs: Methods and applications,'' \emph{arXiv preprint arXiv:1709.05584}, 2017.

\bibitem{liao2019efficient}
R.~Liao, Y.~Li, Y.~Song, S.~Wang, W.~Hamilton, D.~K. Duvenaud, R.~Urtasun, and R.~Zemel, ``Efficient graph generation with graph recurrent attention networks,'' \emph{Advances in neural information processing systems}, vol.~32, 2019.

\bibitem{you2018graphrnn}
J.~You, R.~Ying, X.~Ren, W.~Hamilton, and J.~Leskovec, ``Graphrnn: Generating realistic graphs with deep auto-regressive models,'' in \emph{International conference on machine learning}.\hskip 1em plus 0.5em minus 0.4em\relax PMLR, 2018, pp. 5708--5717.

\bibitem{zhao2025survey}
K.~Zhao, L.~Li, K.~Ding, N.~Z. Gong, Y.~Zhao, and Y.~Dong, ``A survey of model extraction attacks and defenses in distributed computing environments,'' \emph{arXiv preprint arXiv:2502.16065}, 2025.

\bibitem{guan2024realistic}
F.~Guan, T.~Zhu, H.~Tong, and W.~Zhou, ``A realistic model extraction attack against graph neural networks,'' \emph{Knowledge-Based Systems}, p. 112144, 2024.

\bibitem{cheng2025atom}
Z.~Cheng, B.~Shen, T.~Sha, Y.~Gao, S.~Li, and Y.~Dong, ``Atom: A framework of detecting query-based model extraction attacks for graph neural networks,'' \emph{arXiv preprint arXiv:2503.16693}, 2025.

\bibitem{jia2025sigfinger}
J.~Jia, R.~Li, C.~Wu, S.~Ma, L.~Wang, and R.~H. Deng, ``Sigfinger: A subtle and interactive gnn fingerprinting scheme via spatial structure inference perturbation,'' \emph{IEEE Transactions on Dependable and Secure Computing}, 2025.

\bibitem{li2022towards}
H.~Li, J.~Zhang, S.~Gao, L.~Wu, W.~Zhou, and R.~Wang, ``Towards query-limited adversarial attacks on graph neural networks,'' in \emph{2022 IEEE 34th International Conference on Tools with Artificial Intelligence (ICTAI)}, 2022, pp. 516--521.

\bibitem{chen2025vgfl}
Y.~Chen and B.~Zhou, ``Vgfl-sa: Vertical graph federated learning structure attack based on contrastive learning,'' \emph{arXiv preprint arXiv:2502.16793}, 2025.

\bibitem{juuti2019prada}
M.~Juuti, S.~Szyller, S.~Marchal, and N.~Asokan, ``Prada: protecting against dnn model stealing attacks,'' in \emph{2019 IEEE European Symposium on Security and Privacy (EuroS\&P)}, 2019, pp. 512--527.

\bibitem{wang2023making}
H.~Wang, Z.~Zhang, M.~Chen, and S.~He, ``Making watermark survive model extraction attacks in graph neural networks,'' in \emph{ICC 2023-IEEE International Conference on Communications}, 2023, pp. 57--62.

\bibitem{zhang2024imperceptible}
L.~Zhang, M.~Xue, L.~Y. Zhang, Y.~Zhang, and W.~Liu, ``An imperceptible and owner-unique watermarking method for graph neural networks,'' in \emph{Proceedings of the ACM Turing Award Celebration Conference-China 2024}, 2024, pp. 108--113.

\bibitem{zhao2021watermarking}
X.~Zhao, H.~Wu, and X.~Zhang, ``Watermarking graph neural networks by random graphs,'' in \emph{2021 9th International Symposium on Digital Forensics and Security (ISDFS)}, 2021, pp. 1--6.

\bibitem{bachina2024genie}
V.~S.~P. Bachina, A.~Gangwal, A.~A. Sharma, and C.~Sharma, ``Genie: Watermarking graph neural networks for link prediction,'' \emph{arXiv preprint arXiv:2406.04805}, 2024.

\bibitem{JIANG2022109309}
M.~Jiang, Z.~Li, P.~Fu, W.~Cai, M.~Cui, G.~Xiong, and G.~Gou, ``Accurate mobile-app fingerprinting using flow-level relationship with graph neural networks,'' \emph{Computer Networks}, vol. 217, p. 109309, 2022.

\bibitem{waheed2023using}
A.~Waheed, ``On using embeddings for ownership verification of graph neural networks,'' Master's thesis, University of Waterloo, 2023.

\bibitem{zhang2024survey}
Y.~Zhang, Y.~Zhao, Z.~Li, X.~Cheng, Y.~Wang, O.~Kotevska, S.~Y. Philip, and T.~Derr, ``A survey on privacy in graph neural networks: Attacks, preservation, and applications,'' \emph{IEEE Transactions on Knowledge and Data Engineering}, 2024.

\bibitem{gosch2024adversarial}
L.~Gosch, S.~Geisler, D.~Sturm, B.~Charpentier, D.~Z{\"u}gner, and S.~G{\"u}nnemann, ``Adversarial training for graph neural networks: Pitfalls, solutions, and new directions,'' \emph{Advances in Neural Information Processing Systems}, vol.~36, 2024.

\bibitem{wang2019adversarial}
S.~Wang, Z.~Chen, J.~Ni, X.~Yu, Z.~Li, H.~Chen, and P.~S. Yu, ``Adversarial defense framework for graph neural network,'' \emph{arXiv preprint arXiv:1905.03679}, 2019.

\bibitem{kumar2020adversary}
C.~Kumar, R.~Ryan, and M.~Shao, ``Adversary for social good: Protecting familial privacy through joint adversarial attacks,'' in \emph{Proceedings of the AAAI conference on artificial intelligence}, vol.~34, no.~07, 2020, pp. 11\,304--11\,311.

\bibitem{liao2020graph}
P.~Liao, H.~Zhao, K.~Xu, T.~S. Jaakkola, G.~Gordon, S.~Jegelka, and R.~Salakhutdinov, ``Graph adversarial networks: Protecting information against adversarial attacks,'' \emph{arXiv preprint}, 2020.

\bibitem{zheng2024improve}
X.~Zheng, B.~Wu, A.~X. Zhang, and W.~Li, ``Improving robustness of gnn-based anomaly detection by graph adversarial training,'' in \emph{Proceedings of the 2024 Joint International Conference on Computational Linguistics, Language Resources and Evaluation (LREC-COLING 2024)}, 2024, pp. 8902--8912.

\bibitem{10646643}
A.~Waheed, V.~Duddu, and N.~Asokan, ``Grove: Ownership verification of graph neural networks using embeddings,'' in \emph{2024 IEEE Symposium on Security and Privacy (SP)}, 2024, pp. 2460--2477.

\bibitem{ennadir2023unboundattack}
S.~Ennadir, A.~Alkhatib, G.~Nikolentzos, M.~Vazirgiannis, and H.~Bostr{\"o}m, ``Unboundattack: Generating unbounded adversarial attacks to graph neural networks,'' in \emph{International Conference on Complex Networks and Their Applications}, 2023, pp. 100--111.

\bibitem{wu2021adapting}
B.~Wu, X.~Yang, S.~Pan, and X.~Yuan, ``Adapting membership inference attacks to gnn for graph classification: Approaches and implications,'' in \emph{2021 IEEE International Conference on Data Mining (ICDM)}, 2021, pp. 1421--1426.

\bibitem{xu2024query}
Y.~Xu, B.~Fang, M.~Li, X.~Liu, and Z.~Tian, ``Query-efficient model inversion attacks: An information flow view,'' \emph{IEEE Transactions on Information Forensics and Security}, 2024.

\bibitem{hu2022membership}
H.~Hu, Z.~Salcic, L.~Sun, G.~Dobbie, P.~S. Yu, and X.~Zhang, ``Membership inference attacks on machine learning: A survey,'' \emph{ACM Computing Surveys (CSUR)}, vol.~54, no. 11s, pp. 1--37, 2022.

\bibitem{he2021node}
X.~He, R.~Wen, Y.~Wu, M.~Backes, Y.~Shen, and Y.~Zhang, ``Node-level membership inference attacks against graph neural networks,'' \emph{arXiv preprint arXiv:2102.05429}, 2021.

\bibitem{jnaini2022powerful}
A.~Jnaini, A.~Bettar, and M.~A. Koulali, ``How powerful are membership inference attacks on graph neural networks?'' in \emph{Proceedings of the 34th International Conference on Scientific and Statistical Database Management}, 2022, pp. 1--4.

\bibitem{niuimproving}
P.~Niu, C.~Pan, S.~Chen, and O.~Milenkovic, ``Improving defense mechanisms for subgraph-structure membership inference attacks,'' 2025.

\bibitem{conti2022label}
M.~Conti, J.~Li, S.~Picek, and J.~Xu, ``Label-only membership inference attack against node-level graph neural networks,'' in \emph{Proceedings of the 15th ACM Workshop on Artificial Intelligence and Security}, 2022, pp. 1--12.

\bibitem{olatunji2021membership}
I.~E. Olatunji, W.~Nejdl, and M.~Khosla, ``Membership inference attack on graph neural networks,'' in \emph{2021 Third IEEE International Conference on Trust, Privacy and Security in Intelligent Systems and Applications (TPS-ISA)}.\hskip 1em plus 0.5em minus 0.4em\relax IEEE, 2021, pp. 11--20.

\bibitem{yang2023membership}
J.~Yang, H.~Li, W.~Fan, X.~Zhang, and M.~Hao, ``Membership inference attacks against the graph classification,'' in \emph{GLOBECOM 2023-2023 IEEE Global Communications Conference}, 2023, pp. 6729--6734.

\bibitem{anand2024gradient}
D.~Anand~Sinha, Y.~Liu, R.~Du, and Y.~Shen, ``Gradient inversion attack on graph neural networks,'' \emph{arXiv e-prints}, pp. arXiv--2411, 2024.

\bibitem{xiao2024fedgig}
T.~Xiao, Y.~Li, Y.~Qi, H.~Wang, and R.~Li, ``Fedgig: Graph inversion from gradient in federated learning,'' \emph{arXiv preprint arXiv:2412.18513}, 2024.

\bibitem{song2023gnnbleed}
Z.~Song, E.~Kabir, and S.~Mehnaz, ``Gnnbleed: Inference attacks to unveil private edges in graphs with realistic access to gnn models,'' \emph{arXiv preprint arXiv:2311.16139}, 2023.

\bibitem{wang2023link}
X.~Wang and W.~H. Wang, ``Link membership inference attacks against unsupervised graph representation learning,'' in \emph{Proceedings of the 39th Annual Computer Security Applications Conference}, 2023, pp. 477--491.

\bibitem{guan2025topology}
F.~Guan, T.~Zhu, W.~Zhou, and P.~S. Yu, ``Topology-based node-level membership inference attacks on graph neural networks,'' \emph{IEEE Transactions on Big Data}, 2025.

\bibitem{shaikhelislamov2024study}
D.~Shaikhelislamov, K.~Lukyanov, N.~Severin, M.~Drobyshevskiy, I.~Makarov, and D.~Turdakov, ``A study of graph neural networks for link prediction on vulnerability to membership attacks,'' \emph{Journal of Mathematical Sciences}, pp. 1--11, 2024.

\bibitem{wang2024subgraph}
X.~Wang and W.~H. Wang, ``Subgraph structure membership inference attacks against graph neural networks,'' \emph{Proceedings on Privacy Enhancing Technologies}, 2024.

\bibitem{lin2024stealing}
M.~Lin, E.~Dai, J.~Xu, J.~Jia, X.~Zhang, and S.~Wang, ``Stealing training graphs from graph neural networks,'' \emph{arXiv preprint arXiv:2411.11197}, 2024.

\bibitem{dai2025graph}
J.~Dai and Y.~Lu, ``Graph-level label-only membership inference attack against graph neural networks,'' \emph{arXiv preprint arXiv:2503.19070}, 2025.

\bibitem{olatunji2021releasing}
I.~E. Olatunji, T.~Funke, and M.~Khosla, ``Releasing graph neural networks with differential privacy guarantees,'' \emph{arXiv preprint arXiv:2109.08907}, 2021.

\bibitem{xu2023mdp}
W.~Xu, B.~Shi, J.~Zhang, Z.~Feng, T.~Pan, and B.~Dong, ``Mdp: Privacy-preserving gnn based on matrix decomposition and differential privacy,'' in \emph{2023 IEEE International Conference on Joint Cloud Computing (JCC)}, 2023, pp. 38--45.

\bibitem{tran2022heterogeneous}
K.~Tran, P.~Lai, N.~Phan, I.~Khalil, Y.~Ma, A.~Khreishah, M.~T. Thai, and X.~Wu, ``Heterogeneous randomized response for differential privacy in graph neural networks,'' in \emph{2022 IEEE International Conference on Big Data (Big Data)}, 2022, pp. 1582--1587.

\bibitem{wei2024poincare}
Y.~Wei, H.~Yuan, X.~Fu, Q.~Sun, H.~Peng, X.~Li, and C.~Hu, ``Poincar{\'e} differential privacy for hierarchy-aware graph embedding,'' in \emph{Proceedings of the AAAI Conference on Artificial Intelligence}, vol.~38, no.~8, 2024, pp. 9160--9168.

\bibitem{dai2019adversarial}
Q.~Dai, X.~Shen, L.~Zhang, Q.~Li, and D.~Wang, ``Adversarial training methods for network embedding,'' in \emph{The world wide web conference}, 2019, pp. 329--339.

\bibitem{zhang2021multi}
X.~Zhang, L.~Zhang, B.~Jin, and X.~Lu, ``A multi-view confidence-calibrated framework for fair and stable graph representation learning,'' in \emph{2021 IEEE International Conference on Data Mining (ICDM)}, 2021, pp. 1493--1498.

\bibitem{zhou2023strengthening}
Z.~Zhou, C.~Zhou, X.~Li, J.~Yao, Q.~Yao, and B.~Han, ``On strengthening and defending graph reconstruction attack with markov chain approximation,'' in \emph{International Conference on Machine Learning}, 2023, pp. 42\,843--42\,877.

\bibitem{jin2020graph}
W.~Jin, Y.~Ma, X.~Liu, X.~Tang, S.~Wang, and J.~Tang, ``Graph structure learning for robust graph neural networks,'' in \emph{Proceedings of the 26th ACM SIGKDD international conference on knowledge discovery \& data mining}, 2020, pp. 66--74.

\bibitem{boratto2024robustness}
L.~Boratto, F.~Fabbri, G.~Fenu, M.~Marras, and G.~Medda, ``Robustness in fairness against edge-level perturbations in gnn-based recommendation,'' in \emph{European Conference on Information Retrieval}, 2024, pp. 38--55.

\bibitem{cai2024privacy}
L.~Cai, J.~Tang, S.~Dang, and G.~Chen, ``Privacy protection and utility trade-off for social graph embedding,'' \emph{Information Sciences}, p. 120866, 2024.

\bibitem{liu2024revisiting}
X.~Liu, Y.~Zhang, M.~Wu, M.~Yan, K.~He, W.~Yan, S.~Pan, X.~Ye, and D.~Fan, ``Revisiting edge perturbation for graph neural network in graph data augmentation and attack,'' \emph{arXiv preprint arXiv:2403.07943}, 2024.

\bibitem{guan2024topology}
F.~Guan, T.~Zhu, H.~Tong, and W.~Zhou, ``Topology modification against membership inference attack in graph neural networks,'' \emph{Knowledge-Based Systems}, vol. 305, p. 112642, 2024.

\bibitem{zheng2021resisting}
J.~Zheng, Y.~Cao, and H.~Wang, ``Resisting membership inference attacks through knowledge distillation,'' \emph{Neurocomputing}, vol. 452, pp. 114--126, 2021.

\bibitem{mazzone2022repeated}
F.~Mazzone, L.~van~den Heuvel, M.~Huber, C.~Verdecchia, M.~Everts, F.~Hahn, and A.~Peter, ``Repeated knowledge distillation with confidence masking to mitigate membership inference attacks,'' in \emph{Proceedings of the 15th ACM Workshop on Artificial Intelligence and Security}, 2022, pp. 13--24.

\bibitem{tang2022mitigating}
X.~Tang, S.~Mahloujifar, L.~Song, V.~Shejwalkar, M.~Nasr, A.~Houmansadr, and P.~Mittal, ``Mitigating membership inference attacks by $\{$Self-Distillation$\}$ through a novel ensemble architecture,'' in \emph{31st USENIX Security Symposium (USENIX Security 22)}, 2022, pp. 1433--1450.

\bibitem{chen2024maskarmor}
C.~Chen, X.~Zhang, H.~Qiu, J.~Lou, Z.~Liu, and X.~Chen, ``Maskarmor: Confidence masking-based defense mechanism for gnn against mia,'' \emph{Information Sciences}, vol. 669, p. 120579, 2024.

\bibitem{zhou2024model}
Z.~Zhou, J.~Zhu, F.~Yu, X.~Li, X.~Peng, T.~Liu, and B.~Han, ``Model inversion attacks: A survey of approaches and countermeasures,'' \emph{arXiv preprint arXiv:2411.10023}, 2024.

\bibitem{chen2022graph}
J.~Chen, G.~Huang, H.~Zheng, S.~Yu, W.~Jiang, and C.~Cui, ``Graph-fraudster: Adversarial attacks on graph neural network-based vertical federated learning,'' \emph{IEEE Transactions on Computational Social Systems}, vol.~10, no.~2, pp. 492--506, 2022.

\bibitem{liu2022membership}
Z.~Liu, X.~Zhang, C.~Chen, S.~Lin, and J.~Li, ``Membership inference attacks against robust graph neural network,'' in \emph{International Symposium on Cyberspace Safety and Security}, 2022, pp. 259--273.

\bibitem{dai2023unified}
E.~Dai, L.~Cui, Z.~Wang, X.~Tang, Y.~Wang, M.~Cheng, B.~Yin, and S.~Wang, ``A unified framework of graph information bottleneck for robustness and membership privacy,'' in \emph{Proceedings of the 29th ACM SIGKDD Conference on Knowledge Discovery and Data Mining}, 2023, pp. 368--379.

\bibitem{gong2021model}
X.~Gong, Q.~Wang, Y.~Chen, W.~Yang, and X.~Jiang, ``Model extraction attacks and defenses on cloud-based machine learning models,'' \emph{IEEE Communications Magazine}, vol.~58, no.~12, pp. 83--89, 2021.

\bibitem{tang2024modelguard}
M.~Tang, A.~Dai, L.~DiValentin, A.~Ding, A.~Hass, N.~Z. Gong, Y.~Chen \emph{et~al.}, ``$\{$ModelGuard$\}$:$\{$Information-Theoretic$\}$ defense against model extraction attacks,'' in \emph{33rd USENIX Security Symposium (USENIX Security 24)}, 2024, pp. 5305--5322.

\bibitem{chen2025towards}
H.~Chen, Y.~Dong, Z.~Wei, H.~Su, and J.~Zhu, ``Towards the worst-case robustness of large language models,'' \emph{arXiv preprint arXiv:2501.19040}, 2025.

\bibitem{yang2024black}
J.~Yang, R.~Ding, J.~Chen, X.~Zhong, H.~Zhao, and L.~Xie, ``Black-box attacks on graph neural networks via white-box methods with performance guarantees,'' \emph{IEEE Internet of Things Journal}, 2024.

\bibitem{zhu2021hermes}
Y.~Zhu, Y.~Cheng, H.~Zhou, and Y.~Lu, ``Hermes attack: Steal $\{$DNN$\}$ models with lossless inference accuracy,'' in \emph{30th USENIX Security Symposium (USENIX Security 21)}, 2021.

\bibitem{olatunji2023does}
I.~E. Olatunji, A.~Hizber, O.~Sihlovec, and M.~Khosla, ``Does black-box attribute inference attacks on graph neural networks constitute privacy risk?'' \emph{arXiv preprint arXiv:2306.00578}, 2023.

\bibitem{wang2024gcl}
X.~Wang and W.~H. Wang, ``Gcl-leak: Link membership inference attacks against graph contrastive learning,'' \emph{Proceedings on Privacy Enhancing Technologies}, 2024.

\bibitem{zhang2022inference}
Z.~Zhang, M.~Chen, M.~Backes, Y.~Shen, and Y.~Zhang, ``Inference attacks against graph neural networks,'' in \emph{31st USENIX Security Symposium (USENIX Security 22)}, 2022, pp. 4543--4560.

\bibitem{sinha2024gradient}
D.~A. Sinha, Y.~Liu, R.~Du, and Y.~Shen, ``Gradient inversion attack on graph neural networks,'' \emph{arXiv preprint arXiv:2411.19440}, 2024.

\bibitem{zhang2025unlearning}
J.~Zhang, Y.~Wang, Z.~Zhang, X.~Liu, and S.~Wang, ``Unlearning inversion attacks for graph neural networks,'' \emph{arXiv preprint arXiv:2506.00808}, 2025.

\bibitem{struppek2022}
L.~Struppek, D.~Hintersdorf, A.~D.~A. Correia, A.~Adler, and K.~Kersting, ``Plug \& play attacks: Towards robust and flexible model inversion attacks,'' \emph{arXiv preprint arXiv:2201.12179}, 2022.

\bibitem{gilbert1959random}
E.~N. Gilbert, ``Random graphs,'' \emph{The Annals of Mathematical Statistics}, vol.~30, no.~4, pp. 1141--1144, 1959.

\bibitem{10646777}
B.~Wu, X.~Yuan, S.~Wang, Q.~Li, M.~Xue, and S.~Pan, ``Securing graph neural networks in mlaas: A comprehensive realization of query-based integrity verification,'' in \emph{2024 IEEE Symposium on Security and Privacy (SP)}, 2024, pp. 2534--2552.

\bibitem{Xue2021CAP}
H.~Xue, K.~Zhou, T.~Chen, K.~Guo, X.~Hu, Y.~Chang, and X.~Wang, ``Cap: Co-adversarial perturbation on weights and features for improving generalization of graph neural networks,'' \emph{ArXiv}, 2021.

\bibitem{Xue2022Adversarial}
H.~Xue, X.~Wang, and Y.~Wang, ``Adversarial training on weights for graph neural networks,'' \emph{Proceedings of the 2022 5th International Conference on Algorithms, Computing and Artificial Intelligence}, 2022.

\bibitem{Gosch2023Adversarial}
L.~Gosch, S.~Geisler, D.~Sturm, B.~Charpentier, D.~Zugner, and S.~Gunnemann, ``Adversarial training for graph neural networks,'' \emph{ArXiv}, 2023.

\bibitem{zhao2025systematic}
K.~Zhao, L.~Li, K.~Ding, N.~Z. Gong, Y.~Zhao, and Y.~Dong, ``A systematic survey of model extraction attacks and defenses: State-of-the-art and perspectives,'' \emph{arXiv preprint arXiv:2508.15031}, 2025.

\bibitem{cheng2025misleader}
X.~Cheng, M.~Zheng, S.~Zhu, and Y.~Dong, ``Misleader: Defending against model extraction with ensembles of distilled models,'' \emph{arXiv preprint arXiv:2506.02362}, 2025.

\bibitem{podhajski2024efficient}
M.~Podhajski, J.~Dubi{\'n}ski, F.~Boenisch, A.~Dziedzic, A.~Pregowska, and T.~Michalak, ``Efficient model-stealing attacks against inductive graph neural networks,'' \emph{arXiv preprint arXiv:2405.12295}, 2024.

\bibitem{xu2025adage}
J.~Xu, F.~Boenisch, and A.~Dziedzic, ``Adage: Active defenses against gnn extraction,'' \emph{arXiv preprint arXiv:2503.00065}, 2025.

\bibitem{bhaila2024local}
K.~Bhaila, W.~Huang, Y.~Wu, and X.~Wu, ``Local differential privacy in graph neural networks: a reconstruction approach,'' in \emph{Proceedings of the 2024 SIAM International Conference on Data Mining (SDM)}, 2024, pp. 1--9.

\bibitem{dwork2014algorithmic}
C.~Dwork, A.~Roth \emph{et~al.}, ``The algorithmic foundations of differential privacy,'' \emph{Foundations and Trends{\textregistered} in Theoretical Computer Science}, vol.~9, no. 3--4, pp. 211--407, 2014.

\bibitem{chen2020vertically}
C.~Chen, J.~Zhou, L.~Zheng, H.~Wu, L.~Lyu, J.~Wu, B.~Wu, Z.~Liu, L.~Wang, and X.~Zheng, ``Vertically federated graph neural network for privacy-preserving node classification,'' \emph{arXiv preprint arXiv:2005.11903}, 2020.

\bibitem{kumarasinghe2022}
U.~Kumarasinghe, M.~Nabeel, K.~De~Zoysa, K.~Gunawardana, and C.~Elvitigala, ``Heteroguard: Defending heterogeneous graph neural networks against adversarial attacks,'' in \emph{2022 IEEE International Conference on Data Mining Workshops (ICDMW)}, 2022, pp. 698--705.

\bibitem{song2024two}
C.~Song, L.~Niu, and M.~Lei, ``Two-level adversarial attacks for graph neural networks,'' \emph{Information Sciences}, vol. 654, p. 119877, 2024.

\bibitem{hasegawa2023membership}
K.~Hasegawa, K.~Yamashita, S.~Hidano, K.~Fukushima, K.~Hashimoto, and N.~Togawa, ``Membership inference attacks against gnn-based hardware trojan detection,'' in \emph{2023 IEEE 22nd International Conference on Trust, Security and Privacy in Computing and Communications (TrustCom)}, 2023, pp. 1222--1229.

\bibitem{tian2025knowledge}
Y.~Tian, S.~Pei, X.~Zhang, C.~Zhang, and N.~V. Chawla, ``Knowledge distillation on graphs: A survey,'' \emph{ACM Computing Surveys}, vol.~57, no.~8, pp. 1--16, 2025.

\bibitem{dibbo2023sok}
S.~V. Dibbo, ``Sok: Model inversion attack landscape: Taxonomy, challenges, and future roadmap,'' in \emph{2023 IEEE 36th Computer Security Foundations Symposium (CSF)}, 2023, pp. 439--456.

\bibitem{duddu2020quantifying}
V.~Duddu, A.~Boutet, and V.~Shejwalkar, ``Quantifying privacy leakage in graph embedding,'' in \emph{MobiQuitous 2020-17th EAI International Conference on Mobile and Ubiquitous Systems: Computing, Networking and Services}, 2020, pp. 76--85.

\bibitem{cong2022sslguard}
T.~Cong, X.~He, and Y.~Zhang, ``Sslguard: A watermarking scheme for self-supervised learning pre-trained encoders,'' in \emph{Proceedings of the 2022 ACM SIGSAC Conference on Computer and Communications Security}, 2022, pp. 579--593.

\bibitem{ding2022data}
K.~Ding, Z.~Xu, H.~Tong, and H.~Liu, ``Data augmentation for deep graph learning: A survey,'' \emph{ACM SIGKDD Explorations Newsletter}, vol.~24, no.~2, pp. 61--77, 2022.

\bibitem{li2022graph}
M.~M. Li, K.~Huang, and M.~Zitnik, ``Graph representation learning in biomedicine and healthcare,'' \emph{Nature Biomedical Engineering}, vol.~6, no.~12, pp. 1353--1369, 2022.

\bibitem{mernyei2020wiki}
P.~Mernyei and C.~Cangea, ``Wiki-cs: A wikipedia-based benchmark for graph neural networks,'' \emph{arXiv preprint arXiv:2007.02901}, 2020.

\bibitem{liu2022federated}
Z.~Liu, L.~Yang, Z.~Fan, H.~Peng, and P.~S. Yu, ``Federated social recommendation with graph neural network,'' \emph{ACM Transactions on Intelligent Systems and Technology (TIST)}, vol.~13, no.~4, pp. 1--24, 2022.

\bibitem{tao2022revisiting}
Y.~Tao, Y.~Li, S.~Zhang, Z.~Hou, and Z.~Wu, ``Revisiting graph based social recommendation: A distillation enhanced social graph network,'' in \emph{Proceedings of the ACM Web Conference 2022}, 2022, pp. 2830--2838.

\bibitem{adamic2005political}
L.~A. Adamic and N.~Glance, ``The political blogosphere and the 2004 us election: divided they blog,'' in \emph{Proceedings of the 3rd international workshop on Link discovery}, 2005, pp. 36--43.

\bibitem{xu2023openp5}
S.~Xu, W.~Hua, and Y.~Zhang, ``Openp5: Benchmarking foundation models for recommendation,'' \emph{arXiv preprint arXiv:2306.11134}, 2023.

\bibitem{chen2022understanding}
Y.~Chen, H.~Yang, Y.~Zhang, K.~Ma, T.~Liu, B.~Han, and J.~Cheng, ``Understanding and improving graph injection attack by promoting unnoticeability,'' \emph{arXiv preprint arXiv:2202.08057}, 2022.

\bibitem{yang2021consisrec}
L.~Yang, Z.~Liu, Y.~Dou, J.~Ma, and P.~S. Yu, ``Consisrec: Enhancing gnn for social recommendation via consistent neighbor aggregation,'' in \emph{Proceedings of the 44th international ACM SIGIR conference on Research and development in information retrieval}, 2021, pp. 2141--2145.

\bibitem{debnath1991structure}
A.~K. Debnath, R.~L. Lopez~de Compadre, G.~Debnath, A.~J. Shusterman, and C.~Hansch, ``Structure-activity relationship of mutagenic aromatic and heteroaromatic nitro compounds. correlation with molecular orbital energies and hydrophobicity,'' \emph{Journal of medicinal chemistry}, vol.~34, no.~2, pp. 786--797, 1991.

\bibitem{toivonen2003statistical}
H.~Toivonen, A.~Srinivasan, R.~D. King, S.~Kramer, and C.~Helma, ``Statistical evaluation of the predictive toxicology challenge 2000--2001,'' \emph{Bioinformatics}, vol.~19, no.~10, pp. 1183--1193, 2003.

\bibitem{wale2008comparison}
N.~Wale, I.~A. Watson, and G.~Karypis, ``Comparison of descriptor spaces for chemical compound retrieval and classification,'' \emph{Knowledge and Information Systems}, vol.~14, pp. 347--375, 2008.

\bibitem{riesen2008iam}
K.~Riesen and H.~Bunke, ``Iam graph database repository for graph based pattern recognition and machine learning,'' in \emph{Structural, Syntactic, and Statistical Pattern Recognition: Joint IAPR International Workshop, SSPR \& SPR 2008, Orlando, USA, December 4-6, 2008. Proceedings}, 2008, pp. 287--297.

\bibitem{borgwardt2005protein}
K.~M. Borgwardt, C.~S. Ong, S.~Sch{\"o}nauer, S.~Vishwanathan, A.~J. Smola, and H.-P. Kriegel, ``Protein function prediction via graph kernels,'' \emph{Bioinformatics}, vol.~21, pp. i47--i56, 2005.

\bibitem{agrawal2018large}
M.~Agrawal, M.~Zitnik, and J.~Leskovec, ``Large-scale analysis of disease pathways in the human interactome,'' in \emph{PACIFIC SYMPOSIUM on BIOCOMPUTING 2018: Proceedings of the Pacific Symposium}, 2018, pp. 111--122.

\bibitem{kumar2018rev2}
S.~Kumar, B.~Hooi, D.~Makhija, M.~Kumar, C.~Faloutsos, and V.~Subrahmanian, ``Rev2: Fraudulent user prediction in rating platforms,'' in \emph{Proceedings of the Eleventh ACM International Conference on Web Search and Data Mining}, 2018, pp. 333--341.

\bibitem{asghar2016yelp}
N.~Asghar, ``Yelp dataset challenge: Review rating prediction,'' \emph{arXiv preprint arXiv:1605.05362}, 2016.

\bibitem{chin2022datasets}
J.~Y. Chin, Y.~Chen, and G.~Cong, ``The datasets dilemma: How much do we really know about recommendation datasets?'' in \emph{Proceedings of the Fifteenth ACM International Conference on Web Search and Data Mining}, 2022, pp. 141--149.

\bibitem{you2022roland}
J.~You, T.~Du, and J.~Leskovec, ``Roland: graph learning framework for dynamic graphs,'' in \emph{Proceedings of the 28th ACM SIGKDD conference on knowledge discovery and data mining}, 2022, pp. 2358--2366.

\bibitem{ribeiro2017struc2vec}
L.~F. Ribeiro, P.~H. Saverese, and D.~R. Figueiredo, ``struc2vec: Learning node representations from structural identity,'' in \emph{Proceedings of the 23rd ACM SIGKDD international conference on knowledge discovery and data mining}, 2017, pp. 385--394.

\bibitem{wei2021pooling}
L.~Wei, H.~Zhao, Q.~Yao, and Z.~He, ``Pooling architecture search for graph classification,'' in \emph{Proceedings of the 30th ACM International Conference on Information \& Knowledge Management}, 2021, pp. 2091--2100.

\bibitem{wu2022nodeformer}
Q.~Wu, W.~Zhao, Z.~Li, D.~P. Wipf, and J.~Yan, ``Nodeformer: A scalable graph structure learning transformer for node classification,'' \emph{Advances in Neural Information Processing Systems}, vol.~35, pp. 27\,387--27\,401, 2022.

\bibitem{reau2023deeprank}
M.~R{\'e}au, N.~Renaud, L.~C. Xue, and A.~M. Bonvin, ``Deeprank-gnn: a graph neural network framework to learn patterns in protein--protein interfaces,'' \emph{Bioinformatics}, vol.~39, no.~1, p. btac759, 2023.

\bibitem{guo2020deep}
Z.~Guo and H.~Wang, ``A deep graph neural network-based mechanism for social recommendations,'' \emph{IEEE Transactions on Industrial Informatics}, vol.~17, no.~4, pp. 2776--2783, 2020.

\bibitem{hu2023towards}
W.~Hu and H.~Fang, ``Towards differential privacy in sequential recommendation: A noisy graph neural network approach,'' \emph{ACM Transactions on Knowledge Discovery from Data}, 2023.

\bibitem{islam2020comparative}
M.~K. Islam, S.~Aridhi, and M.~Smail-Tabbone, ``A comparative study of similarity-based and gnn-based link prediction approaches,'' \emph{arXiv preprint arXiv:2008.08879}, 2020.

\bibitem{guo2022mixed}
Z.~Guo, K.~Yu, A.~Jolfaei, G.~Li, F.~Ding, and A.~Beheshti, ``Mixed graph neural network-based fake news detection for sustainable vehicular social networks,'' \emph{IEEE Transactions on Intelligent Transportation Systems}, vol.~24, no.~12, pp. 15\,486--15\,498, 2022.

\bibitem{lane2013milliseconds}
T.~J. Lane, D.~Shukla, K.~A. Beauchamp, and V.~S. Pande, ``To milliseconds and beyond: challenges in the simulation of protein folding,'' \emph{Current opinion in structural biology}, vol.~23, no.~1, pp. 58--65, 2013.

\bibitem{zitnik2018modeling}
M.~Zitnik, M.~Agrawal, and J.~Leskovec, ``Modeling polypharmacy side effects with graph convolutional networks,'' \emph{Bioinformatics}, vol.~34, no.~13, pp. i457--i466, 2018.

\bibitem{bongini2022biognn}
P.~Bongini, N.~Pancino, F.~Scarselli, and M.~Bianchini, ``Biognn: how graph neural networks can solve biological problems,'' in \emph{Artificial Intelligence and Machine Learning for Healthcare: Vol. 1: Image and Data Analytics}.\hskip 1em plus 0.5em minus 0.4em\relax Springer, 2022, pp. 211--231.

\bibitem{basaad2024bert}
A.~Basaad, S.~Basurra, E.~Vakaj, A.~K. Eldaly, and M.~M. Abdelsamea, ``A bert-gnn approach for metastatic breast cancer prediction using histopathology reports,'' \emph{Diagnostics}, vol.~14, no.~13, p. 1365, 2024.

\bibitem{li2023survey}
X.~Li, L.~Sun, M.~Ling, and Y.~Peng, ``A survey of graph neural network based recommendation in social networks,'' \emph{Neurocomputing}, vol. 549, p. 126441, 2023.

\bibitem{wu2022graph}
S.~Wu, F.~Sun, W.~Zhang, X.~Xie, and B.~Cui, ``Graph neural networks in recommender systems: a survey,'' \emph{ACM Computing Surveys}, vol.~55, no.~5, pp. 1--37, 2022.

\bibitem{wang2023financial}
D.~Wang, Z.~Zhang, Y.~Zhao, K.~Huang, Y.~Kang, and J.~Zhou, ``Financial default prediction via motif-preserving graph neural network with curriculum learning,'' in \emph{Proceedings of the 29th ACM SIGKDD Conference on Knowledge Discovery and Data Mining}, 2023, pp. 2233--2242.

\bibitem{liu2018heterogeneous}
Z.~Liu, C.~Chen, X.~Yang, J.~Zhou, X.~Li, and L.~Song, ``Heterogeneous graph neural networks for malicious account detection,'' in \emph{Proceedings of the 27th ACM international conference on information and knowledge management}, 2018, pp. 2077--2085.

\bibitem{yang2019using}
Y.~Yang, Z.~Wei, Q.~Chen, and L.~Wu, ``Using external knowledge for financial event prediction based on graph neural networks,'' in \emph{Proceedings of the 28th ACM international conference on information and knowledge management}, 2019, pp. 2161--2164.

\bibitem{xu2024idea}
C.~Xu, Q.~Cui, J.~Dong, W.~He, and C.-H. Chang, ``Idea: An inverse domain expert adaptation based active dnn ip protection method,'' \emph{arXiv preprint arXiv:2410.00059}, 2024.

\bibitem{hubscher2022graph}
G.~H{\"u}bscher, V.~Geist, D.~Auer, A.~Ekelhart, R.~Mayer, S.~Nadschl{\"a}ger, and J.~K{\"u}ng, ``Graph-based managing and mining of processes and data in the domain of intellectual property,'' \emph{Information Systems}, vol. 106, p. 101844, 2022.

\bibitem{ma2023graphnei}
Z.~Ma, S.~Zhang, N.~Li, T.~Li, X.~Hu, H.~Feng, Q.~Zhou, F.~Liu, X.~Quan, H.~Wang \emph{et~al.}, ``Graphnei: A gnn-based network entity identification method for ip geolocation,'' \emph{Computer Networks}, vol. 235, p. 109946, 2023.

\bibitem{10.14778/3659437.3659457}
H.~Li, S.~Di, C.~H.~Y. Li, L.~Chen, and X.~Zhou, ``Fight fire with fire: Towards robust graph neural networks on dynamic graphs via actively defense,'' \emph{Proc. VLDB Endow.}, vol.~17, no.~8, p. 2050–2063, 2024.

\bibitem{yasaei2021gnn4ip}
R.~Yasaei, S.-Y. Yu, E.~K. Naeini, and M.~A. Al~Faruque, ``Gnn4ip: Graph neural network for hardware intellectual property piracy detection,'' in \emph{2021 58th ACM/IEEE Design Automation Conference (DAC)}, 2021, pp. 217--222.

\bibitem{wu2022survey}
B.~Wu, J.~Li, J.~Yu, Y.~Bian, H.~Zhang, C.~Chen, C.~Hou, G.~Fu, L.~Chen, T.~Xu \emph{et~al.}, ``A survey of trustworthy graph learning: Reliability, explainability, and privacy protection,'' \emph{arXiv preprint arXiv:2205.10014}, 2022.

\bibitem{ganz2021explaining}
T.~Ganz, M.~H{\"a}rterich, A.~Warnecke, and K.~Rieck, ``Explaining graph neural networks for vulnerability discovery,'' in \emph{Proceedings of the 14th ACM Workshop on Artificial Intelligence and Security}, 2021, pp. 145--156.

\bibitem{platonov2023critical}
O.~Platonov, D.~Kuznedelev, M.~Diskin, A.~Babenko, and L.~Prokhorenkova, ``A critical look at the evaluation of gnns under heterophily: Are we really making progress?'' \emph{arXiv preprint arXiv:2302.11640}, 2023.

\bibitem{xue2023turn}
M.~Xue, L.~Y. Zhang, Y.~Zhang, and W.~Liu, ``Turn passive to active: A survey on active intellectual property protection of deep learning models,'' \emph{arXiv preprint arXiv:2310.09822}, 2023.

\bibitem{galkin2023towards}
M.~Galkin, X.~Yuan, H.~Mostafa, J.~Tang, and Z.~Zhu, ``Towards foundation models for knowledge graph reasoning,'' \emph{arXiv preprint arXiv:2310.04562}, 2023.

\bibitem{ye2023natural}
R.~Ye, C.~Zhang, R.~Wang, S.~Xu, Y.~Zhang \emph{et~al.}, ``Natural language is all a graph needs,'' \emph{arXiv preprint arXiv:2308.07134}, vol.~4, no.~5, p.~7, 2023.

\bibitem{deng2024advances}
S.~Deng, M.~de~Rijke, and Y.~Ning, ``Advances in human event modeling: From graph neural networks to language models,'' in \emph{Proceedings of the 30th ACM SIGKDD Conference on Knowledge Discovery and Data Mining}, 2024, pp. 6459--6469.

\bibitem{yu2024cosmo}
C.~Yu, X.~Liu, J.~Maia, Y.~Li, T.~Cao, Y.~Gao, Y.~Song, R.~Goutam, H.~Zhang, B.~Yin \emph{et~al.}, ``Cosmo: A large-scale e-commerce common sense knowledge generation and serving system at amazon,'' in \emph{Companion of the 2024 International Conference on Management of Data}, 2024, pp. 148--160.

\bibitem{xiao2024fuselinker}
Y.~Xiao, S.~Zhang, H.~Zhou, M.~Li, H.~Yang, and R.~Zhang, ``Fuselinker: Leveraging llm’s pre-trained text embeddings and domain knowledge to enhance gnn-based link prediction on biomedical knowledge graphs,'' \emph{Journal of Biomedical Informatics}, vol. 158, p. 104730, 2024.

\bibitem{yuan2024gnnavi}
S.~Yuan, E.~Nie, M.~F{\"a}rber, H.~Schmid, and H.~Sch{\"u}tze, ``Gnnavi: Navigating the information flow in large language models by graph neural network,'' \emph{arXiv preprint arXiv:2402.11709}, 2024.

\bibitem{wang2023empower}
J.~Wang, M.~Luo, J.~Li, Y.~Lin, Y.~Dong, J.~S. Dong, and Q.~Zheng, ``Empower post-hoc graph explanations with information bottleneck: A pre-training and fine-tuning perspective,'' in \emph{Proceedings of the 29th ACM SIGKDD Conference on Knowledge Discovery and Data Mining}, 2023, pp. 2349--2360.

\bibitem{liu2024git}
P.~Liu, Y.~Ren, J.~Tao, and Z.~Ren, ``Git-mol: A multi-modal large language model for molecular science with graph, image, and text,'' \emph{Computers in biology and medicine}, vol. 171, p. 108073, 2024.

\bibitem{zhang2023adversarial}
B.~Zhang, Y.~Dong, C.~Chen, Y.~Zhu, M.~Luo, and J.~Li, ``Adversarial attacks on fairness of graph neural networks,'' \emph{arXiv preprint arXiv:2310.13822}, 2023.

\end{thebibliography}



\clearpage

\appendices
\section{Detailed Statistics and Descriptions of Benchmark Datasets} \label{appendA}

\subsection{Citation Datasets} \label{appendA1}
\textbf{Cora}\cite{duddu2020quantifying}. Cora dataset includes 2,708 academic papers categorized into seven distinct classes. The citation network is composed of 5,429 citation links between these papers. Each paper is represented by a binary word vector, where each element signifies the presence or absence of a particular word from a dictionary containing 1,433 unique words. 

\noindent \textbf{Citeseer}\cite{hu2022membership}. The CiteSeer dataset comprises 3,312 scientific papers divided into six categories. The citation network includes 4,732 citation links among these papers. Each document is represented by a binary word vector, where each entry denotes the presence or absence of a specific word from a dictionary containing 3,703 distinct words. 

\noindent \textbf{DBLP}\cite{cong2022sslguard}. DBLP dataset is extensively used as a benchmark to evaluate the performance of Graph Neural Networks due to its detailed representation of academic citation dynamics. DBLP comprises 17,716 nodes, each representing a unique publication. The network includes 105,734 edges, illustrating the citation links between these publications. Each node is characterized by 1,639 attributes. DBLP categorizes the publications into four distinct classes, adding another layer of complexity for classification tasks. 

\noindent \textbf{ogbn-arxiv}\cite{ding2022data}. ogbn-arxiv is valuable due to its extensive coverage and rich metadata, encompassing a wide array of computer science disciplines. Each node (paper) in the graph is accompanied by a variety of features, including paper abstracts. The edges between nodes signify citation relationships. ogbn-arxiv consists of 169,343 nodes and 1,166,243 directed edges. The dataset is also categorized into 40 subject areas, providing a multi-class classification challenge for GML models.

\noindent \textbf{PubMed}\cite{li2022graph}. It contains 19,717 academic papers from the PubMed database, all related to diabetes and categorized into three distinct classes. The citation network includes 44,338 citation links connecting these papers. Each paper is represented by a TF/IDF weighted word vector, with the dictionary comprising 500 unique words. 

\noindent \textbf{Wiki-CS}\cite{mernyei2020wiki}. Wiki-CS is a widely used benchmark in citation network analysis. It encompasses 11,701 nodes representing different articles, which are interconnected by 216,123 edges that denote the citation relationships among these articles. The dataset is categorized into 10 distinct classes, providing a rich and varied classification challenge. Each node is characterized by a 300-dimensional attribute vector, capturing various attributes of the articles.

\subsection{Social Network Datasets} \label{appendA2}

\textbf{Facebook}\cite{liu2022federated}. The dataset includes 4,039 nodes, each representing a unique user account on the social network, connected by 88,234 edges. Each user node possesses various attributes such as gender, education, job position, hobbies, location, etc. User information has been anonymized through pseudonymization, and the feature interpretations have been obscured (e.g., the attributes 'male' and 'female' were replaced with 'gender 1' and 'gender 2', respectively).

\noindent \textbf{Flickr}\cite{tao2022revisiting}. Flickr is a popular dataset for social recommendation tasks. This dataset captures social interactions among users on the Flickr platform, where nodes represent users and edges denote the social connections between them. Specifically, the dataset comprises 8,252 nodes and 327,815 edges, forming a complex network of user interactions. It includes 8,358 graphs and is categorized into 81 distinct classes, providing a comprehensive data for evaluating GNN-based recommendation. 

\noindent \textbf{PolBlogs}\cite{adamic2005political}. PolBlogs is a network dataset that captures the hyperlink connections between blogs discussing US politics during the period leading up to the 2004 presidential election. This dataset includes 1,490 blogs labeled as either right-leaning or left-leaning, with a total of 19,090 directed hyperlinks. PolBlogs is notable for its clear division between the two political camps, making it an excellent resource for studying information flow within ideologically homogeneous groups. 

\noindent \textbf{LastFM}\cite{xu2023openp5}. LastFM gathered in March 2020 using the social network's public API, specifically focusing on users from Asian countries. It comprises 7,624 nodes interconnected by 27,806 edges based on mutual follower relationships. Each user node includes attributes such as preferred music genres, favorite artists, geographical location, etc.

\noindent \textbf{Reddit}\cite{chen2022understanding}. Reddit consists of networks formed by hyperlinks present in the titles and bodies of Reddit posts. Each hyperlink creates a directed edge between two sub-reddits, capturing the interaction and information flow within Reddit community. The dataset includes over 50,000 nodes representing individual post and over 860,000 edges, reflecting the dynamic interactions on the platform. It is particularly useful for studying community interaction, conflict, and the spread of information across different topics.

\noindent \textbf{Twitter}\cite{yang2021consisrec,guo2020deep}. Twitter dataset is a compact user-relation network derived from Twitter social media platform. In the dataset, nodes represent individual user profiles, while edges signify the connections between users. The dataset encompasses node features that include user profile information, and it contains detailed data on circles and ego networks, which reflect the sub-communities and direct connections of users, respectively.

\subsection{Molecular \& Protein Networks Datasets} \label{appendA3}

\textbf{MUTAG}\cite{debnath1991structure}. MUTAG comprises 188 mutagenic aromatic and heteroaromatic nitro compounds, each labeled based on whether they exhibit a mutagenic effect on the Gram-negative bacterium Salmonella typhimurium. The dataset is widely utilized in the fields of chemical and molecular graph learning. The compounds are represented as graphs where nodes denote atoms, and edges represent chemical bonds. Key features of this dataset include molecular structures and properties that correlate with mutagenicity, such as molecular orbital energies and hydrophobicity. 

\noindent \textbf{PTC}\cite{toivonen2003statistical}. Predictive Toxicology Challenge (PTC) dataset, is a well-known benchmark in the field of bioinformatics. PTC comprises 344 chemical compounds labeled according to their carcinogenicity for male and female rats. It includes four subsets: PTC-MR, PTC-FR, PTC-MM, and PTC-FM, each focusing on different biological contexts. In PTC dataset, each compound is represented as a graph where nodes correspond to atoms, and edges represent chemical bonds. This dataset is valuable for evaluating the performance of graph-based models in toxicity prediction. 

\noindent \textbf{NCT1 \& NCI109}\cite{wale2008comparison}. NCI1 and NCI109 are subsets of NCI series. The two differ primarily in the number of compounds they include. NCI1 has 4,110 compounds, whereas NCI109 has slightly more with 4,127 compounds. These chemical compounds screened for their ability to inhibit the growth of human tumor cell lines. Each compound is represented as a graph where nodes correspond to atoms and edges represent chemical bonds. NCI1 and NCI109 provide extensive molecular descriptors and structural features, making them valuable for chemical compound retrieval and predicting biological activities.

\noindent \textbf{AIDS}\cite{riesen2008iam}. AIDS dataset is constructed from the AIDS Antiviral Screen Database of Active Compounds. It includes molecular structures specifically screened for activity against the human immunodeficiency virus (HIV). It comprises 43,467 compounds, each represented as a graph where nodes correspond to atoms and edges represent chemical bonds. The dataset includes detailed features such as atom types, bond types, and molecular descriptors.

\noindent \textbf{ENZYMES}\cite{zhang2022model}. ENZYMES dataset is derived from the BRENDA enzyme database and includes protein tertiary structures. The dataset consists of 600 protein structures, categorized into six different enzyme classes, with each class containing 100 structures. Each protein is represented as a graph where nodes correspond to secondary structure elements (SSEs) and edges represent spatial or sequential proximities between these elements. The dataset includes detailed features such as the types of SSEs, the lengths of SSEs, and spatial distances between them. ENZYMES is valuable for benchmarking graph-based models in protein classification tasks.

\noindent \textbf{PROTEINS} \cite{borgwardt2005protein}. PROTEINS dataset is extensively used in graph learning for bioinformatics. PROTEINS comprises protein structures classified into two categories: enzymes and non-enzymes. Each protein is represented as a graph where nodes correspond to amino acids, and edges are formed between nodes if the corresponding amino acids are within 6 Angstroms of each other in the 3D space. The dataset includes 1,113 protein graphs, with detailed features for each node, such as amino acid types and their properties. 

\noindent \textbf{Disease Pathway (DP)}\cite{agrawal2018large}. DP describes a system of interacting human proteins whose malfunction collectively leads to a variety of diseases. Nodes in the network represent human proteins and edges indicate protein-protein interactions. The task is to predict for every protein node what diseases (i.e., labels/classes) that protein might cause. The dataset has 73-dimensional continuous node features representing graphlet-orbit counts (i.e., the number of occurrences of higher-order network motifs).

\subsection{E-Commerce Platform Datasets} \label{appendA4}

\textbf{Computers}\cite{kumar2018rev2}. Computers is an Amazon co-purchase graph specifically focusing on computer-related merchandise. In the dataset, nodes represent individual products, and edges indicate that two products are frequently purchased together, forming a co-purchase relationship. The dataset includes detailed node features extracted from product reviews, such as review text, ratings, and other metadata. Additionally, each product is labeled with its corresponding category. The dataset contains 13,752 nodes and 287,209 edges, making it a rich resource for studying the relationships between different merchandise. 

\noindent \textbf{Photo}\cite{kumar2018rev2}. Photo is another graph learning dataset derived from Amazon's co-purchase data, specifically focusing on photography-related products. In this dataset, nodes represent individual products, such as cameras, lenses, and accessories, while edges indicate that two products are frequently purchased together, forming co-purchase relationships. The Photo dataset comprises 7,487 nodes and 119,043 edges. Each product is labeled with its respective category, which aids in classification tasks.

\noindent \textbf{Yelp}\cite{asghar2016yelp}. Originating from the Yelp Dataset Challenge, this dataset includes rich information about local businesses, user reviews, and ratings. In the graph representation, nodes denote users and businesses, while edges denote interactions such as reviews and ratings. Yelp dataset includes over 1.6 million reviews and 70,000 businesses. It also contains detailed features for both users and businesses, including review text, star ratings, business categories, and user metadata.

\noindent \textbf{ML-1M}\cite{chin2022datasets}. ML-1M is a widely used dataset in graph learning for recommendation systems. The dataset originates from the MovieLens project and includes 1 million ratings of 3,900 movies made by about 6,000 users. In the graph representation, nodes represent users and movies, and edges indicate the ratings given by users to movies, forming a user-item interaction graph. 

\noindent \textbf{Bitcoin-Alpha}\cite{you2022roland}. Bitcoin-Alpha is a prominent dataset retrieved from online trading platforms. The dataset captures the "who-trusts-whom" relationships among users trading on the Bitcoin-Alpha platform. In this network, nodes represent users, and edges indicate the trust ratings given by one user to another, forming a trust graph. The dataset includes 3,783 nodes and 24,186 edges, providing a comprehensive view of the trust dynamics within the Bitcoin-Alpha trading community. 

\subsection{Datasets of Other Domain Applications} \label{appendA5}

\textbf{Brazil}\cite{ribeiro2017struc2vec}. Brazil Airport Traffic Network dataset is commonly used for studying traffic networks. The dataset consists of 131 nodes, each representing an airport, and 1,038 edges, each representing a flight between two airports. The label of each airport denotes the level of the airport. The dataset captures the structural information of the airport network, illustrating the traffic flow between different airports in Brazil.

\noindent \textbf{USA}\cite{ribeiro2017struc2vec}. USA Airport Traffic Network dataset, similar to the Brazil dataset, provides a detailed representation of air traffic between airports within the United States. The dataset includes 1,190 nodes, each representing an airport, and 13,599 edges, each representing a flight route between two airports. Each airport is labeled according to its level, which provides additional context. The USA dataset, being denser than many other datasets, is ideal for studying network optimization, route planning, and traffic management.

\noindent \textbf{COLLAB}\cite{xu2023watermarking}. COLLAB dataset is derived from scientific collaboration networks, where each graph represents the collaboration relationships of a single researcher. COLLAB includes a total of 5,000 graphs, with each graph containing an average of 74 nodes and 245 edges, reflecting the typical collaboration patterns among researchers. The nodes represent individual researchers, and the edges indicate co-authorship relationships. 

\noindent \textbf{IMDB-BINARY}\cite{wei2021pooling}. IMDB-BINARY dataset consists of ego-networks derived from actor collaborations in movies, where each graph represents the collaboration network of actors who have co-appeared in movies. The primary task associated with this dataset is to predict the genre of the movie, specifically distinguishing between Action and Romance genres. IMDB-BINARY includes 1,000 graphs, evenly split between the two genres. Nodes represent actors, and edges indicate co-appearance in the same movie.

\section{General IP Protection Performance Evaluation Criteria} \label{appendB}

We summarize the general performance evaluation criteria of Graph Learning IP Protection, which can be categorized accordingly into three aspects, as shown in Table~\ref{tab2}. It is noticeable that the evaluation metrics of a specific IP protection task vary on a case-by-case basis, depending on the nature of the problem and the method employed, yet they fall within the three aspects outlined in Table~\ref{tab2}.

\begin{table*}[!htbp]
\centering
\footnotesize
\caption{Performance Evaluation Metrics of Graph Learning IP Protection and Descriptions.}
\label{tab2}
\begin{tabular}{c|c|p{8cm}}
\toprule
\multicolumn{1}{c}{\textbf{Type}} & \multicolumn{1}{c}{\textbf{Criteria}} & \multicolumn{1}{c}{\textbf{Description}} \\ \midrule
 & Model Accuracy & The IP protection method should preserve the prediction accuracy of the GML model on its original tasks. \\ \cline{2-3} 
Model Fidelity& Model Efficiency & It evaluates the impact of IP construction on the efficiency of GML model inference and training, including introducing extra IP components, optimization objectives, training data, and training processes. \\ \cline{2-3} 
 & Adversarial Robustness & The robustness of a GML model against adversarial attacks should not be greatly affected by constructing IP identifiers. \\ \cline{2-3} 
& No Potential Risks & IP construction should not induce unexpected mispredictions by which attackers can manipulate predictions deterministically. \\ \hline
& IP Robustness & IP identifiers should be robust against legal model modifications (e.g., graph simplification, graph compression, transfer learning) and various malicious attacks (e.g., IP detection, evasion, removal, ambiguity, etc.). \\ \cline{2-3} 
& IP Capacity & IP Capacity indicates the bit number of IP identifiers' valid information payload and the theoretical upper bound. \\ \cline{2-3} 
Quality Of IP (QoI) & Imperceptibility & Attackers cannot discover the IP identifiers hidden in protected GML models. \\ \cline{2-3} 
& Unforgeability & Even if an attacker knows the constructed IP messages, he cannot convince a third party that he owns the model IP by creating fake secret keys. \\ \cline{2-3} 
& Anti-overwriting & An attacker cannot reconstruct an IP identifier to claim the model IP, even if he knows the protection algorithm. \\ \hline
& Scalability & Graph learning IP schemes should be adapted to various scenarios and consider diverse IP functions. It can be employed in downstream-task versions of the original pre-trained model, covering various learning paradigms. \\ \cline{2-3} 
Efficiency Of IP (EoI)& Agility of Construction & IP defenders can construct and verify IP identifiers with minimal overhead, enabling a defender to easily create his own IP identifier. \\ \cline{2-3} 
& Slowness of Removal & IP attackers have to remove or forge IP identifiers at a much higher cost and performance loss, to degrade attackers’ incentives to crack protected models. \\ 
\bottomrule
\end{tabular}
\end{table*}

\noindent \textbf{1.Model Fidelity}: IP identifiers are usually constructed by modifying the original model, such as fine-tuning model parameters or adding some extra watermark components to the model. One basic requirement is to ensure that the protected model and the original model have close performance. Fidelity is the aspect used to measure the performance similarity between the protected and the original models, including Including Model Accuracy, Model Efficiency, Adversarial Robustness, No Potential Risks.

\noindent \textbf{2.Quality-of-IP (QoI)}: QoI means that IP owners/verifiers have enough information and confidence to claim the model IP. Including but not limited to: IP Robustness, IP Capacity, Imperceptibility, Unforgeability, and Anti-Overwriting.

\noindent \textbf{3.Efficiency-of-IP (EoI):} The IP construction and verification in Graph Learning IP Protection should be efficient for IP defenders but inefficient for IP attackers, which mainly includes Scalability, Agility of Construction, and Slowness of Removal.

\end{document}